# Denotational Results in Compiler Correctness and Comparisons with Operational Approaches

Compiling lexically scoped functions with recursive types through closure conversion

JEREMY G. SIEK, Indiana University, USA
MATTHEW HEIMERDINGER, Indiana University, USA
TIANYU CHEN, Indiana University, USA
DARSHAL SHETTY, Indiana University, USA

Anyone setting out to prove a compiler correct must choose a style of semantics. The CakeML and CompCert projects, among others, show that operational semantics such as abstract machines and interpreters are viable choices. Their correctness proofs reason about program execution using simulations and induction on evaluation. In this paper we explore whether denotational semantics can also be used in proofs of compiler correctness and evaluate the tradeoffs. We focus on a compiler from FPC, a language with lexically scoped functions and recursive types, to a first-order low-level intermediate language. This compiler includes closure conversion, but no prior proof of correctness with respect to denotational semantics covers closure conversion for a Turing-complete language. Compiler correctness requires that the denotation of a program be both preserved and reflected by the compiler. Without reflection, a compiler would be allowed to translate a diverging program into one that halts. For closure conversion, reflection is the hard direction. We present the first denotational proofs of semantic reflection for closure conversion applied to a Turing-complete language, one proof with respect to a graph model and one with respect to Scott domains. For the graph-model proof we develop a new logical relation, the *approximant logical relation*, and for the Scott-domain proof we adapt a logical relation of Reynolds.

As a first step in evaluating the tradeoffs, we compare mechanized correctness proofs of an eight-pass compiler, from FPC, a typed $\lambda$-calculus with recursive types, down to a typed low-level intermediate language, in five semantic styles. Two of the styles are denotational (graph model and Scott domain) and three are operational (interpreter, reduction semantics, and abstract machine). We find that no one semantics yields the shortest proofs for every compiler pass, though the interpreter proof is the smallest overall and the graph-model proof is the second smallest. We identify two factors that influence proof size. For operational semantics, proof size tends to grow with how much the runs of the source and target differ. For denotational semantics, the proofs are succinct when the correctness of a pass is an equation between meanings, as for a source optimizer, and lengthy when a pass needs a logical relation, as for the closure-conversion pass.

## 1 Introduction

This paper furthers our understanding of how the choice of language semantics impacts proofs of compiler correctness and delivers two novel denotational proofs of correctness for closure conversion. But before explaining our results, let us review the state of the art. Thanks to major compiler verification projects such as CompCert [Leroy 2009] and CakeML [Kumar et al. 2014; Owens et al. 2016] we have a fairly good understanding of the use of operational semantics in proofs of compiler correctness. CompCert employs small-step semantics and relates them using simulations [Leroy et al. 2016]. CakeML uses interpreters and reasons about them using induction on evaluation and value relations. Both proofs were large undertakings. The proof of CompCert is 100,000 lines of Coq and took 6 person-years, and the proof of the CakeML compiler backend is about 100,000 lines of HOL4 [Tan et al. 2019], by six people over two years [Myreen 2016].

Authors' Contact Information: Jeremy G. Siek, Indiana University, Bloomington, Indiana, USA, jsiek@iu.edu; Matthew Heimerdinger, Indiana University, Bloomington, Indiana, USA, matthew.heimerdinger@gmail.com; Tianyu Chen, Indiana University, Bloomington, USA, chen512@iu.edu; Darshal Shetty, Indiana University, Bloomington, USA, dcshetty@iu.edu.

In contrast, the literature is less complete when it comes to the use of denotational semantics in compiler correctness proofs, but there is evidence that denotational semantics can yield simpler proofs due to equational and compositional reasoning [Morris 1973; Stoy 1977; Zakowski et al. 2021]. This paper fills a fundamental gap in the literature, establishing the first proofs of correctness with respect to denotational semantics for a compiler from a Turing-complete, higher-order language to a first-order language. Why is there such a fundamental gap in the literature? We suspect two reasons: (1) the denotational semantics for Turing-complete languages are more complex than those for strongly-normalizing languages, and (2) compilers for higher-order languages that target a first-order language must go through closure conversion, which involves a change of representation, and that is more challenging to reason about in denotational semantics. Recall that *closure conversion* replaces each function with a pair of code and the values of its free variables, and replaces each function call with code that takes the pair apart [Cardelli 1984; Steele 1978] and calls its first component with the second component as an extra argument.

To date, every proof of compiler correctness with respect to a denotational semantics stops short of closure conversion for a Turing-complete language, in several different ways. Nielsen [2000] proved that defunctionalization, a close cousin of closure conversion, preserves the denotations of programs. However, compiler correctness requires that the denotation of programs is also reflected. Otherwise a compiler could translate a diverging source program into a converging target. Nielsen [2000] says “Eventually, we want to prove full correctness, i.e., semantic equivalence of the two programs.” Banerjee et al. [2001] proved both directions, but gave an operational semantics to the source language. They report that they “know of no other proofs of correctness (that is, both directions) for ordinary or CFA-based defunctionalization in the presence of recursion.” Chlipala [2007] proved correctness of a compiler from the simply typed $\lambda$-calculus (STLC) to assembly, including closure conversion, but the STLC is not Turing-complete, which let him interpret each type as a Coq type and each function as a total Coq function. Other denotational proofs of compiler correctness avoid closure conversion, because either their source language is first-order [Cheng et al. 2026; Polak 1981; Xia et al. 2019] or their target language is still higher-order [Benton and Hur 2009; Clinger 1984; Schweimeier and Jeffrey 1999]. When Chlipala [2010] proved correct a compiler for a language with general recursion, he switched to operational semantics, explaining that the denotational approach “cannot be applied directly to Turing-complete programming languages.”

A natural setting to study the compilation of higher-order languages is Plotkin’s FPC language [Plotkin 1985], a call-by-value $\lambda$-calculus with sums, products, and iso-recursive types. The recursive types of FPC enable the fixed-point combinator, so FPC has general recursion and is Turing-complete. That being said, FPC lacks many of the features of modern programming languages, such as mutable state, exceptions, and multi-threading, to name a few. The claims in this paper are specific to the compilation of FPC. We make no claims about the impact of semantic styles on proofs of correctness for compilers of other programming languages. Our proofs of correctness, like most of those cited above, are for whole programs. They relate the meaning of a closed source program to the meaning of its compiled code. Compositional compiler correctness [Perconti and Ahmed 2014; Ramananandro et al. 2015; Stewart et al. 2015], which addresses linking with separately compiled code or code written in another language, is just as important. Cheng et al. [2026] achieve compositional compiler correctness with a denotational semantics of open modules and a semantic linking operator, for imperative first-order languages, including the front end of CompCert. Extending our denotational proofs to open terms and linking, through closure conversion, would be a good next step.

The compiler for FPC that we study begins with a source optimization pass based on the cp0 pass of Chez Scheme [Dybvig 2006; Waddell and Dybvig 1997], includes three passes for closure conversion, and four back-end passes that target a typed low-level intermediate language that is on the way to x86 assembly. For this compiler we seek to answer three research questions:

**RQ1** Can we prove correctness of this compiler with respect to a denotational semantics?
**RQ2** How do the sizes of mechanized correctness proofs for this compiler compare across different operational and denotational semantic styles, pass by pass and overall?
**RQ3** Which properties of a pass determine the size of its correctness proof in each style?

We answer RQ1 affirmatively, in two denotational models (Theorems 4.1 and 4.2). Our first model is a graph model, in which a program denotes a set of inductively defined graph elements [Engeler 1981; Plotkin 1993; Scott 1976]. Each element is a finite description of the program's behavior, and in particular, the semantics of a function is given by a set of elements that represent input–output entries. To prove semantic reflection for closure conversion we introduce a new kind of relation between denotations, the *approximant logical relation*. Our key insight is to relate a finite set of target graph elements (a finite approximation) to a possibly infinite source denotation, and to define the relation by recursion on the depth of the elements in the finite set (Lemma 4.6 in Section 4.2). Our second model uses Scott domains, in which a program denotes an element of a universal domain of tagged values [Scott 1982]. There we define the relation by recursion on the stages of the bilimit that constructs the domain (Lemma G.5 in Section 4.3), as Reynolds [1974] did to relate the direct and continuation semantics of an untyped $\lambda$-calculus. Other styles of denotational semantics remain to be evaluated for compiler correctness and closure conversion, among them synthetic domain theory [Birkedal et al. 2011; Hyland 1991; Møgelberg and Paviotti 2016; Rosolini 1986], guarded interaction trees [Frumin et al. 2024], categorical semantics [Fiore and Plotkin 1994], the partiality monad [Danielsson 2012], and game semantics [Hyland and Ong 2000].

On top of our denotational results for closure conversion, we develop two correctness proofs for the eight-pass compiler from FPC to a typed low-level intermediate language. For seven of the eight passes, correctness is expressed as equality of denotations (Lemma 4.3) and the proofs are by induction on terms and use equational reasoning. The exception is closure conversion, in particular the **split closure** pass that separates the code of each closure from its environment, whose semantic reflection proof requires the logical relations we discussed above.

To answer RQ2 and RQ3, we compare the two denotational proofs of correctness for our FPC compiler with three more proofs of correctness, each with respect to an operational semantics (Propositions F.1–F.3) and proved using the standard technique for its style: clocked interpreters with relations on values, abstract machines with simulations, and reduction semantics with simulations. We mechanize all five compiler correctness proofs in the Agda proof assistant [Agda Team 2026]. Earlier work that compares proofs of compiler correctness focuses on a single phase of a compiler [Kahn et al. 2025; Knothe and Bringmann 2026; Owens et al. 2016]; ours is the first on a whole compiler for a higher-order language. Each of the five correctness results is with respect to a different semantics, so one might wonder if the semantics and therefore the results are equivalent. To check that our mechanized semantics agree, we prove the graph model adequate with respect to the interpreters, the abstract machines, and the reduction semantics, and the Scott-domain semantics adequate with respect to the interpreters (Proposition 5.1), so the five results are equivalent.

No one semantic style gives the smallest proof for every pass (RQ2), but for every pass, the smallest proof is either the graph-model proof or the interpreter proof. Not counting language-independent infrastructure, the interpreters give the smallest proof, at 35,860 lexical tokens of Agda, and the graph model gives the second smallest, at 40,905. The sizes split by the kind of pass (RQ3). The denotational proofs tend to be the smallest for passes that preserve the meaning of a program exactly but change how it runs, such as the source optimizer **simplify**. The operational proofs are several times smaller for **split closure**, which changes the representation of closures (Section 6).

Section 2 presents the compiler and Section 3 gives background on the graph model. Section 4 presents the correctness of each pass wrt. the graph model and the proof for **split closure** wrt.

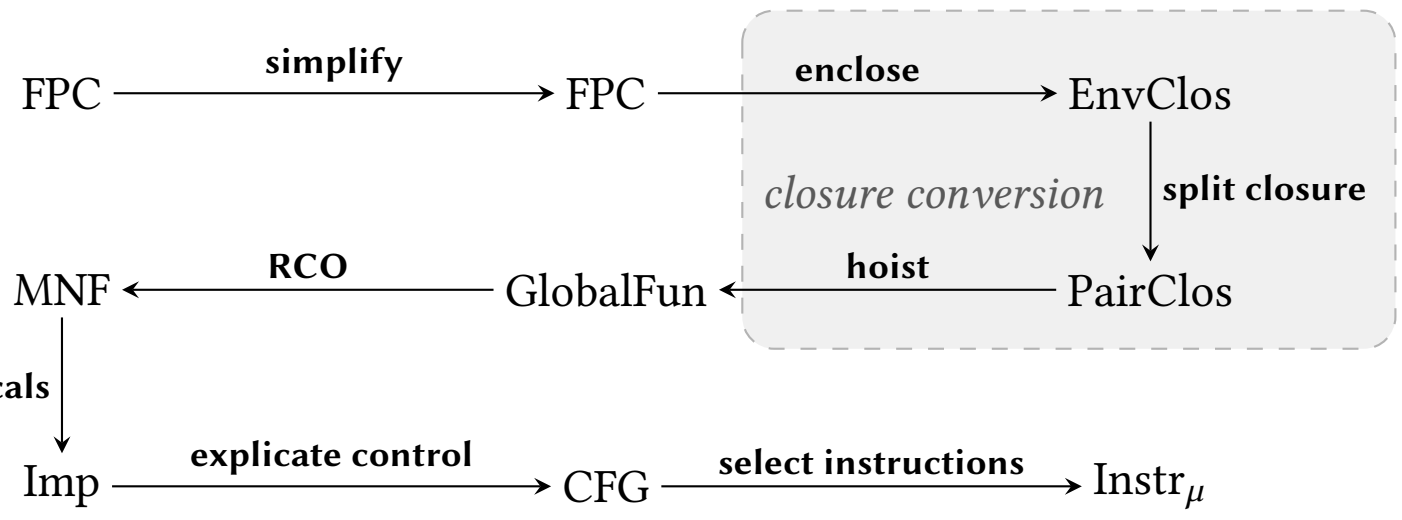

Fig. 1. The passes of the compiler and the languages between them.

| | | |
|---|---|---|
| | Variables | $x, y, z$ (from Imp on, slots of a stack frame) |
| | Labels | $l$ (the number of a block, $t_l$ or $s_l$, in a function def. of CFG and Instr$_\mu$) |
| | Base types | $B ::= \mathsf{Nat} \mid \mathsf{Int} \mid \mathsf{Bool} \mid \mathsf{Unit} \mid \mathsf{Void}$ |
| | Types | $A, C ::= B \mid A \to C \mid \mathsf{tup}(A_1, \ldots, A_n) \mid A + C \mid \mu X.\, A \mid X \mid \mathsf{code}(\Delta; A \to C)$ |
| | | $\Delta, \Phi ::= A_1, \ldots, A_n$ |
| FPC | Terms | $K, L, M, N ::= x \mid \lambda x.\, N \mid L\ M \mid c \mid \mathsf{tuple}\ \vec{M} \mid \mathsf{get}_i\ M \mid \mathsf{inl}\ M \mid \mathsf{inr}\ M$ |
| | | $\mid\ \mathsf{case}\ L\ (x.\, M)\ (y.\, N) \mid \mathsf{prim}_p\ L\ M \mid \mathsf{let}\ x = M\ \mathsf{in}\ N \mid \mathsf{fold}\ M \mid \mathsf{unfold}\ M$ |
| EnvClos | Terms | $K, L, M, N ::= \ldots \mid \mathsf{clos}\ (z, y.\, N)\ \vec{M}$ (in place of $\lambda x.\, N$) |
| PairClos | Terms | $K, L, M, N ::= \ldots \mid \mathsf{code}\ (z, y.\, N) \mid \mathsf{pair}\ L\ M \mid \mathsf{callc}\ L\ M$ (in place of clos and $L\ M$) |
| GlobalFun | Terms | $K, L, M, N ::= \ldots \mid f_k$ (in place of code) |
| | Function Def. | $d ::= \mathsf{code}\ (z, y.\, N) \qquad P ::= (d_0, \ldots, d_{m-1};\ M)$ |
| MNF | Atoms | $a ::= x \mid c \mid f_k$ |
| | Operations | $o ::= \mathsf{pair} \mid \mathsf{callc} \mid \mathsf{tuple} \mid \mathsf{get}_i \mid \mathsf{inl} \mid \mathsf{inr} \mid \mathsf{prim}_p \mid \mathsf{fold} \mid \mathsf{unfold}$ |
| | Terms | $M, N ::= a \mid o\ \vec{a} \mid \mathsf{case}\ a\ (x.\, M)\ (y.\, N) \mid \mathsf{let}\ x = M\ \mathsf{in}\ N \qquad P ::= (\vec{d};\ M)$ |
| Imp | Atoms | $a ::= x \mid c \mid f_k \qquad rhs ::= a \mid o\ \vec{a}$ |
| | Expressions | $e ::= rhs \mid x := e_1;\ e_2 \mid \mathsf{case}\ a\ x\ y\ e_1\ e_2$ |
| | Function Def. | $d ::= (\Phi;\ e) \qquad P ::= (\vec{d};\ d)$ |
| CFG | Tails | $t ::= \mathsf{return}\ rhs \mid x := rhs;\ t \mid \mathsf{goto}\ l \mid \mathsf{case}\ a\ x\ y\ l_1\ l_2$ |
| | Function Def. | $d ::= (\Phi;\ t_0, \ldots, t_{j-1};\ t) \qquad P ::= (\vec{d};\ d)$ |
| Instr$_\mu$ | Operands | $a ::= x \mid c \mid f_k \mid \%r \mid \mathsf{fold}\ a \mid \mathsf{unfold}\ a \qquad dst ::= \%r \mid x \qquad r ::= \mathsf{rax} \mid \mathsf{rdi} \mid \mathsf{rsi}$ |
| | Constructors | $con ::= \mathsf{pair}(a_1, a_2) \mid \mathsf{tuple}(\vec{a}) \mid \mathsf{inl}(a) \mid \mathsf{inr}(a)$ |
| | Instructions | $i ::= \mathsf{movq}\ a, dst \mid \mathsf{subq}\ a, \%r \mid \mathsf{cmpq}\ a_2, a_1 \mid \mathsf{sete}\ dst \mid \mathsf{alloc}\ con, dst$ |
| | | $\mid\ \mathsf{load}\ \mathsf{get}_i\ a, dst \mid \mathsf{fst}\ x, \%r \mid \mathsf{snd}\ x, \%r \mid \mathsf{callq}\ a$ |
| | Sequences | $s ::= i;\ s \mid \mathsf{retq} \mid \mathsf{jmp}\ l \mid \mathsf{case}\ x\ y\ z\ l_1\ l_2$ |
| | Function Def. | $d ::= (\Phi;\ s_0, \ldots, s_{j-1};\ s) \qquad P ::= (\vec{d};\ d)$ |

Fig. 2. The grammars of the compiler's languages.

Scott domains. Section 5 gives the adequacy theorems for the denotational semantics, Section 6 compares the five proofs, Section 7 discusses related work, and Section 8 concludes. Appendix F.5 defines the Scott domains. For background on domain theory, we refer the reader to Scott [1982], Winskel [1993a], and Abramsky and Jung [1994]. The other appendices define the semantics of every language and compiler pass. Our Agda development is in the supplementary material, which can be accessed using the following link:

https://github.com/cty12/denot-compiler-supplementary

## 2 The Compiler and Its Languages

Our compiler translates FPC, the call-by-value $\lambda$-calculus with sums, products, and iso-recursive types of Plotkin [1985], extended with integers, to $\text{Instr}_\mu$, a typed low-level intermediate language, in eight passes through six intermediate languages. Figure 1 shows the sequence of passes and languages. Figure 2 presents the grammars of all the languages and Appendix B gives their typing rules.[1] Appendix C lists the full definition of each compiler pass.

Throughout Figure 2, $c$ ranges over constants, $p$ over the primitives `sub` and `eq`, and $f_k$ over the names of a program's global definitions, where $f_k$ refers to the definition $d_k$. In the languages prior to Imp, $x$, $y$, and $z$ range over variables. From Imp on, they range over fixed locations in a function's stack frame. We write binders as one or more names followed by a period, and the scope of the binders is the term after the period. Every language has a graph-model semantics (Section 3), given in full in Appendix B.

*The source language.* FPC has integers, $n$-ary tuples, binary sums, the primitives `sub` and `eq`, and iso-recursive types. The primitive `eq` returns `inl` $\langle\rangle$ for true and `inr` $\langle\rangle$ for false, so `case` serves as the conditional. A recursive type $\mu X.\, A$ is isomorphic to its unfolding $A[\mu X.\, A/X]$, and the two coercions of this isomorphism are explicit. The `fold` operation turns a term of the unfolding into a term of the recursive type, and `unfold` turns it back:

$$\frac{\Gamma \vdash M : A[\mu X.\, A/X]}{\Gamma \vdash \texttt{fold}\; M : \mu X.\, A} \qquad \frac{\Gamma \vdash M : \mu X.\, A}{\Gamma \vdash \texttt{unfold}\; M : A[\mu X.\, A/X]}$$

Neither coercion does anything at run time. Recursive types make self-application typable, so a program can define the call-by-value fixed-point combinator to write recursive functions.

*The optimizer.* The first pass, **simplify** (§C.1), translates from FPC to FPC. It is a simplified version of the optimizer cp0 of Chez Scheme [Waddell and Dybvig 1997]. It propagates known values, folds constants, reduces a `case`, projection, or `unfold` of a known value, and inlines small known functions.

*Closure conversion.* The next three passes perform closure conversion. The **enclose** pass (§C.2) translates from FPC to EnvClos. It turns each $\lambda$-abstraction into a closure `clos` $(z, y.\, N)\; \vec{M}$, which captures the values $\vec{M}$ of the variables that the code $N$ uses. The closure binds two variables in its code, the environment $z$ and the parameter $y$, and the code, which has no other free variables, reads the $i$th captured value as $\texttt{get}_i\; z$.[2] The **split closure** pass (§C.3) then separates a closure into its code and its environment, translating from EnvClos to PairClos. In this target IR, `code` $(z, y.\, N)$ is closed code of type $\texttt{code}(\Delta; A \to C)$, which expects an environment of type $\texttt{tup}(\Delta)$ and an argument of type $A$. The form `pair` $L\; M$ builds a closure, which is a pair of code $L$ and an environment $M$ of the same $\Delta$. Its type, $A \to C$, hides $\Delta$, as in the existential type $\exists\Delta.\, \texttt{code}(\Delta; A \to C) \times \texttt{tup}(\Delta)$ of typed closure conversion [Minamide et al. 1996], so closures with different environments have the same type. Thus `pair` packs the existential, but its dynamic semantics is the usual for a pair. The form `callc` $L\; M$ calls the code of the closure $L$ with its environment and the argument $M$.

$$\frac{z : \texttt{tup}(\Delta),\; y : A \vdash N : C}{\Gamma \vdash \texttt{code}\; (z, y.\, N) : \texttt{code}(\Delta; A \to C)} \qquad \frac{\Gamma \vdash L : \texttt{code}(\Delta; A \to C) \quad \Gamma \vdash M : \texttt{tup}(\Delta)}{\Gamma \vdash \texttt{pair}\; L\; M : A \to C} \qquad \frac{\Gamma \vdash L : A \to C \quad \Gamma \vdash M : A}{\Gamma \vdash \texttt{callc}\; L\; M : C}$$

[1] Every language is intrinsically typed in our Agda development, so every term is well typed by construction and no pass needs a precondition about scoping or typing.

[2] Our Agda development implements **enclose** as three passes, which capture every variable in scope, drop the captured values that the code does not use, and store the rest in a tuple. The proof of **enclose** composes their proofs, and Table 1 reports the sum of their sizes.

Finally, the **hoist** pass (§C.4) (similar to the hoisting phase of Morrisett et al. [1999]) translates from PairClos to GlobalFun. It turns the code of every function into a global definition $d_k$, and each occurrence of the code into a reference $f_k$ to it, so a GlobalFun program is a list of function definitions and a main term.

*The back end.* The pass **remove complex operands** (**RCO**, §C.5) translates from GlobalFun to MNF, a *monadic normal form* [Hatcliff and Danvy 1994]. It binds each operand that is not atomic to a variable with `let`, so the operands of every operation in MNF are atomic. The atomic terms are variables, constants, and function names. The **lift locals** pass (§C.6) translates from MNF to Imp. It gives each variable its own location in the function's stack frame, called a *slot*. A function body declares its frame $\Phi$, which lists the types of its slots. The type system tracks which slots are initialized. The **explicate control** pass (§C.7) translates from Imp to CFG. It turns the nested `case`s into a control-flow graph. A CFG program has a function body for each global definition and for the main program. The name of the tail non-terminal is short for tail position. A function body consists of numbered blocks and a starting block, each of which is a tail.

Finally, **select instructions** (§C.8) translates from CFG to $\text{Instr}_\mu$, a typed low-level language in the style of TAL [Morrisett et al. 1999], with an x86 flavor. $\text{Instr}_\mu$ has three registers (`rax`, `rdi`, and `rsi`), a stack frame of slots, and an abstract heap. Programs have the same structure as in CFG. Instruction operands $a$ are slot numbers, constants, names of function definitions, or registers. Instructions write to a destination $dst$, a register or a slot.

The pass translates each simple expression of CFG into instructions that write its value to a destination, using the registers to stage arithmetic and calls. A closure (`pair`), a tuple, or an injection becomes the instruction `alloc` $con, dst$, which allocates on the heap the result of the constructor $con$. A call `callc` $x\ a$ of the closure in slot $x$ on the argument $a$ becomes the instruction sequence `fst` $x$, `%rax`; `snd` $x$, `%rdi`; `movq` $a$, `%rsi`; `callq %rax`. The first two instructions load the code and the environment of the closure, the third puts the argument in place, and the call leaves its result in `rax`. The type system allows only one closure to be open at a time, so that a call never combines the code of one closure with the environment of another. The coercions `fold` and `unfold` become annotations on operands, `fold` $a$ and `unfold` $a$, that do nothing.

$\text{Instr}_\mu$ abstracts from a real instruction set. Its heap is abstract, with no addresses or memory layout, and control flow within a function is acyclic. Neither abstraction concerns closure conversion, because in $\text{Instr}_\mu$, as in machine code, a closure is a heap object that pairs a function's code pointer with a tuple of captured values, and the caller passes the tuple, along with the argument, in argument-passing registers before it calls the code pointer.

## 3 Graph Models

A denotational semantics maps each program to a mathematical object that describes its behavior. In a graph model, that object is a set of finite pieces of evidence about the behavior, which we call *elements*. Our model is a variant of the model of Engeler [1981], which we extend with constants, pairs, tuples, and disjoint sums. Figure 3 defines elements, several semantic operators on sets of elements, and the semantics of FPC. We first explain how the model represents functions, then introduce the elements, and then the operators. We review how the model handles recursion and divergence, and why we require sets of elements to be consistent. We end this section with the properties of the semantics that our proofs rely on, including a $\beta$-law.

*Functions as tables.* One can describe a function by tabulating it. Each row of the table pairs an input with the corresponding output, and applying the function means looking up a row. Two problems arise with this simple approach. First, a function on an infinite domain (such as the integers) needs an infinite table. Second, a function that is applied to itself (such as the Y combinator)

**Graph elements** $u, v, w \in \text{Element}$

$$u, v, w ::= c \mid V \mapsto w \mid \nu \mid \langle\!|u, v|\!\rangle \mid \mathsf{tup}_i(v) \mid \mathsf{left}(v) \mid \mathsf{right}(v) \mid \langle\rangle$$

**Finite lists of elements** $U, V, W \in \mathsf{List}(\text{Element})$

**Denotations (sets of elements)** $D, E \subseteq \text{Element}$

**Indices** $i, j \in \mathbb{N}$

**Consistency (Appendix A.1)** $u \sim v \qquad \mathsf{con}(V)$

$$\mathsf{con}(V) \iff u \sim v \text{ for all } u, v \in V$$

**Operators on denotations** $op(D_1, \ldots, D_n) \subseteq \text{Element}$

$$
\begin{aligned}
\Lambda(F) &= \{\, V \mapsto w \mid V \neq [\,],\ \mathsf{con}(V),\ w \in F(V) \,\} \cup \{\nu\} \quad (F \text{ a function on sets, e.g. } \boldsymbol{\lambda} X.\, E)\\
D \bullet E &= \{\, w \mid V \mapsto w \in D,\ V \neq [\,],\ V \subseteq E \,\}\\
\mathsf{pair}(D, E) &= \{\, \langle\!|u, v|\!\rangle \mid u \in D,\ v \in E \,\}\\
\mathsf{car}(D) &= \{\, u \mid \langle\!|u, v|\!\rangle \in D \,\}\\
\mathsf{cdr}(D) &= \{\, v \mid \langle\!|u, v|\!\rangle \in D \,\}\\
\mathcal{T}(D_1, \ldots, D_n) &= \{\, \mathsf{tup}_i(v) \mid v \in D_i \,\} \quad \text{if } D_1, \ldots, D_n \neq \emptyset, \text{ and } \emptyset \text{ otherwise}\\
\mathsf{inl}(D) &= \{\, \mathsf{left}(v) \mid v \in D \,\} \qquad \mathsf{inr}(D) = \{\, \mathsf{right}(v) \mid v \in D \,\}\\
\pi_t(D) &= \{\, v \mid t(v) \in D \,\} \quad \text{for } t \in \{\mathsf{tup}_i, \mathsf{left}, \mathsf{right}\}\\
C(D, F, G) &= \{\, w \mid V \neq [\,],\ \mathsf{con}(V),\ \mathsf{left}(V) \subseteq D,\ w \in F(V) \,\}\\
&\quad \cup \{\, w \mid V \neq [\,],\ \mathsf{con}(V),\ \mathsf{right}(V) \subseteq D,\ w \in G(V) \,\}\\
\mathsf{left}(V) &= \{\, \mathsf{left}(v) \mid v \in V \,\} \qquad \mathsf{right}(V) = \{\, \mathsf{right}(v) \mid v \in V \,\}\\
\mathsf{prim}_p(D, E) &= \{\, \delta(p, n, m) \mid n \in D,\ m \in E \text{ integers} \,\}\\
\delta(\mathsf{sub}, n, m) &= n - m \qquad \delta(\mathsf{eq}, n, m) = \mathsf{left}(\langle\rangle) \text{ if } n = m, \text{ and } \mathsf{right}(\langle\rangle) \text{ otherwise}
\end{aligned}
$$

**Semantics of** FPC $[\![M]\!]\rho = D$

$$
\begin{aligned}
[\![x]\!]\rho &= \rho(x) & [\![c]\!]\rho &= \{c\}\\
[\![\lambda x.\, N]\!]\rho &= \Lambda(\boldsymbol{\lambda} X.\, [\![N]\!]\rho[x{:=}X]) & [\![L\ M]\!]\rho &= [\![L]\!]\rho \bullet [\![M]\!]\rho\\
[\![\mathtt{tuple}\ (M_1, \ldots, M_n)]\!]\rho &= \mathcal{T}([\![M_1]\!]\rho, \ldots, [\![M_n]\!]\rho) & [\![\mathtt{get}_i\ M]\!]\rho &= \pi_{\mathsf{tup}_i}([\![M]\!]\rho)\\
[\![\mathtt{inl}\ M]\!]\rho &= \mathsf{inl}([\![M]\!]\rho) & [\![\mathtt{inr}\ M]\!]\rho &= \mathsf{inr}([\![M]\!]\rho)\\
[\![\mathtt{prim}_p\ L\ M]\!]\rho &= \mathsf{prim}_p([\![L]\!]\rho, [\![M]\!]\rho) & [\![\mathtt{fold}\ M]\!]\rho &= [\![M]\!]\rho\\
[\![\mathtt{let}\ x = M\ \mathtt{in}\ N]\!]\rho &= \Lambda(\boldsymbol{\lambda} X.\, [\![N]\!]\rho[x{:=}X]) \bullet [\![M]\!]\rho & [\![\mathtt{unfold}\ M]\!]\rho &= [\![M]\!]\rho
\end{aligned}
$$

$$[\![\mathtt{case}\ L\ (x.\, M)\ (y.\, N)]\!]\rho = C([\![L]\!]\rho,\ \boldsymbol{\lambda} X.\, [\![M]\!]\rho[x{:=}X],\ \boldsymbol{\lambda} Y.\, [\![N]\!]\rho[y{:=}Y])$$

Fig. 3. The graph model (elements, denotations, and operators) and the semantics of FPC.

would need a table that contains a copy of itself. Graph models solve both problems. For the first problem, a terminating program applies each function to only finitely many arguments, so finite tables suffice [Plotkin 1993; Scott 1976]. However, to handle all possible contexts, the meaning of a function is the set of all its finite tables, or equivalently the set of all their rows. For the second problem, a row applies to an argument when the row's input is *included* in the argument, rather than equal to it. A table then needs only a smaller copy of itself as an input entry, and elements, defined as an inductive data type, can provide that.

*Graph elements.* An element is a finite and partial description of a term's behavior (Figure 3). A row of a table is represented by an element $V \mapsto w$, called an *entry*, whose input $V$ is a list of elements[3] and whose output $w$ is an element.[4] The element $\nu$ records that a term evaluates to a function, without describing any of the function's rows. The other elements describe the data of our languages: a constant $c$, a pair $\llparenthesis u, v \rrparenthesis$, a tuple element $\mathsf{tup}_i(v)$ which says that component $i$ of a tuple is described by $v$, the empty tuple $\langle\rangle$, and the injections $\mathsf{left}(v)$ and $\mathsf{right}(v)$ of a sum. A denotation is a set of elements.

*Operators.* Each case of the semantics of FPC (Figure 3) applies an operator to the denotations of the sub terms. In the semantics, an environment $\rho$ maps each variable to a set of elements, $\rho[x{:=}D]$ maps $x$ to $D$ and every other variable as $\rho$ does, and $\boldsymbol{\lambda}$ denotes a mathematical function.

The meaning of a function, given by the operator $\Lambda$, is the set of all entries $V \mapsto w$ such that the output $w$ is in the meaning of the function's body in an environment that maps its parameter to the input $V$, together with $\nu$. The input $V$ must be nonempty and consistent, written $\mathsf{con}(V)$, which we explain at the end of this section. For example, $[\![\lambda x.\, x + 1]\!]$ contains the entry $[n] \mapsto n{+}1$ for every integer $n$. The meaning of the twice function $[\![\lambda f.\, \lambda x.\, f\ (f\ x)]\!]$ contains the entries

$$[\,[0] \mapsto 1,\ [1] \mapsto 2\,] \mapsto ([0] \mapsto 2) \quad \text{and} \quad [\,[2] \mapsto 4,\ [4] \mapsto 8\,] \mapsto ([2] \mapsto 8)$$

whose input has one entry for each call to $f$. Because of $\nu$, a $\lambda$-abstraction has a nonempty meaning, as befits a value in a call-by-value language, even if it is never applied.

The operator $D \bullet E$ gives the meaning of application. It collects the outputs of the entries of $D$ whose input is contained in the argument $E$. For example, we apply the increment function to 3 to get 4, $[\![(\lambda x.\, x + 1)\ \ (2 + 1)]\!] = \{4\}$. because the input of entry $[3] \mapsto 4$ is contained in $[\![2 + 1]\!] = \{3\}$. Likewise, applying the twice function to $(\lambda x.\, x + 1)$ and then to 0 gives $\{2\}$ because $\{[0] \mapsto 1,\ [1] \mapsto 2\}$ is contained in $[\![\lambda x.\, x + 1]\!]$.

The other operators build and take apart data. The operator $\mathsf{pair}$ builds pairs, and $\mathsf{car}$ and $\mathsf{cdr}$ take their components. The operator $\mathcal{T}$ builds tuples, and $\pi_{\mathsf{tup}_i}$ takes component $i$. The operators $\mathsf{inl}$ and $\mathsf{inr}$ build injections, and $C$ gives the meaning of a `case`. Like $\Lambda$, it binds the variable of a branch to a nonempty, consistent list, here of the payloads of the injections in the scrutinee. The operator $\mathsf{prim}_p$ applies a primitive to the integers in its operands. The empty tuple $\langle\rangle$ also fills the positions of an environment that hold no variable, because we represent environments as total functions whose positions are nonempty. The environment $\rho_0$ of a closed program maps every variable to $\{\langle\rangle\}$, and we define $[\![M]\!] = [\![M]\!]\rho_0$.

*Recursion and divergence.* Recursion in FPC comes from self-application, which the recursive types of FPC make typable. The constructs `fold` and `unfold` mean the identity, so recursive types cost nothing in the semantics. Because entries apply to arguments that include their input, the self-applied function in a fixed-point combinator such as $Z$ has a meaning whose tables contain smaller copies of themselves, like a set of nested dolls [Siek 2017].

A program that diverges produces no evidence, so its meaning is the empty set. For example, let $\Omega = \Delta\ (\texttt{fold}\ \Delta)$ with $\Delta = (\lambda x.\, (\texttt{unfold}\ x)\ x)$. Then $[\![\Omega]\!] = \emptyset$ because an output of $\Omega$ would need an entry of $[\![\Delta]\!]$ whose input contains a strictly smaller entry of $[\![\Delta]\!]$ with the same output, and so on forever, which is impossible for inductively defined elements.

*Consistency.* We write $u \sim v$ when the elements $u$ and $v$ are consistent, which means that they could describe the same value. Appendix A.1 gives the definition. The operator $\Lambda$ builds only entries whose inputs are consistent, which rules out entries such as $[3, 5] \mapsto 6$ in the meaning of $\lambda x.\, x + 1$, because the input $[3, 5]$ of that entry describes no single argument, i.e. 3 conflicts with 5. Our model requires consistency because the proof for the **split closure** pass needs it (Section 4.2). The meaning of a call in PairClos takes the curried function from one pair and the environment from another, so it may combine the function of one pair with the environment of a different pair. Consistency ensures that the environments of these pairs are consistent with each other.

[3]We sometimes use a list of elements where a set of elements is expected, and assume the natural conversion takes place.
[4]The input of an entry could instead be a single element, as in filter models [Barendregt et al. 1983], but a list makes the metatheory cleaner because the argument of a function is often a function, and $V$ can contain one entry for each call to it.

*The properties of the semantics.* The meaning of a term is monotone in its environment (Lemma A.1), and it is continuous in the sense that each element of the meaning depends on only finitely many elements of the environment. These properties, together with consistency, give the graph model the following $\beta$ laws. Its premises come from the definitions of $\Lambda$ and application, which use only entries whose input is nonempty and consistent. A function $F$ on sets is called *finitary* at $D$ if every $w \in F(D)$ is in $F(V)$ for some nonempty $V \subseteq D$, and similarly for a function of two sets.

Lemma 3.1 ($\beta$ laws).

(1) *If $F$ is monotone and finitary at $D$ and $D$ is consistent, then* $\Lambda(F) \bullet D = F(D)$.
(2) *If $F$ is finitary at $T$ and $E$, and both $T$ and $E$ are consistent, then* $F(T, E) \subseteq (\Lambda(\boldsymbol{\lambda} X.\, \Lambda(\boldsymbol{\lambda} Y.\, F(X, Y))) \bullet T) \bullet E$.

## 4 Denotational Compiler Correctness

This section proves the following two theorems about the FPC compiler.

Theorem 4.1 (Compiler correctness, graph model). *Let $M$ be a closed term of* FPC *of type* `Int`. *For every constant $c$, $c \in [\![M]\!]$ if and only if $c \in [\![\textbf{compile}(M)]\!]$. Moreover, $[\![M]\!] \neq \emptyset$ if and only if $[\![\textbf{compile}(M)]\!] \neq \emptyset$.*

Theorem 4.2 (Compiler correctness, Scott domains). *Assume propositional truncation, function extensionality, and propositional extensionality. Let $M$ be a closed term of* FPC *of type* `Int`. *For every constant $c$, $[\![M]\!]\rho_0 = \lfloor (c, ()) \rfloor$ if and only if $[\![\textbf{compile}(M)]\!] = \lfloor (c, ()) \rfloor$. Moreover, $[\![M]\!]\rho_0$ is defined if and only if $[\![\textbf{compile}(M)]\!]$ is defined.*

Theorems 4.1 and 4.2, whose proofs we mechanize in Agda, answer RQ1 affirmatively. The same statements hold for closed terms of every type; we state them for `Int`, the type of a whole program's result, to align them with the operational correctness results (Section 6). Each proof is a composition of one correctness lemma per pass. We first show how most of the passes can be proved correct by equational reasoning (Section 4.1). We then prove that the **split closure** pass is correct using a logical relation (Section 4.2). Finally, we sketch the corresponding proof of correctness for Scott domains (Section 4.3).

### 4.1 Passes Proved by Equational Reasoning

For seven of the eight passes, correctness is an equation between the meaning of a program and the meaning of its translation.

Lemma 4.3 (Passes proved by an equation). *Let $p$ be a pass other than **split closure**, and let $M$ be a closed term of the source language of $p$. Then $[\![M]\!] = [\![p(M)]\!]$.*

Each proof is by induction on the term, generalizing Lemma 4.3 to open terms in a source and a target environment in which each variable means the same as its translation, with side conditions such as nonemptiness and consistency. Every semantics is compositional and the passes translate most forms homomorphically, so most cases apply the same operator to equal arguments. The real work is in the few forms that a pass changes, where an equation about the operators of the graph model turns the source form into the target form. Below we give these calculations for **simplify** and for the last step of **enclose**. We skip over side conditions about non-empty and consistent denotations but of course they are present in the Agda mechanization.

Appendix D gives the rest of the equational proofs, including the proof for **select instructions**, whose key case turns a call into `fst`, `snd`, `movq`, and `callq`. Where a step needs a nonempty set, we show the nonempty case; otherwise both sides are empty.

*Simplify.* The pass **simplify** keeps a static environment $E$ (Appendix C.1). It maps each variable to the term that replaces it and, when the variable is bound to a statically known function, to that function as a $\lambda$-abstraction $\langle \lambda x.\, N, E_f \rangle$ paired with the static environment $E_f$ where it was defined. The main lemma for **simplify** carries the invariant that $E$ agrees with the source environment $\rho$ and the target environment $\rho'$. If $E$ replaces a variable by the term $M'$, the variable means $[\![M']\!]\rho'$, and if $E$ maps it to the function $\langle \lambda x.\, N, E_f \rangle$, it means $[\![\lambda x.\, N]\!]\rho_f$ for some environment $\rho_f$ such that $E_f$ agrees with $\rho_f$ and $\rho'$. The key case is inlining. Suppose $L$ is statically known to be the function $\langle \lambda x.\, N, E_f \rangle$, the operand $M$ simplifies to a small value $M'$, and the body simplifies to $N'$. Then

$$
\begin{aligned}
[\![L\ M]\!]\rho &= [\![L]\!]\rho \bullet [\![M]\!]\rho && \text{(meaning of application)}\\
&= [\![\lambda x.\, N]\!]\rho_f \bullet [\![M']\!]\rho' && \text{(invariant for } L\text{, induction hypothesis for } M)\\
&= [\![N]\!]\rho_f[x{:=}[\![M']\!]\rho'] && \text{(Lemma 3.1, part 1)}\\
&= [\![N']\!]\rho' && \text{(induction hypothesis for } N)
\end{aligned}
$$

The last step applies because the inlined body is simplified in $E_f$ extended to replace $x$ by $M'$, and this static environment agrees with $\rho_f[x{:=}[\![M']\!]\rho']$ and $\rho'$. The invariant for $L$ gives the agreement of $E_f$, and both $x$ and its replacement $M'$ mean $[\![M']\!]\rho'$.

*Enclose.* The **enclose** pass works in three steps, *capture*, *prune*, and *concretize* (§C.2). The first two produce closures of an intermediate language, Clos, written $\texttt{clos}\ (\vec{x}, y.\, N)\ \vec{M}$, which bind a name $x_i$ for each captured value $M_i$. The last step, $\textit{concretize}_\varphi$, stores the captured values in a tuple. Each variable $x$ satisfies $[\![x]\!]\rho = [\![\varphi(x)]\!]\rho'$, where its reference $\varphi(x)$ is either $x$ or $\texttt{get}_i\ z$. Let $M'_i = \textit{concretize}_\varphi(M_i)$, $D_i = [\![M_i]\!]\rho$, $D'_i = [\![M'_i]\!]\rho'$, and $N' = \textit{concretize}_{\varphi'}(N)$:

$$
\begin{aligned}
[\![\texttt{clos}\ (\vec{x}, y.\, N)\ \vec{M}]\!]\rho &= \mathrm{guard}(\vec{D};\ \Lambda(\boldsymbol{\lambda} Y.\, [\![N]\!]\rho_0[\vec{x}{:=}\vec{D}, y{:=}Y])) && \text{(meaning of a Clos closure)}\\
&= \Lambda(\boldsymbol{\lambda} Y.\, [\![N]\!]\rho_0[\vec{x}{:=}\vec{D}, y{:=}Y]) && \text{(each } D_i \text{ is nonempty)}\\
&= \Lambda(\boldsymbol{\lambda} Y.\, [\![N]\!]\rho_0[\vec{x}{:=}\vec{D}', y{:=}Y]) && \text{(induction hypothesis for } \vec{M})\\
&= \Lambda(\boldsymbol{\lambda} Y.\, [\![N']\!]\rho_0[z{:=}\mathcal{T}(\vec{D}'), y{:=}Y]) && \text{(induction hypothesis for } N)\\
&= \Lambda(\boldsymbol{\lambda} X.\, \Lambda(\boldsymbol{\lambda} Y.\, [\![N']\!]\rho_0[z{:=}X, y{:=}Y])) \bullet \mathcal{T}(\vec{D}') && \text{(Lemma 3.1, part 1)}\\
&= [\![\texttt{clos}\ (z, y.\, N')\ \vec{M}']\!]\rho' && \text{(meaning of an EnvClos closure)}
\end{aligned}
$$

The induction hypothesis for $N$ applies because the code reads $x_i$ as $\texttt{get}_i\ z$, and $\pi_{\mathsf{tup}_i}(\mathcal{T}(\vec{D}')) = D'_i$.

Each interesting case of the seven passes is a short chain of equations like these, which is why their denotational proofs are small (RQ3, Section 6). The exception is **split closure**, whose correctness is not an equation (Section 4.2).

## 4.2 Correctness of split closure and the Approximant Logical Relation

The **split closure** pass separates the code of each closure from its environment, making the closure a pair. Its proof is the hardest part of the compiler proof. This section shows why semantic reflection for **split closure** resisted our earlier attempts and how the approximant logical relation proves it.

In EnvClos, the source of **split closure**, a closure $\texttt{clos}\ (z, y.\, N)\ \vec{M}$ still means a function. Parameter $z$ is for the environment and $y$ is for the argument. The meaning of $N$ is a curried function of the environment $X$ and then the argument $Y$. The meaning of the closure is this

$$\boxed{v \sqsubseteq D'}$$

$$\begin{aligned}
(U \mapsto w) \sqsubseteq D' &= (\exists u, t.\ \langle\!| u, t |\!\rangle \in D') \ \wedge\ \forall E'.\ \mathsf{con}(E') \wedge U \sqsubseteq^* E' \Rightarrow w \sqsubseteq (\mathsf{car}(D') \bullet \mathsf{cdr}(D')) \bullet E' \\
v \sqsubseteq D' &= \exists u, t.\ \langle\!| u, t |\!\rangle \in D' \\
t(v) \sqsubseteq D' &= v \sqsubseteq \pi_t(D') \qquad \text{for } t \in \{\mathsf{tup}_i, \mathsf{left}, \mathsf{right}\} \\
v \sqsubseteq D' &= v \in D' \qquad \text{for the other elements}
\end{aligned}$$

$$\boxed{U \sqsubseteq^* E'}$$

$$\begin{aligned}
[\,] \sqsubseteq^* E' &= \mathsf{true} \\
(u :: U) \sqsubseteq^* E' &= u \sqsubseteq E' \wedge U \sqsubseteq^* E'
\end{aligned}$$

Fig. 4. The relation $v \sqsubseteq D'$ ("$v$ is included in the behavior of $D'$") for the preservation half of **split closure**. It relates a source element $v$ (of EnvClos) to a set $D'$ of target elements (of PairClos) and is defined by recursion on $v$. The relation $U \sqsubseteq^* E'$ extends it to a finite list $U$.

function applied to the captured values $\vec{M}$. An application $L\ M$ simply applies the function:

$$\begin{aligned}
[\![\mathtt{clos}\ (z, y. N)\ \vec{M}]\!]\rho &= \Lambda(\boldsymbol{\lambda} X. \Lambda(\boldsymbol{\lambda} Y. [\![N]\!]\rho_0[z{:=}X, y{:=}Y])) \ \bullet\ \mathcal{T}([\![\vec{M}]\!]\rho) \\
[\![L\ M]\!]\rho &= [\![L]\!]\rho \bullet [\![M]\!]\rho
\end{aligned}$$

In PairClos, the target language, a closure is a pair of code and environment, and a call takes the pair apart. The meaning of code $(z, y. N)$ is the curried function. The meaning of `pair` $L\ M$ is the set of pairs $\langle\!| u, v |\!\rangle$ with $u$ from the code $L$ and $v$ from the environment $M$. A call applies the first components of the pairs to their second components and then to the argument:

$$\begin{aligned}
[\![\mathtt{code}\ (z, y. N)]\!]\rho &= \Lambda(\boldsymbol{\lambda} X. \Lambda(\boldsymbol{\lambda} Y. [\![N]\!]\rho_0[z{:=}X, y{:=}Y])) \\
[\![\mathtt{pair}\ L\ M]\!]\rho &= \{\, \langle\!| u, v |\!\rangle \mid u \in [\![L]\!]\rho,\ v \in [\![M]\!]\rho \,\} \\
[\![\mathtt{callc}\ L\ M]\!]\rho &= (\mathsf{car}([\![L]\!]\rho) \bullet \mathsf{cdr}([\![L]\!]\rho)) \bullet [\![M]\!]\rho
\end{aligned}$$

The **split closure** pass translates each closure into a pair and each application into a `callc`. It translates every other construct homomorphically:

$$\begin{aligned}
\mathbf{split}(\mathtt{clos}\ (z, y. N)\ \vec{M}) &= \mathtt{pair}\ (\mathtt{code}\ (z, y. \mathbf{split}(N)))\ (\mathtt{tuple}\ \mathbf{split}(\vec{M})) \\
\mathbf{split}(L\ M) &= \mathtt{callc}\ \mathbf{split}(L)\ \mathbf{split}(M)
\end{aligned}$$

*Why preservation is easy and reflection is not.* The correctness of **split closure** is not an equation, because the pass changes the denotation of a function. In EnvClos, a closure denotes the graph of the function. In PairClos, it denotes pairs of the graph of the function curried over the values of its free variables and the tuple of those values. Lemma 4.6 therefore states only that a closed program and its translation produce the same constants and that one produces something exactly when the other does. The forward half, preservation, requires only a relation between a *single* source element and the *set* of target elements, defined by recursion on the source element (Figure 4). We read $v \sqsubseteq D'$ as "$v$ is included in the behavior of $D'$". The source element is finite, so the recursion is structural, and preservation is a routine induction on $M$.

The backward half, reflection, must show that the source accounts for every element that the target produces. Here the representation of closures gets in the way. Consider the function $(\lambda y. x{+}y)$ in an environment where $x$ is 9. In EnvClos, its meaning includes entries $[1] \mapsto 10$ and $[6] \mapsto 15$. After **split closure**, its code takes the environment as an extra parameter. The meaning of the code is a table of *curried entries*, which map an environment to an entry. It has entries for *every* environment, such as the following two:

$$[9] \mapsto ([1] \mapsto 10) \qquad [2] \mapsto ([1] \mapsto 3) \qquad \ldots$$

The program pairs this table only with environment [9], so the entry for [2] cannot be related to source function. Worse, the denotation of a call takes the table from car and the environment from cdr *separately*, so it may combine the table of one pair with the environment of another.

*Design alternatives: which sides to relate.* Before we define our relation for reflection, we explain how we chose its shape, because the obvious designs fail. A relation for reflection relates a source side to a target side, and each side can be a single graph element, a finite list of elements, or a set that may be infinite. Of these nine designs, the following reasons rule out all but one (Appendix E gives examples and the variants we tried in Agda).

- *A single target element*, against any source side, fails because a call combines the curried entry of one pair with the environment of another, so no single pair accounts for its result.
- *A single source element or a finite source list*, against any target side, fails in the function case, because infinitely many related source argument sets have no element in common, so no finite source side has an entry for each of them.
- *Whole denotations on both sides* handle the mixing, but the function case uses the relation in negative position, where nothing decreases. A well-founded definition would need a later modality or a construction by stages, as in our Scott-domain proof (Section 4.3).
- *A target set of bounded depth* against the source denotation is well founded, but the target denotation is unbounded, and the outputs of a call can come from curried entries of any depth.

The remaining design, which we adopt, relates a finite list of target elements to the whole source denotation. We also had to choose what the definition recurses on. Recursion on types, as in the proofs of Chlipala [2007] and Nielsen [2000], is not an option. Because `fold` and `unfold` mean the identity, the relation at $\mu X. A$ would be the relation at $A[\mu X. A/X]$, which is not a smaller type. Our relation instead recurses on the depth of the target elements.

*The key insight.* When a call $(\mathsf{car}(D') \bullet \mathsf{cdr}(D')) \bullet E'$ produces a result $w$, it uses only finitely many pairs of $D'$. One supplies a curried entry $U_e \mapsto (U_a \mapsto w)$, and others supply the environments in $U_e$. These pairs form a finite list $V \subseteq D'$ that contains both halves of the call, so the relation matches the curried functions and the environments *inside* $V$. The relation lets a call use exactly the curried entries whose environments $V$ itself provides, which excludes the junk. The finite list also makes the relation well founded.

*The approximant logical relation.* Figure 5 defines the approximant logical relation $D \trianglerighteq V$, which relates a finite list $V$ of target elements to a source set $D$. We read $D \trianglerighteq V$ as "$V$ approximates $D$". Its building block is the *usable entry* of a list. A target closure is a set of pairs $\langle\!| u, t |\!\rangle$ of a curried function $u$ and an environment $t$, so the entries of $u$ have the form $U_e \mapsto (U_a \mapsto w)$, which maps the environment $U_e$ to a function that maps the arguments $U_a$ to $w$. We write $U_a \mapsto w \mathrel{\dot{\in}} V$, read "$U_a \mapsto w$ is a usable entry of $V$", when such an entry is among the curried functions $\mathsf{car}(V)$ *and* the environments $U_e$ are among the environments $\mathsf{cdr}(V)$ that $V$ itself provides. The second condition makes the relation sound in the presence of junk. A curried entry for an environment that the program never supplies, such as the entry for [2] in the example above, is not usable. A call on $D'$ takes the curried function from $\mathsf{car}(D')$ and the environment from $\mathsf{cdr}(D')$. In the same way, the curried function and the environment of a usable entry may come from different pairs of $V$.

We write $V \cdot U \rightsquigarrow W$, read "$V$ maps $U$ to $W$", when every argument in $U$ is taken by some usable entry of $V$ (the first conjunct, $U \subseteq \mathsf{args}(V)$) and each result $w$ in $W$ is produced by a usable entry of $V$ whose arguments are among $U$. We call this a *self-contained call*, because $V$ supplies both the curried entries and the environments.

The relation $D \trianglerighteq_{k+1} V$ has four clauses, which use $\trianglerighteq_k$ at the recursive occurrences:

**Usable entries** $\boxed{U_a \mapsto w \mathrel{\dot\in} V}$

$$U_a \mapsto w \mathrel{\dot\in} V \iff \exists U_e.\ (U_e \mapsto (U_a \mapsto w)) \in \mathsf{car}(V) \ \wedge\ U_e \neq [\,] \ \wedge\ U_e \subseteq \mathsf{cdr}(V)$$

**Arguments of usable entries** $\boxed{\mathsf{args}(V) = D}$

$$\mathsf{args}(V) \ = \ \{ u \mid U_a \mapsto w \mathrel{\dot\in} V,\ u \in U_a \}$$

**Self-contained calls** $\boxed{V \cdot U \rightsquigarrow W}$

$$V \cdot U \rightsquigarrow W \iff U \subseteq \mathsf{args}(V) \ \wedge\ (\forall w \in W.\ \exists U_a.\ U_a \mapsto w \mathrel{\dot\in} V \wedge U_a \neq [\,] \wedge U_a \subseteq U)$$

**The approximant logical relation** $\boxed{D \trianglerighteq_k V \qquad D \trianglerighteq V}$

$$\begin{aligned}
D \trianglerighteq_0 V &\iff \mathsf{true} \\
D \trianglerighteq_{k+1} V &\iff (\forall c.\ c \in V \Rightarrow c \in D) \\
&\quad\ \wedge \ (V \neq [\,] \Rightarrow D \neq \emptyset) \\
&\quad\ \wedge \ (\forall U, W, E.\ V \cdot U \rightsquigarrow W \wedge \mathsf{con}(E) \wedge \mathsf{con}(U) \wedge E \trianglerighteq_k U \Rightarrow (D \bullet E) \trianglerighteq_k W) \\
&\quad\ \wedge \ (\forall \pi, W.\ W \subseteq \pi(V) \Rightarrow \pi(D) \trianglerighteq_k W) \\
D \trianglerighteq V &\iff D \trianglerighteq_{1+\mathsf{depth}(V)} V
\end{aligned}$$

Fig. 5. The approximant logical relation $D \trianglerighteq V$ between a set $D$ of source (EnvClos) elements and a finite list $V$ of target (PairClos) elements. Here $c$ ranges over constants and $\pi$ over the projections of tuple components and sum injections; car and cdr are as in Figure 3, con is consistency (Section 3), and $\mathsf{depth}(V)$ is the largest depth of an element of $V$ (Appendix A.2).

(1) every constant in $V$ is in $D$;
(2) if $V$ is nonempty, then $D$ is nonempty;
(3) whenever $V$ maps $U$ to $W$ and $E$ is a consistent set related to $U$, the set $D \bullet E$ is related to $W$; and
(4) every sublist of a projection of $V$ is related to the same projection of $D$.

Appendix E works clause 3 through an example.

*Termination of the Approximant Logical Relation.* The relation $D \trianglerighteq V$ is defined by recursion on the depth of $V$. Every recursive occurrence, including the negative one at the arguments of a call, is at a list shallower than $V$. The arguments $U$ are shallower than $V$ because the first conjunct of a self-contained call requires $U \subseteq \mathsf{args}(V)$; without that conjunct, $U$ could contain elements deeper than $V$. The results $W$ are shallower than $V$ because each of them is the output of an entry in $V$. Finally, the projections of $V$ are shallower than $V$ because they consist of components of the elements of $V$.

The index $k$ of $\trianglerighteq_k$ serves only to ensure termination. Any index that bounds the depth of $V$ gives the same relation, and therefore $D \trianglerighteq V$ is equivalent to the four clauses with $\trianglerighteq$ itself at the recursive occurrences (Lemma 4.4), and the proofs never mention the index.

Lemma 4.4 (Depth irrelevance and unfolding).

(1) *If every element in $V$ has depth less than $k$ and less than $j$, then $D \trianglerighteq_k V$ if and only if $D \trianglerighteq_j V$.*
(2) *$D \trianglerighteq V$ if and only if the four clauses of $D \trianglerighteq_{k+1} V$ hold with $\trianglerighteq$ in place of $\trianglerighteq_k$.*

The relation is also monotone in $D$ and antitone in $V$, that is, enlarging the source set or taking a sublist of the target list preserves it.

*The fundamental lemma.* A source environment $\rho$ and a target environment $\rho'$ are related, written $\rho \trianglerighteq \rho'$ and read "$\rho'$ approximates $\rho$", when $\rho(x) \trianglerighteq V$ for every variable $x$ and every $V \subseteq \rho'(x)$. We also require that both environments be consistent and that $\rho$ map every variable to a nonempty set.

Lemma 4.5 (Fundamental lemma). *Let $M$ be a term of EnvClos. If $\rho \trianglerighteq \rho'$ and $V \subseteq [\![\mathbf{split}(M)]\!]\rho'$, then $[\![M]\!]\rho \trianglerighteq V$.*

The proof is by induction on $M$. Throughout, the second part of Lemma 4.4 lets us pass between $D \trianglerighteq V$ and its four clauses, with $\trianglerighteq$ at the recursive occurrences. The case of a variable follows from $\rho \trianglerighteq \rho'$, and the cases of `fold`, `unfold`, tuples, injections, `let`, and `case` follow from the induction hypotheses for their subterms. Two cases carry the weight.

*Case $M = L\ K$.* Then $\mathbf{split}(M) = \texttt{callc}\,\mathbf{split}(L)\,\mathbf{split}(K)$, and we write $W$ for the lemma's list $V$, so $W \subseteq (\mathrm{car}(D') \bullet \mathrm{cdr}(D')) \bullet E'$ for $D' = [\![\mathbf{split}(L)]\!]\rho'$ and $E' = [\![\mathbf{split}(K)]\!]\rho'$. Each element of $W$ is produced by a curried entry from $\mathrm{car}(D')$, finitely many environments from $\mathrm{cdr}(D')$, and finitely many arguments from $E'$. We collect the pairs of $D'$ that supply these curried entries and environments into a finite list $V \subseteq D'$, and the arguments into a finite list $U \subseteq E'$. Then $V$ maps $U$ to $W$. The induction hypotheses give $[\![L]\!]\rho \trianglerighteq V$ and $[\![K]\!]\rho \trianglerighteq U$, and both $[\![K]\!]\rho$ and $U$ are consistent. Clause 3 for $[\![L]\!]\rho \trianglerighteq V$, applied to this self-contained call with $E = [\![K]\!]\rho$, gives $[\![L]\!]\rho \bullet [\![K]\!]\rho \trianglerighteq W$. That is, $[\![M]\!]\rho \trianglerighteq W$.

*Case $M = \texttt{clos}\,(z, y.\,N)\,\vec{K}$.* Then $\mathbf{split}(M) = \texttt{pair}\,(\texttt{code}\,(z, y.\,\mathbf{split}(N)))\,(\texttt{tuple}\,\mathbf{split}(\vec{K}))$, so $V$ is a list of pairs whose environments lie in $T' = \mathcal{T}([\![\mathbf{split}(\vec{K})]\!]\rho')$. Let $D = [\![M]\!]\rho$ and $T = \mathcal{T}([\![\vec{K}]\!]\rho)$. We prove the four clauses of $D \trianglerighteq V$. Clauses 1 and 4 hold trivially because $V$ contains only pairs.

For clause 2, suppose $\langle\!|u, t'|\!\rangle \in V$. Then $t' \in T'$, so $v_i \in [\![\mathbf{split}(K_i)]\!]\rho'$ for some $v_i$ at each $i$. The induction hypothesis for $K_i$, with $\rho \trianglerighteq \rho'$ and $[v_i] \subseteq [\![\mathbf{split}(K_i)]\!]\rho'$, gives $[\![K_i]\!]\rho \trianglerighteq [v_i]$, and clause 2 of the $\trianglerighteq$ relation makes $[\![K_i]\!]\rho$ nonempty. Hence $t \in T$ for some $t$. Every $\Lambda(F)$ contains $\nu$, so the curried function of the source closure contains the entry $[t] \mapsto \nu$, and applying that to $t \in T$ gives $\nu \in D$.

Clause 3 is the heart of the proof. Suppose that $V$ maps $U$ to $W$ and $E \trianglerighteq U$. The bodies of the source and target closures run in the environments $\rho_N = \rho_0[z{:=}T, y{:=}E]$ and $\rho'_N = \rho_0[z{:=}\mathrm{cdr}(V), y{:=}U]$. We show the two premises of the induction hypothesis for the body $N$. First, $W \subseteq [\![\mathbf{split}(N)]\!]\rho'_N$. Each $w \in W$ comes from a curried entry $U_e \mapsto (U_a \mapsto w)$ in $\mathrm{car}(V)$ with $U_e \subseteq \mathrm{cdr}(V)$ and $U_a \subseteq U$, so $w \in [\![\mathbf{split}(N)]\!]\rho_0[z{:=}U_e, y{:=}U_a] \subseteq [\![\mathbf{split}(N)]\!]\rho'_N$ by monotonicity (Lemma A.1). Second, $\rho_N \trianglerighteq \rho'_N$. We have $T \trianglerighteq \mathrm{cdr}(V)$ by the induction hypothesis for $\vec{K}$, because $\mathrm{cdr}(V) \subseteq T'$, and $E \trianglerighteq U$. With the two premises in hand, the induction hypothesis for $N$ gives us $[\![N]\!]\rho_N \trianglerighteq W$. Finally, Lemma 3.1 yields $[\![N]\!]\rho_N \subseteq D \bullet E$, and $\trianglerighteq$ is monotone in its source set, so $D \bullet E \trianglerighteq W$.

Lemma 4.6 (Correctness of **split closure**). *Let $M$ be a closed term of EnvClos. For every constant $c$, $c \in [\![M]\!]$ if and only if $c \in [\![\mathbf{split}(M)]\!]$. Moreover, $[\![M]\!] \neq \emptyset$ if and only if $[\![\mathbf{split}(M)]\!] \neq \emptyset$.*

Because $\rho_0 \trianglerighteq \rho_0$, reflection follows from Lemma 4.5. If $c \in [\![\mathbf{split}(M)]\!]$, the lemma applied to the one-element list $[c]$ gives $[\![M]\!] \trianglerighteq [c]$, and clause 1 gives $c \in [\![M]\!]$. If $[\![\mathbf{split}(M)]\!]$ contains an element $v$, the lemma applied to $[v]$ and clause 2 show that $[\![M]\!]$ is nonempty. Preservation follows in the same way from the routine induction on $M$ mentioned above, which shows that if $v \in [\![M]\!]\rho$ then $v \sqsubseteq [\![\mathbf{split}(M)]\!]\rho'$, for environments related pointwise by $\sqsubseteq$. For a constant, $c \sqsubseteq D'$ means $c \in D'$, and $v \sqsubseteq D'$ implies $D' \neq \emptyset$.

### 4.3 Proof of Compiler Correctness with Scott Domains

We also proved the compiler correct with Scott domains (Appendix F.5). There a value is an element of a universal domain $\mathbb{D}$ of tagged values, divergence is its least element $\bot$, and **split closure** is again the only pass that needs a relation. For space, this section only summarizes that proof; Appendix G gives it in full. We define the relation by recursion on the stages $\mathbb{D}_0, \mathbb{D}_1, \ldots$ of the bilimit that constructs $\mathbb{D}$, as Reynolds [1974] did. Let $d$ be a source value and $d'$ a target value, and write $\pi_n^\infty(d) \in \mathbb{D}_n$ for the approximation of $d$ at stage $n$. For preservation, the stage relation $\pi_n^\infty(d)\ \mathcal{R}_n^+\ d'$ relates the approximation of $d$ to the whole of $d'$. We write $d \trianglelefteq^+ d'$, read "$d'$ preserves $d$", when it holds at every stage. For reflection, the roles swap, and $d' \trianglelefteq^- d$, read "$d$ reflects $d'$", holds when $\pi_n^\infty(d')\ \mathcal{R}_n^-\ d$ at every stage (Figure 7 in Appendix G). The clauses for closures are

$$\begin{aligned}(\mathit{fun}, h)\ \mathcal{M}_n^+\ (\mathit{pr}, (c, e)) &\iff \forall v, a.\ v\ \mathcal{R}_n^+\ a \Rightarrow h(v)\ \mathcal{R}_n^+\ \mathit{call}(c, e, a)\\ (\mathit{pr}, (c, e))\ \mathcal{M}_{m+2}^-\ (\mathit{fun}, g) &\iff \forall v, a.\ v\ \mathcal{R}_m^-\ a \Rightarrow \mathit{call}_m((c, e), v)\ \mathcal{R}_m^-\ g'(a)\end{aligned}$$

where $\mathit{call}(c, e, a)$ applies the curried function $c$ to the environment $e$ and then to the argument $a$. The preservation clause has the shape of the relation that Nielsen [2000] used for defunctionalization. The reflection clause relates calls across stages, because a call of a pair at stage $m + 2$ applies its curried function twice and shows its result only at stage $m$ ($\mathit{call}_m$). Because the closure clause looks three levels into the target domain but only one into the source, the method of Pitts [1996] does not apply directly. The result, Lemma G.5, is the counterpart of Lemma 4.6 and the key step in proving Theorem 4.2.

## 5 Adequacy

Each of our five correctness results speaks of a different semantics. As due diligence on our mechanized definitions, we prove the graph model of FPC adequate with respect to the clocked interpreters, the CEK machines, and the reduction semantics, and we prove the Scott-domain semantics of FPC adequate with respect to the clocked interpreters. Each proof adapts a proof from the literature [Plotkin 1977, 1985; Siek 2017; Winskel 1993b]. Here is the statement of adequacy of the graph model for FPC with respect to the interpreter.

Proposition 5.1 (Adequacy). *Let $M$ be a closed term of* FPC. *For every constant $c$, $c \in [\![M]\!]$ iff the interpreter of* FPC *terminates on $M$ returning $c$, and $[\![M]\!]$ is empty iff the interpreter diverges on $M$, that is, times out at every clock.*

We have also proved that the graph model of $\mathrm{Instr}_\mu$ is adequate with respect to the interpreter for $\mathrm{Instr}_\mu$. These adequacy results therefore cover every semantics in our comparison and make the five correctness results equivalent.

## 6 Evaluation: Comparison with Other Semantics

To compare the sizes of the proofs across the five styles (RQ2) and to explain the differences (RQ3), we proved the same compiler correct three more times using several styles of operational semantics. This section summarizes the five developments, compares the sizes of their proofs pass by pass, and explains the differences.

### 6.1 Setup

All five developments prove the correctness of the same eight-pass compiler (Section 2), from the same Agda definitions of the languages and passes, and each is our best effort in its style.

- *Graph model.* The semantics of each language is the graph-model semantics of Appendix B, and we prove each pass as in Section 4.
- *Scott domains.* Meanings are continuous functions into a universal domain of tagged values (Appendix F.5), built as a bilimit with the domain theory of TypeTopology [Escardó 2026].

- *Abstract machines and simulations*, in the style of CompCert [Leroy 2009]. Each language has a small-step CEK machine (Appendix F.3), and we prove each pass by a forward simulation. The backward direction follows from forward simulation and determinism.
- *Clocked interpreters and value relations*, in the style of CakeML [Kumar et al. 2014]. (Clocked interpreters go back to Boyer and Moore [1984].) Each language has a total interpreter with a clock that ticks each time the interpreted program calls a function (Appendix F.2), and we prove each pass by showing that if one interpreter returns a result at some clock, the other returns a related result at some clock.
- *Reduction semantics and simulations.* We define a reduction semantics for each language, using substitution for function application (Appendix F.4), and we prove each pass by a simulation. We include reduction semantics in this comparison because of its central role in the theory of programming languages [Pierce 2002; Wright and Felleisen 1994].

*The correctness results and their proofs.* Each development proves the compiler correct with respect to its own semantics, for closed programs of integer type. The denotational theorems state that a program and its compilation produce the same integer and that one has a meaning exactly when the other does; the operational propositions state that both terminate with the same integer or both diverge, for three different operational semantics. We do not regard the operational or denotational formulation of correctness as more fundamental than the other. Scott and Strachey [1971] proposed a mathematical semantics as “the standard against which to judge an implementation”, while Plotkin [1977] judged denotational semantics by their closeness to an operational one. Our adequacy proofs (Section 5) make the choice immaterial. For programs of integer type, the five results are equivalent. We therefore count each development up to its own correctness result and report adequacy separately.

We measure each proof in lexical tokens of Agda (identifiers, operators, keywords, and delimiters, as Agda’s lexer sees them), not counting comments or imports. A count of lines depends on layout [Park 1992], and a count of lexical tokens, essentially the program length of Halstead [1977], does not. The counts include the semantics of the languages, the proof of each pass, and the composition of the passes. They exclude the code that all five proofs use unchanged: the intrinsically typed syntax of the languages and the functions that implement the passes (17,929 lexical tokens). We report separately the language-independent infrastructure of each style.

*AI Use.* All of the Agda code of the five developments was written by Claude Code (Opus 5.5 and Sonnet 5.5) under our direction. Claude started from our earlier Agda mechanizations of graph models of the $\lambda$-calculus and ISWIM [Landin 1966], which included adequacy proofs and a partial proof of correctness for closure conversion. The results do not depend on trusting the AI. Agda checks every development in safe mode, with no postulates. However, the Scott-domain theorems take three standard axioms as hypotheses because it depends on the TypeTopology library. The authors checked all the definitions and every statement of the theorems and propositions.

### 6.2 Results

Table 1 gives the sizes of the proofs for each pass of the compiler, with one column per semantic style, with the totals in the last rows. The proof with clocked interpreters is the smallest, at 35,860 lexical tokens, followed by the graph model at 40,905, abstract machines at 50,631, and reduction semantics at 68,243. One pass decides the graph model’s place. The proof of **split closure** alone is 8,672, more than a fifth of the total, against 773 to 2,534 in the operational styles. Without **split closure**, the graph model would be the smallest, at 32,233 against 35,087 for the interpreters. The second denotational proof, with Scott domains, is 54,972, between the abstract machines and reduction semantics: 1.3 times the graph model, and larger in every row but the semantics of the languages. Counting the infrastructure (Table 2) changes the order. The interpreters remain the smallest at 42,401, followed by the machines at 57,582, the graph model at 68,890, reduction semantics at 88,456, and Scott domains at 149,416. For comparison, Benton et al. [2010] report 4,400 lines of Coq for their domain theory, against 6,327 lines for TypeTopology's.

| | denotational | | operational | | |
|---|---|---|---|---|---|
| | graph model | Scott domains | machines | interpreters | reduction |
| semantics of the languages | 9,602 | **6,026** | 10,601 | 7,773 | 15,168 |
| lemmas shared by the passes | — | 3,186 | 2,971 | 2,289 | 2,289 |
| **simplify** | **7,044** | 9,393 | 11,880 | 11,339 | 14,589 |
| **enclose** | 3,462 | 4,355 | 6,728 | **3,221** | 11,716 |
| **split closure** | 8,672 | 15,211 | 1,650 | **773** | 2,534 |
| **hoist** | **2,796** | 3,282 | 4,740 | 3,387 | 4,275 |
| **remove complex operands** | 2,580 | 3,608 | 3,244 | **2,018** | 4,064 |
| **lift locals** | 1,924 | 2,016 | 2,315 | **1,597** | 7,137 |
| **explicate control** | 2,158 | 3,875 | 3,084 | **1,941** | 3,084 |
| **select instructions** | 2,153 | 3,428 | 3,067 | **1,144** | 3,067 |
| composition | 514 | 592 | 351 | 378 | **320** |
| total | 40,905 | 54,972 | 50,631 | **35,860** | 68,243 |

Table 1. Lexical tokens of Agda in the five correctness proofs. The smallest number in each row is in bold.

| | graph model | Scott domains | machines | interpreters | reduction |
|---|---|---|---|---|---|
| our infrastructure | 27,985 | 25,506 | 6,951 | **6,541** | 20,213 |
| TypeTopology's domain theory | — | 68,938 | — | — | — |
| total with Table 1 | 68,890 | 149,416 | 57,582 | **42,401** | 88,456 |

Table 2. Lexical tokens of Agda in the infrastructure that each proof relies on, which Table 1 excludes. For Scott domains, TypeTopology's domain theory is the modules on dcpos and lifting that the proof imports. Not counted: the Agda standard library and TypeTopology's general foundations (224,533 lexical tokens).

### 6.3 Analysis

We identify two tendencies in the proof sizes of Table 1 (RQ3). The size of a denotational proof tends to depend most on whether the pass preserves the meaning of a program exactly. When it does, the proof is an induction whose cases are congruences and a few laws of the model (Section 4.1), largely independent of how the pass changes the way the program runs. The size of an operational proof tends to depend instead on how closely the run of the target follows the run of the source. On the denotational side, the correctness of seven of the eight passes is an equation between the meaning of a program and the meaning of its translation; only **split closure** is the exception. On the operational side, the runs of the source and target correspond closely for most passes, including **split closure**, but differ more for **simplify** and, to a lesser degree, **hoist**. Combining the two tendencies gives three groups of passes, which we discuss in turn.

*Equal meanings, similar runs: the interpreter proofs tend to be smaller.* For **enclose**, **remove complex operands**, **lift locals**, **explicate control**, and **select instructions**, both the graph-model and the interpreter proofs are short. The interpreter proofs are smaller, but by only 7% for **enclose** and 10% for **explicate control**. The Scott-domain proofs of these passes are 1.3 to 3.0 times the interpreter proofs, for the reasons given at the end of this section. The congruence cases take one line per construct in both styles. The difference lies in the obligations of the graph model. Because a term denotes a set of values, some operators require their sets to be nonempty or consistent, and some laws have premises, such as the $\beta$-law of Lemma 3.1 and the commuting conversions that move a `let` or `case` past a continuation. An interpreter computes a single value, so these obligations do not arise. These obligations cost the most in the graph-model proof of **select instructions**, whose function prologue needs two uses of Lemma 3.1, and which is about twice the interpreter proof. The commuting conversion that the continuation-passing semantics of Imp needs makes the graph-model proof of **lift locals** 1.2 times the interpreter proof. The tendency concerns the interpreters only. The proofs in the reduction semantics, which works by substitution, are often several times larger even for these passes, as for **enclose** and **lift locals**.

*Equal meanings, different runs: the denotational proofs tend to be smallest.* **simplify** is the clearest case, at 7,044 lexical tokens in the graph model and 9,393 with Scott domains, against 11,339 to 14,589 for the operational proofs. This pass reduces expression under binders and inlines calls, so the target takes fewer steps than the source. Each operational proof must replay those reductions as steps of the source program: on open terms in the reduction semantics, or as a run-ahead of the source machine or interpreter, with stuttering where the target does more. For the interpreters, the target needs fewer clock ticks than the source, which breaks the usual clock-for-clock argument. The **hoist** pass fits the pattern less cleanly, at 2,796 in the graph model and 3,282 with Scott domains, against 3,387 to 4,740 for the operational proofs. The interpreter proof is clock for clock; the runs differ only in where the code of a function lives, because in the target it is a reference into a table of definitions. The interpreter proof relates the two tables value by value, while the denotational proof shows that the meaning of a reference equals the meaning of the code it replaces.

*Changed representations: the operational proofs are much smaller.* **Split closure** changes the representation of closures, so its correctness is not an equation, yet the run of the target follows the run of the source closely. Its denotational proofs take 8,672 lexical tokens in the graph model and 15,211 with Scott domains, against 773 to 2,534 in the operational styles. For the graph model, most of the difference has three sources. First, the denotational proof needs two directions, while a simulation proof gets the backward direction for free from forward simulation and the determinism of the target. The graph model has an analogue of determinism, the consistency of denotations, and it does part of the job. If the source produces a constant, the forward direction puts that constant in the target, and consistency rules out any other. But the graph model represents divergence as the empty set, the absence of evidence, and a forward relation cannot carry absence. Showing that a diverging program compiles to a diverging program is therefore a separate argument, and that argument is reflection. Second, reflection needs the finite lists, the self-contained calls, and the lemma that extracts them from a target call. Third, the $\beta$-law of Lemma 3.1 holds only under side conditions of nonemptiness, consistency, and continuity, which a machine does not need. The Scott-domain proof shares the first source, and its stage relations replace the other two with the costs discussed next.

*The graph model is smaller than Scott domains.* The two denotational proofs have the same structure, yet the Scott-domain proof is 1.3 times the graph model, larger in every row but the semantics: 1.05 to 1.8 times for the equational passes, and 15,211 lexical tokens against 8,672 for **split closure**. Three properties of Scott domains explain much of the difference. First, a Scott domain has infinite elements, so the relation for **split closure** is built by stages (Section 4.3). The proof shows that each stage is a proposition, downward closed, admissible, and preserved by the maps between stages. Second, the relation for a call at one stage refers to an earlier stage, and relating calls across stages is the most expensive part of the proof. Third, the universal domain is a sum of tagged elements. The primitives and the tags need their own lemmas, and the shared lemmas are mostly laws of strictness and definedness, which replace nonemptiness. In exchange, the proof needs no consistency. The graph-model relations recurse on finite elements, and neither of the first two costs arises. The infrastructure widens the gap further (Table 2).

### 6.4 Threats to Validity

We measure the size of the proofs, which is only a proxy for the effort of writing, checking, and maintaining them. Lexical tokens, lines, and non-whitespace characters rank the totals alike, but lines depend on layout. Size also says little about difficulty, which depends on whether a proof has precedent in the literature or its authors must invent it. Because experience carries over from one development to the next, we gave every development a final cleanup pass under the same rules (shorten and clarify the proofs but change no statement and move nothing into the uncounted infrastructure). The cleanup did not change the order of the four other proofs; the Scott-domain proof, written last, shrank by more than a quarter, moving it ahead of reduction semantics.

The comparison covers one compiler, and the totals depend on its selection of passes. The denotational proofs tend to be smaller for optimizations such as **simplify** and several times larger for passes that change the representation of values such as **split closure**, so a compiler with more of one kind or the other could rank the styles differently. The sizes also depend on choices of semantics, for example the continuation-passing semantics of Imp, which makes **explicate control** an equation, and on proof engineering that could still shrink any of the five developments. Like our graph model, the Scott-domain semantics uses one universal domain of tagged values, as Milner [1978] and MacQueen et al. [1986] did. A domain for each type, more usual for typed languages [Benton et al. 2009; Dockins 2014], would avoid the case analysis on tags, a tenth of the Scott-domain proof of **split closure**, but would need a domain equation for each recursive type.

The sizes also depend on the proof assistant. Agda's main automation is normalization, which probably favors the interpreters, whose results it computes directly; with the tactics of Rocq or Lean, the relative sizes of the five styles could differ.

## 7 Related Work

The introduction discusses the work closest to ours. This section covers compiler proofs by semantic style, operational proofs of closure conversion, relations between denotations, and graph models.

*Semantic styles in compiler correctness.* The idea that a compiler is correct when a diagram of meanings commutes goes back to Morris [1973] and Thatcher et al. [1981], and Milne and Strachey [1976] related semantics at different levels of abstraction with inclusive predicates. Among proofs of particular compilers, those of Pardo et al. [2018] and Siek [2017] are denotational for both the source and the target of each pass. Pardo et al. [2018] give clocked denotations to a first-order imperative language and a stack language, and Siek [2017] proves an inliner correct by equality of graph-model denotations. Other proofs mix the styles, with a denotational semantics on only one side. Benton and Hur [2009] relate a domain-theoretic semantics of a typed language with recursion to an operational SECD machine, using step-indexed biorthogonal relations; the target machine has primitive closures. Œuf [Mullen et al. 2018] gives a denotation only to its total source language and proves every pass operationally. Most compiler proofs, finally, are operational on both sides. The CLI stack [Moore 1989; Young 1989] runs the source for $n$ steps and the target for a number of steps given by a clock function, and shows that the target reproduces every run of the source that ends without error, but says nothing about divergence. CompCert [Leroy 2009] and CakeML [Kumar et al. 2014], discussed in the introduction, use simulations and clocked big-step semantics, respectively. The calculated compilers of Bahr and Hutton [2015] derive a compiler and its virtual machine from a definitional interpreter; the virtual machine keeps closures as values. Our back end adapts the operational proofs of Siek [2025] for a first-order compiler.

*Operational proofs of closure conversion.* Typed closure conversion represents a closure as a package of an existential type, as our PairClos does. Minamide et al. [1996] prove it correct with logical relations between source and target terms, Ahmed and Blume [2008] show that it preserves observational equivalence, and Bowman and Ahmed [2018] extend it to the calculus of constructions. Dargaye [2009] and Hur and Dreyer [2011] prove closure conversion correct within compilers for core ML. Pottier and Gauthier [2006] prove that defunctionalization preserves convergence and divergence with a simulation. The simulation proof of Wand and Steckler [1994] covers only terminating programs, and Neis et al. [2015] use parametric inter-language simulations. Ghica and Zanasi [2027] present closure conversion for the $\lambda$-calculus with string diagrams and show that a closure-converted $\beta$-redex reduces to the same result as the original. That lemma would be the main case of an operational proof by simulation.

*Categorical accounts of closure conversion.* Schweimeier and Jeffrey [1999] justify closure conversion for an ML-like language with references and recursive functions with a categorical model presented as flow graphs. They prove that every program is equal, by the equations of these categories, to a program whose functions are all at top level. Their closures keep a function type, with the code and the environment inseparable, so the target is still higher-order, and they place the transformation "part-way between closure conversion and lambda-lifting." Their equations hold in every model of these categories. A statement about divergence therefore requires a particular model that is adequate for the execution of programs, a step the paper leaves open. In a blog post, New [2017] gives a categorical account of the parametricity argument of Ahmed and Blume [2008], deriving the isomorphism between $A \to B$ and $\exists\Gamma.\, \Gamma \times (\Gamma \times A \to B)$ from the coYoneda lemma.

*Relations between denotations.* Relating two denotational semantics is an old problem. Our Scott-domain relation for **split closure** (Section 4.3) follows Reynolds [1974], who related the direct and continuation semantics of a language by a relation built stage by stage along the inverse-limit construction of the domains [Scott 1971]. His index ranges over the stages of both domains, and his relation is the intersection over all stages. Ours recurses on the stages of one side only and relates each approximation to the whole other value. The approximant logical relation (Section 4.2) has a similar shape, but its index bounds the depth of finite elements of the target side, and depth irrelevance (Lemma 4.4) makes an intersection unnecessary. Nielsen [2000] relates the denotational meanings of a program and its defunctionalization with a type-indexed relation whose function clause, like that of our preservation relation, relates a function to a closure whose calls give related results. His source has no recursive types, so recursion on types suffices, and he proves only preservation. Later work related two semantics with inclusive predicates [Milne and Strachey 1976; Stoy 1977], finitary projections [Mulmuley 1987], or types [Filinski 2007; Meyer and Wand 1985]. Pitts [1996] showed that a relation specified by a mixed-variance equation over a recursively defined domain exists and is unique, but his method does not apply directly to **split closure** (Section 4.3).

*Graph models and consistency.* Our model descends from the graph models of Scott [1976], Engeler [1981], and Plotkin [1993] and is related to the filter models of Barendregt et al. [1983]. Our consistency is pairwise, as in the coherent domains of Plotkin [1978], and our rule for entries is the function-space consistency of information systems [Larsen and Winskel 1984; Scott 1982] in the pairwise form of Schwichtenberg [2006].

## 8 Conclusion

Our main results are denotational proofs, one for a graph model and another for Scott domains, that closure conversion preserves and reflects the meaning of programs, divergence included, for a Turing-complete language. The proof of semantic reflection for the graph model rests on our new *approximant logical relation*. We extend the proofs to an eight-pass compiler and mechanize them in Agda. Comparing the size of these proofs to operational proofs of the same compiler shows a split by the kind of pass. Where a pass keeps the representation of values, its denotational proof is an equation between meanings, and the graph-model proofs are smaller than the machine and reduction proofs. For the source optimizer, the graph-model proof is the smallest of the five. Where a pass changes that representation, as **split closure** does, proving semantic reflection requires a relation between denotations, and the operational proofs, which follow the run of the target and get reflection from determinism, are several times smaller. Not counting infrastructure, the clocked interpreters give the smallest proof, the graph model the second, and Scott domains the fourth.

## Data-Availability Statement

The artifact is our Agda development: the five correctness proofs of the compiler, the adequacy proofs of Section 5, the shared definitions of the languages and passes, and the script that computes the proof sizes of Section 6. It checks with Agda 2.8 and its standard library 2.4 in safe mode, with no postulates, in about two minutes from scratch on a desktop workstation (Intel Core Ultra 9 285K, 64 GB). The Scott-domain proofs also need the modules of TypeTopology that it imports, which the artifact includes, lightly patched, from a pinned commit. With them, the Scott-domain proofs check in safe mode in about six and a half more minutes, and its theorems take propositional truncation, function extensionality, and propositional extensionality as hypotheses. A Dockerfile builds and checks everything. The README of the artifact lists the commands that check each development and regenerate Tables 1 and 2, and it maps each claim of the paper to the file that states it. An anonymized version is provided as supplementary material, and we will submit it for Artifact Evaluation.

The appendices complete the definitions of the language semantics and compiler passes used in the paper. Appendix A completes the graph model of Section 3; Appendix B gives the type system and graph-model semantics of each language; Appendix C defines the compiler passes of Section 2; Appendix D expands the equational proofs in Section 4.1 to more compiler passes; Appendix E works through an example of the approximant logical relation of Section 4.2; and Appendix F gives the semantics of the source and target languages in the four other semantic styles of Section 6, with their correctness results; and Appendix G gives the Scott-domain proof of **split closure**. Each section begins with a line, marked "Agda:", that names the modules of the artifact that contain the section's definitions and proofs, with paths relative to its `agda/` directory.

## A The Graph Model

This appendix completes the definition of the graph model of Section 3: consistency, which $\Lambda$ and $C$ require of their inputs and the proof of **split closure** relies on; the depth of an element, on which the approximant logical relation recurses; and the operators of the semantics not shown in Figure 3.

### A.1 Consistency

Agda: `Framework/GraphModel/Domain.agda, proof/Framework/GraphModel/Domain.agda`

Consistency is defined on elements, and from it on sets and lists (Section 3). Two constants are consistent when they are equal, two entries are consistent when consistent inputs lead to consistent outputs, and elements that describe different kinds of values, such as $\mathsf{left}(u)$ and $\mathsf{right}(v)$, are not consistent. A set $D$ is consistent, written $\mathsf{con}(D)$, when its elements are pairwise consistent, and two lists are consistent with each other, written $V \approx V'$, when every element of one is consistent with every element of the other.

**Consistency of elements** $\boxed{u \sim v}$

$$\begin{aligned} c \sim c' &\iff c = c' \\ (V \mapsto w) \sim (V' \mapsto w') &\iff (V \approx V' \Rightarrow w \sim w') \qquad (V \mapsto w) \sim \nu, \quad \nu \sim \nu \\ \langle\!| u, v |\!\rangle \sim \langle\!| u', v' |\!\rangle &\iff u \sim u' \land v \sim v' \\ \mathsf{tup}_i(v) \sim \mathsf{tup}_j(v') &\iff i = j \Rightarrow v \sim v' \\ \mathsf{left}(v) \sim \mathsf{left}(v') &\iff v \sim v' \qquad \mathsf{right}(v) \sim \mathsf{right}(v') \iff v \sim v' \end{aligned}$$

and no other pairs of elements are consistent.

**Consistency of sets and lists** $\boxed{\mathsf{con}(D) \qquad V \approx V'}$

$$\begin{aligned} \mathsf{con}(D) &\iff u \sim v \text{ for all } u, v \in D \\ V \approx V' &\iff u \sim v \text{ for all } u \in V \text{ and } v \in V' \end{aligned}$$

Fig. 6. Consistency of elements, sets, and lists.

### A.2 Depth

Agda: `proof/Framework/GraphModel/DomainUtilBase.agda`

The approximant logical relation of Section 4.2 is defined by recursion on the depth of a list of elements. The depth of an element is the height of its syntax tree, not counting constants, $\nu$ and $\langle\rangle$;

the depth of a finite list is the largest depth of its elements (zero for the empty list):

$$\begin{aligned}
\mathrm{depth}(V \mapsto w) &= 1 + \max(\mathrm{depth}(V), \mathrm{depth}(w)) \\
\mathrm{depth}(\langle\!| u, v |\!\rangle) &= 1 + \max(\mathrm{depth}(u), \mathrm{depth}(v)) \\
\mathrm{depth}(t(v)) &= 1 + \mathrm{depth}(v) \quad \text{for } t \in \{\mathrm{tup}_i, \mathrm{left}, \mathrm{right}\} \\
\mathrm{depth}(v) &= 0 \quad \text{for } v \in \{c, \nu, \langle\rangle\} \\
\mathrm{depth}(V) &= \max_{v \in V} \mathrm{depth}(v)
\end{aligned}$$

## A.3 Further Operators

Agda: `Framework/GraphModel/Ops.agda, Semantics/Graph/Prim.agda`

Besides the operators of Figure 3, the semantics of the languages use the following ones. The empty tuple means $\mathcal{T}() = \{\langle\rangle\}$. A guard makes a set strict in other sets:

$$\mathrm{guard}(D_1, \ldots, D_n;\ E) = E \text{ if } D_1, \ldots, D_n \text{ are nonempty, and } \emptyset \text{ otherwise.}$$

## A.4 Monotonicity

Agda: `FPC/proof/Semantics/GraphModel/{FPC,EnvClos,PairClos}.agda`

Lemma A.1 (Monotonicity). *Let $M$ be a term of* FPC, *EnvClos, or PairClos. If $\rho(x) \subseteq \rho'(x)$ for every variable $x$, then $[\![M]\!]\rho \subseteq [\![M]\!]\rho'$.*

# B The Languages

This appendix gives, for each language of Section 2 in the order of the pipeline, its type system and its semantics in the graph model. In this Appendix, we write variables by name. In the Agda development, we instead write variables as de Bruijn indices, and every term is well scoped by construction. The type systems are those of our intrinsically typed Agda definitions, so every pass maps well-typed terms to well-typed terms. The semantics are the ones that Theorem 4.1 and the proofs of Section 4 speak of. Figure 2 lists the base types. The result type $out(p)$ of a primitive is `Int` for `sub` and `tup() + tup()` for `eq`.

In the semantics, an environment $\rho$ maps each variable to a set of elements, and $\rho[x{:=}X]$ maps $x$ to $X$ and every other variable as $\rho$ does. Every position of an environment must be nonempty, so the environment $\rho_0$ of a closed program maps every variable to $\{\langle\rangle\}$, and so do the positions of an environment that hold no variable. The meaning of a closed program $M$ is $[\![M]\!] = [\![M]\!]\rho_0$. The operators are those of Figure 3 and Appendix A.3.

## B.1 The FPC Language

Agda: `FPC/Language/FPC.agda, FPC/Semantics/GraphModel/FPC.agda`

FPC is the source language, and the first pass, **simplify**, maps it to itself. A judgment $\Gamma \vdash M : A$ types a term in a context $\Gamma$ of variable types.

**Typing of** FPC $\boxed{\Gamma \vdash M : A}$

$$\frac{x{:}A \in \Gamma}{\Gamma \vdash x : A} \quad \frac{\Gamma, x{:}A \vdash N : C}{\Gamma \vdash \lambda x.\, N : A \to C} \quad \frac{\Gamma \vdash L : A \to C \quad \Gamma \vdash M : A}{\Gamma \vdash L\, M : C} \quad \frac{c \text{ a constant of base type } B}{\Gamma \vdash c : B}$$

$$\frac{\overrightarrow{\Gamma \vdash M_i : A_i}^{\,i}}{\Gamma \vdash \mathtt{tuple}\ \vec{M} : \mathtt{tup}(\vec{A})} \quad \frac{\Gamma \vdash M : \mathtt{tup}(\vec{A}) \quad 1 \le i \le |\vec{A}|}{\Gamma \vdash \mathtt{get}_i\, M : A_i}$$

$$\frac{\Gamma \vdash M : A}{\Gamma \vdash \mathtt{inl}\ M : A + C} \quad \frac{\Gamma \vdash M : C}{\Gamma \vdash \mathtt{inr}\ M : A + C} \quad \frac{\Gamma \vdash L : A + C \quad \Gamma, x{:}A \vdash M : D \quad \Gamma, y{:}C \vdash N : D}{\Gamma \vdash \mathtt{case}\ L\ (x.\, M)\ (y.\, N) : D}$$

$$\frac{\Gamma \vdash L : \mathtt{Int} \quad \Gamma \vdash M : \mathtt{Int}}{\Gamma \vdash \mathtt{prim}_p\ L\ M : out(p)} \quad \frac{\Gamma \vdash M : A \quad \Gamma, x{:}A \vdash N : C}{\Gamma \vdash \mathtt{let}\ x = M\ \mathtt{in}\ N : C}$$

$$\frac{\Gamma \vdash M : A[\mu X.\, A/X]}{\Gamma \vdash \mathtt{fold}\ M : \mu X.\, A} \quad \frac{\Gamma \vdash M : \mu X.\, A}{\Gamma \vdash \mathtt{unfold}\ M : A[\mu X.\, A/X]}$$

Figure 3 gives the semantics of FPC.

## B.2 The Clos and EnvClos Languages

Agda: `FPC/Language/Clos.agda`, `FPC/Language/EnvClos.agda`, `FPC/Semantics/GraphModel/Clos.agda`, `FPC/Semantics/GraphModel/EnvClos.agda`

Clos is the intermediate language between the steps of **enclose**, whose closures bind a name for each captured value. EnvClos, the target of **enclose** and the source of **split closure**, stores the captured values of a closure in one tuple. Both type and interpret their other forms as FPC does. A closure's type does not show the types $\Delta = A_1, \ldots, A_n$ of its captured values.

**Typing of closures** $\boxed{\Gamma \vdash M : A}$

$$\frac{\overrightarrow{\Gamma \vdash M_i : A_i}^{\,i} \qquad x_1{:}A_1, \ldots, x_n{:}A_n, y{:}A \vdash N : C}{\Gamma \vdash \mathtt{clos}\ (\vec{x}, y.\, N)\ \vec{M} : A \to C}\ (\text{Clos})$$

$$\frac{\overrightarrow{\Gamma \vdash M_i : A_i}^{\,i} \qquad z{:}\mathtt{tup}(\Delta),\ y{:}A \vdash N : C}{\Gamma \vdash \mathtt{clos}\ (z, y.\, N)\ \vec{M} : A \to C}\ (\text{EnvClos})$$

Below, $[\![\vec{M}]\!]\rho$ stands for $[\![M_1]\!]\rho, \ldots, [\![M_n]\!]\rho$. A Clos closure is the function of its parameter $y$ whose code runs with the captured values for $x_1, \ldots, x_n$; closures are strict, so it is guarded by the captured values. An EnvClos closure applies the curried meaning of its code to the tuple of the captured values (Section 4.2).

**Semantics of closures** $\boxed{[\![M]\!]\rho = D}$

$$[\![\mathtt{clos}\ (\vec{x}, y.\, N)\ \vec{M}]\!]\rho = \mathsf{guard}([\![M_1]\!]\rho, \ldots, [\![M_n]\!]\rho;\ \Lambda(\boldsymbol{\lambda} Y.\, [\![N]\!]\rho_0[\vec{x}{:=}[\![\vec{M}]\!]\rho, y{:=}Y]))\ (\text{Clos})$$

$$[\![\mathtt{clos}\ (z, y.\, N)\ \vec{M}]\!]\rho = \Lambda(\boldsymbol{\lambda} X.\, \Lambda(\boldsymbol{\lambda} Y.\, [\![N]\!]\rho_0[z{:=}X, y{:=}Y])) \bullet \mathcal{T}([\![\vec{M}]\!]\rho) \quad (\text{EnvClos})$$

## B.3 The PairClos Language

Agda: `FPC/Language/PairClos.agda`, `FPC/Semantics/GraphModel/PairClos.agda`

PairClos, the target of **split closure**, replaces closures and application with code, pairs, and calls, and types and interprets its other forms as FPC does.

**Typing of PairClos** $\boxed{\Gamma \vdash M : A}$

$$\frac{z{:}\mathtt{tup}(\Delta),\ y{:}A \vdash N : C}{\Gamma \vdash \mathtt{code}\ (z, y.\, N) : \mathtt{code}(\Delta; A \to C)} \quad \frac{\Gamma \vdash L : \mathtt{code}(\Delta; A \to C) \quad \Gamma \vdash M : \mathtt{tup}(\Delta)}{\Gamma \vdash \mathtt{pair}\ L\ M : A \to C}$$

$$\frac{\Gamma \vdash L : A \to C \quad \Gamma \vdash M : A}{\Gamma \vdash \mathtt{callc}\ L\ M : C}$$

**Semantics of PairClos** $\boxed{[\![M]\!]\rho = D}$

$$[\![\texttt{code}\ (z, y.\, N)]\!]\rho = \Lambda(\boldsymbol{\lambda} X.\, \Lambda(\boldsymbol{\lambda} Y.\, [\![N]\!]\rho_0[z{:=}X, y{:=}Y])) \qquad [\![\texttt{pair}\ L\ M]\!]\rho = \mathsf{pair}([\![L]\!]\rho, [\![M]\!]\rho)$$

$$[\![\texttt{callc}\ L\ M]\!]\rho = (\mathsf{car}([\![L]\!]\rho) \bullet \mathsf{cdr}([\![L]\!]\rho)) \bullet [\![M]\!]\rho$$

## B.4 The GlobalFun and MNF Languages

Agda: `FPC/Language/GlobalFun.agda, FPC/Language/Mon.agda, FPC/Semantics/GraphModel/GlobalFun.agda`

GlobalFun, the target of **hoist**, makes the code of every function a global definition. A GlobalFun program has a table $\Sigma = \Sigma_0, \dots, \Sigma_{m-1}$ of the signatures $\Sigma_k = (\Delta_k; A_k \to C_k)$ of its definitions. The $k$-th definition, referred to by $f_k$, has environment types $\Delta_k$, argument type $A_k$, and result type $C_k$. A term is typed by the table and its context, $\Sigma; \Gamma \vdash M : A$, and a program by its table, $\Sigma \vdash P : A$. The other forms are typed as in PairClos.

**Typing of GlobalFun** $\boxed{\Sigma; \Gamma \vdash M : A \qquad \Sigma \vdash P : A}$

$$\frac{\Sigma_k = (\Delta_k; A_k \to C_k)}{\Sigma; \Gamma \vdash f_k : \mathsf{code}(\Delta_k; A_k \to C_k)}$$

$$\frac{\overrightarrow{d_k = \texttt{code}\ (z, y.\, N_k) \quad \Sigma_k = (\Delta_k; A_k \to C_k) \quad \Sigma; z{:}\mathsf{tup}(\Delta_k), y{:}A_k \vdash N_k : C_k}^{\,k} \qquad \Sigma; \emptyset \vdash M : A}{\Sigma \vdash (\vec{d};\ M) : A}$$

The premises under the arrow stand for one copy for each index $k$ of a definition $d_k$ of $\vec{d}$. A term means a set given a table $T$ that gives each definition its meaning, $[\![M]\!]_T\rho$; the other forms are interpreted as in PairClos, passing $T$ along. A definition means the curried function of its code, given the table $T_k$ of the definitions older than it ($T_k(j) = T(j)$ for $j < k$, and $T_k(j) = \emptyset$ otherwise).

**Semantics of GlobalFun** $\boxed{[\![M]\!]_T\rho = D \qquad T(k) = D}$

$$[\![f_k]\!]_T\rho = T(k) \qquad T(k) = \Lambda(\boldsymbol{\lambda} X.\, \Lambda(\boldsymbol{\lambda} Y.\, [\![N_k]\!]_{T_k}\rho_0[z{:=}X, y{:=}Y]))$$

$$[\![(d_0, \dots, d_{m-1};\ M)]\!] = [\![M]\!]_T\rho_0$$

MNF, the target of **remove complex operands**, has the syntax of Figure 2, is typed as GlobalFun, and has the meaning of its embedding into GlobalFun, because an atom and an operation on atoms are GlobalFun terms already.

## B.5 The Imp Language

Agda: `FPC/Language/Frame.agda, FPC/Language/Imp.agda, FPC/Semantics/GraphModel/Imp.agda`

Imp, the target of **lift locals**, replaces variables by slots of a stack frame. A function body $(\Phi; e)$ declares its frame, the list $\Phi = A_0, \dots, A_{n-1}$ of the types of its slots. We write $x : A \in \Phi$ when slot $x$ has type $A$, which it keeps for the whole body. A function body is typed by the types $\Phi_p$ of its parameters, $\Sigma; \Phi_p \vdash d : A$. Its frame starts with the slots of the parameters, $0, \dots, |\Phi_p| - 1$, and its expression starts with exactly those slots initialized. An expression is typed by the set $I$ of the slots that are initialized where it starts, and a slot $x$ can be read only where it is initialized. An assignment $x := e_1;\ e_2$ writes the value of $e_1$ to slot $x$, and `case` $a\ x\ y\ e_1\ e_2$ writes the payload of the injection $a$ to slot $x$ and runs $e_1$, or to slot $y$ and runs $e_2$. Both write a slot that is not initialized, so a write never overwrites a variable that is in scope. A constant $c$, a function reference $f_k$, and an operation on atoms $o\ \vec{a}$ are typed as in MNF, with the atoms of an operation typed by the rules here.

**Typing of Imp** $\boxed{\Sigma; \Phi; I \vdash e : C \qquad \Sigma; \Phi_p \vdash d : A \qquad \Sigma \vdash P : A}$

$$\frac{x : A \in \Phi \quad x \in I}{\Sigma; \Phi; I \vdash x : A} \qquad \frac{x : A \in \Phi \quad x \notin I \quad \Sigma; \Phi; I \vdash e_1 : A \quad \Sigma; \Phi; I \cup \{x\} \vdash e_2 : C}{\Sigma; \Phi; I \vdash x := e_1;\ e_2 : C}$$

$$\frac{x : A_1 \in \Phi \quad y : A_2 \in \Phi \quad \Sigma; \Phi; I \vdash a : A_1 + A_2 \quad x, y \notin I \quad \Sigma; \Phi; I \cup \{x\} \vdash e_1 : C \quad \Sigma; \Phi; I \cup \{y\} \vdash e_2 : C}{\Sigma; \Phi; I \vdash \mathtt{case}\ a\ x\ y\ e_1\ e_2 : C}$$

$$\frac{\Phi = \Phi_p, \Phi' \quad \Sigma; \Phi; \{0, \ldots, |\Phi_p| - 1\} \vdash e : A}{\Sigma; \Phi_p \vdash (\Phi; e) : A}$$

$$\frac{\overrightarrow{\Sigma_k = (\Delta_k; A_k \to C_k) \quad \Sigma; \mathtt{tup}(\Delta_k), A_k \vdash d_k : C_k}^{k} \quad \Sigma; \emptyset \vdash d : A}{\Sigma \vdash (\vec{d};\ d) : A}$$

The premises under the arrow stand for one copy for each index $k$ of a definition $d_k$ of $\vec{d}$. A definition has two parameters, its environment in slot 0 and its argument in slot 1; the main function body has none, so it starts with no slot initialized. The semantics is in continuation-passing style. The frame $\sigma$ maps each slot to a set, and the continuation $\kappa$ takes the meaning of the expression. A write updates its slot in place. An atom means $\mathcal{A}[\![x]\!]\sigma = \sigma(x)$, $\mathcal{A}[\![c]\!]\sigma = \{c\}$, or $\mathcal{A}[\![f_k]\!]\sigma = T(k)$, and a simple expression $\mathcal{S}[\![o\ \vec{a}]\!]\sigma$ applies the GlobalFun operator of $o$ to the meanings of its atoms. An assignment is strict in the value it assigns, like a `let`.

**Semantics of Imp** $\quad [\![e]\!]_T\sigma\kappa = D$

$$
\begin{aligned}
[\![rhs]\!]_T\sigma\kappa &= \kappa(\mathcal{S}[\![rhs]\!]\sigma) \\
[\![x := e_1;\ e_2]\!]_T\sigma\kappa &= [\![e_1]\!]_T\sigma\,(\boldsymbol{\lambda} D.\,\Lambda(\boldsymbol{\lambda} X.\,[\![e_2]\!]_T\sigma[x{:=}X]\kappa) \bullet D) \\
[\![\mathtt{case}\ a\ x\ y\ e_1\ e_2]\!]_T\sigma\kappa &= C(\mathcal{A}[\![a]\!]\sigma,\ \boldsymbol{\lambda} X.\,[\![e_1]\!]_T\sigma[x{:=}X]\kappa,\ \boldsymbol{\lambda} X.\,[\![e_2]\!]_T\sigma[y{:=}X]\kappa) \\
T(k) &= \Lambda(\boldsymbol{\lambda} X.\,\Lambda(\boldsymbol{\lambda} Y.\,[\![e_k]\!]_{T_k}\rho_0[0{:=}X, 1{:=}Y]\ (\boldsymbol{\lambda} D.\,D))) \\
[\![(\vec{d};\ (\Phi; e))]\!] &= [\![e]\!]_T\rho_0\,(\boldsymbol{\lambda} D.\,D)
\end{aligned}
$$

## B.6 The CFG Language

Agda: `FPC/Language/CFG.agda`, `FPC/Semantics/GraphModel/CFG.agda`

CFG, the target of **explicate control**, organizes each function body as a control-flow graph. A function body $(\Phi; \vec{t};\ t)$ declares its frame $\Phi$, as in Imp, and has numbered blocks $\vec{t}$ and a starting tail $t$. A tail has the result type $A$ of its function body, written $\Sigma; \Phi; \vec{I}; I \vdash t : A$, and is typed by the list $\vec{I}$ of the sets of slots that are initialized where the blocks it may jump to start, one set $I_l$ for each block $t_l$, and the set $I$ of the slots that are initialized where it starts. As in Imp, a function body is typed by the types $\Phi_p$ of its parameters, $\Sigma; \Phi_p \vdash d : A$, and its starting tail starts with exactly the slots of the parameters initialized. A program is typed by the same rule as in Imp, $\Sigma \vdash P : A$, with CFG function bodies in place of Imp ones, so the figure omits it. A label $l$ is an index into $\vec{I}$. Block $t_l$ is typed over the blocks older than it, whose sets are the first $l$ sets $\vec{I}_{<l}$ of $\vec{I}$, so control flow is acyclic. Imp tracks the initialized slots by scope, and CFG tracks them along the flow of control. An assignment $x := rhs;\ t$ initializes slot $x$, and its tail $t$ may start with fewer initialized slots ($I' \subseteq I \cup \{x\}$), which forgets the slots of variables whose scope has ended. Join points need this. **explicate control** (Appendix C.7) translates an assignment of a `case` to slot $x$ into a `case` whose branches each assign $x$ and jump to one block for the rest. Each branch has also initialized its own payload slot, but `goto` requires the initialized slots to be exactly those of the block it jumps to. So each branch's assignment forgets its payload slot, and both branches reach the block with the same initialized slots. The statement `case` $a\ x\ y\ l_1\ l_2$ writes the payload of the injection $a$ to slot $x$ and jumps to block $l_1$, or to slot $y$ and jumps to block $l_2$. The judgment $\Sigma; \Phi; I \vdash rhs : C$ types a simple expression $rhs$, an atom or an operation on atoms $o\ \vec{a}$, by the rules for atoms and operations on atoms in Imp.

**Typing of CFG** $\quad \Sigma; \Phi; \vec{I}; I \vdash t : A \qquad \Sigma; \Phi_p \vdash d : A$

$$\frac{\Sigma; \Phi; I \vdash rhs : A}{\Sigma; \Phi; \vec{I}; I \vdash \mathtt{return}\ rhs : A} \qquad \frac{x : C \in \Phi \quad \Sigma; \Phi; I \vdash rhs : C \quad I' \subseteq I \cup \{x\} \quad \Sigma; \Phi; \vec{I}; I' \vdash t : A}{\Sigma; \Phi; \vec{I}; I \vdash x := rhs;\ t : A}$$

$$\frac{I_l = I}{\Sigma;\Phi;\vec{I};I \vdash \texttt{goto}\ l : A} \qquad \frac{x : A_1 \in \Phi \quad y : A_2 \in \Phi \quad \Sigma;\Phi;I \vdash a : A_1 + A_2 \quad I_{l_1} = I \cup \{x\} \quad I_{l_2} = I \cup \{y\}}{\Sigma;\Phi;\vec{I};I \vdash \texttt{case}\ a\ x\ y\ l_1\ l_2 : A}$$

$$\frac{\Phi = \Phi_p, \Phi' \quad \overrightarrow{\Sigma;\Phi;\vec{I}_{<l};I_l \vdash t_l : A}^{\,l} \quad \Sigma;\Phi;\vec{I};\{0,\ldots,|\Phi_p|-1\} \vdash t : A}{\Sigma;\Phi_p \vdash (\Phi;\vec{t};\ t) : A}$$

The premises under the arrow stand for one copy for each index $l$ of a block $t_l$ of $\vec{t}$. A tail means a set, given its frame $\sigma$ and the meanings $B$ of the blocks it can see. A write updates its slot in place.

**Semantics of CFG** $\boxed{[\![t]\!]_{T,B}\sigma = D}$

$$\begin{aligned}
[\![\texttt{return}\ rhs]\!]_{T,B}\sigma &= \mathcal{S}[\![rhs]\!]\sigma \\
[\![x := rhs;\ t]\!]_{T,B}\sigma &= \Lambda(\boldsymbol{\lambda} X.\ [\![t]\!]_{T,B}\,\sigma[x{:=}X]) \bullet \mathcal{S}[\![rhs]\!]\sigma \\
[\![\texttt{goto}\ l]\!]_{T,B}\sigma &= B(l)(\sigma) \\
[\![\texttt{case}\ a\ x\ y\ l_1\ l_2]\!]_{T,B}\sigma &= C(\mathcal{A}[\![a]\!]\sigma,\ \boldsymbol{\lambda} X.\, B(l_1)(\sigma[x{:=}X]),\ \boldsymbol{\lambda} X.\, B(l_2)(\sigma[y{:=}X]))
\end{aligned}$$

The blocks of a function body $(\Phi;\vec{t};\ t)$ mean $B(l)(\sigma) = [\![t_l]\!]_{T,B_l}\sigma$, where $B_l$ gives the blocks older than $l$, and the function body means $[\![t]\!]_{T,B}$. Definitions and programs are given meaning as in Imp, with the function body in place of the expression and no continuation.

## B.7 The Instr$_\mu$ Language

Agda: `FPC/Language/X86.agda, FPC/Semantics/GraphModel/X86.agda`

Instr$_\mu$ is the target language, produced by **select instructions**; Section 2 describes its instructions informally. A function body declares its frame $\Phi$, as in CFG. An instruction is typed by the type of the machine state it starts in and the one it leaves. A state type $\psi = \langle\Theta; I; O; F\rangle$ gives the types $\Theta$ of the registers, the set $I$ of the initialized slots, the open closure $O$, and whether the flag $F$ is set. The open closure $O$ is either none, written $\cdot$, or the initialized slot $x$, of a function type, that holds the closure that `fst` opened. The register types $\Theta$ map a register $r$ to either nothing, indicating an unset register, or the type $\Theta(r)$ of what it holds: a value of a type $A$, the code $\mathsf{opencode}(A \longrightarrow C)$ of the open closure, or the environment $\mathsf{openenv}$ of the open closure; $\emptyset$ maps every register to nothing. The type $\mathsf{opencode}(A \longrightarrow C)$ is the type of the function reference of the open closure, which names the global function definition that takes an argument of type $A$ and returns a $C$. $\mathsf{openenv}$ is the type of the closing environment of the open closure, the tuple of the values it captured. Neither names the types of the environment, which the function type $A \longrightarrow C$ of the closure hides. We write $\tau$ for any of these types of what a register holds, $\tau ::= A \mid \mathsf{opencode}(A \longrightarrow C) \mid \mathsf{openenv}$; operands and destinations are typed at $\tau$, because they may be registers. The operands of an instruction are typed by these rules, with the atoms typed as in Imp, over the initialized slots of the state.

**Typing of operands** $\boxed{\Sigma;\Phi;\psi \vdash a : \tau}$

$$\frac{\Sigma;\Phi;I \vdash a : A}{\Sigma;\Phi;\langle\Theta;I;O;F\rangle \vdash a : A} \qquad \frac{\Theta(r) = \tau}{\Sigma;\Phi;\langle\Theta;I;O;F\rangle \vdash \%r : \tau}$$

$$\frac{\Sigma;\Phi;\psi \vdash a : A[\mu X.A/X]}{\Sigma;\Phi;\psi \vdash \texttt{fold}\ a : \mu X.A} \qquad \frac{\Sigma;\Phi;\psi \vdash a : \mu X.A}{\Sigma;\Phi;\psi \vdash \texttt{unfold}\ a : A[\mu X.A/X]}$$

A destination, where an instruction writes a value, is either a register or a slot of the frame. Writing a value of type $\tau$ to a destination $dst$ changes the state $\psi$ to a new state $\psi'$, written $\Phi \vdash dst : \psi \Rightarrow_\tau \psi'$. A register can hold a value of any type $\tau$ (DST-REG), but a slot $x$ holds only a value of the type $A$ that the frame gives it, $x : A \in \Phi$ (DST-SLOT). A write to a slot initializes it, lets the rest start with fewer initialized slots ($I' \subseteq I \cup \{x\}$), as in CFG (Appendix B.6), and closes the open closure.

**Typing of destinations** $\boxed{\Phi \vdash dst : \psi \Rightarrow_\tau \psi'}$

$$\Phi \vdash \%r : \langle \Theta; I; O; F \rangle \Rightarrow_\tau \langle \Theta[r{:}\tau]; I; O; F \rangle \ \textsc{dst-reg} \qquad \frac{x : A \in \Phi \quad I' \subseteq I \cup \{x\}}{\Phi \vdash x : \langle \Theta; I; O; F \rangle \Rightarrow_A \langle \Theta; I'; \cdot; F \rangle} \ \textsc{dst-slot}$$

A constructor *con*, which `alloc` allocates, builds a closure, a tuple, or an injection from operands typed in the state $\psi$. A closure $\texttt{pair}(a_1, a_2)$ pairs the code of a definition with the tuple of its environment, whose types $\Delta$ its function type hides.

**Typing of constructors** $\boxed{\Sigma; \Phi; \psi \vdash con : A}$

$$\frac{\Sigma; \Phi; \psi \vdash a_1 : \mathsf{code}(\Delta; A \to C) \quad \Sigma; \Phi; \psi \vdash a_2 : \mathsf{tup}(\Delta)}{\Sigma; \Phi; \psi \vdash \texttt{pair}(a_1, a_2) : A \to C} \qquad \frac{\overrightarrow{\Sigma; \Phi; \psi \vdash a_i : A_i}^{\,i}}{\Sigma; \Phi; \psi \vdash \texttt{tuple}(\vec{a}) : \mathsf{tup}(\vec{A})}$$

$$\frac{\Sigma; \Phi; \psi \vdash a : A}{\Sigma; \Phi; \psi \vdash \texttt{inl}(a) : A + C} \qquad \frac{\Sigma; \Phi; \psi \vdash a : C}{\Sigma; \Phi; \psi \vdash \texttt{inr}(a) : A + C}$$

With $\psi = \langle \Theta; I; O; F \rangle$, the instructions are typed as follows, using the typing of operands, constructors, and destinations above.

**Typing of instructions** $\boxed{\Sigma; \Phi \vdash i : \psi \Rightarrow \psi'}$

$$\frac{\Sigma; \Phi; \psi \vdash a : \tau \quad \Phi \vdash dst : \psi \Rightarrow_\tau \psi'}{\Sigma; \Phi \vdash \mathsf{movq}\ a, dst : \psi \Rightarrow \psi'} \ \textsc{ins-movq} \qquad \frac{\Sigma; \Phi; \psi \vdash a : \mathsf{Int} \quad \Theta(r) = \mathsf{Int}}{\Sigma; \Phi \vdash \mathsf{subq}\ a, \%r : \psi \Rightarrow \psi} \ \textsc{ins-subq}$$

$$\frac{\Sigma; \Phi; \psi \vdash a_1 : \mathsf{Int} \quad \Sigma; \Phi; \psi \vdash a_2 : \mathsf{Int}}{\Sigma; \Phi \vdash \mathsf{cmpq}\ a_2, a_1 : \psi \Rightarrow \langle \Theta; I; O; \mathsf{set} \rangle} \ \textsc{ins-cmpq}$$

$$\frac{\Phi \vdash dst : \langle \Theta; I; O; \mathsf{set} \rangle \Rightarrow_{\mathsf{Bool}} \psi'}{\Sigma; \Phi \vdash \mathsf{sete}\ dst : \langle \Theta; I; O; \mathsf{set} \rangle \Rightarrow \psi'} \ \textsc{ins-sete} \qquad \frac{\Sigma; \Phi; \psi \vdash con : A \quad \Phi \vdash dst : \psi \Rightarrow_A \psi'}{\Sigma; \Phi \vdash \mathsf{alloc}\ con, dst : \psi \Rightarrow \psi'} \ \textsc{ins-alloc}$$

$$\frac{\Sigma; \Phi; \psi \vdash a : \mathsf{tup}(\vec{A}) \quad 1 \le i \le |\vec{A}| \quad \Phi \vdash dst : \psi \Rightarrow_{A_i} \psi'}{\Sigma; \Phi \vdash \mathsf{load}\ \mathsf{get}_i\ a, dst : \psi \Rightarrow \psi'} \ \textsc{ins-load}$$

$$\frac{x : A \to C \in \Phi \quad x \in I}{\Sigma; \Phi \vdash \mathsf{fst}\ x, \%r : \psi \Rightarrow \langle \mathit{wipe}(\Theta)[r{:}\mathsf{opencode}(A \to C)]; I; x; F \rangle} \ \textsc{ins-fst}$$

$$\Sigma; \Phi \vdash \mathsf{snd}\ x, \%r : \langle \Theta; I; x; F \rangle \Rightarrow \langle \Theta[r{:}\mathsf{openenv}]; I; x; F \rangle \ \textsc{ins-snd}$$

$$\frac{\Sigma; \Phi; \psi \vdash a : \mathsf{opencode}(A \to C) \quad \Theta(\mathsf{rdi}) = \mathsf{openenv} \quad \Theta(\mathsf{rsi}) = A}{\Sigma; \Phi \vdash \mathsf{callq}\ a : \psi \Rightarrow \langle \mathsf{rax}{:}C; I; \cdot; \mathsf{unset} \rangle} \ \textsc{ins-callq}$$

The rules for `fst`, `snd`, and `callq` ensure that a call combines the code and the environment of one closure, as Section 2 requires. The types $\mathsf{opencode}(A \to C)$ and `openenv` do not name the closure, so two constraints keep its code and environment together. First, $\mathit{wipe}(\Theta)$ in INS-FST unsets the registers that hold the code or the environment of an earlier open closure. Second, a write to a slot closes the open closure (DST-SLOT), so `snd` runs only while slot $x$ still holds the closure that `fst` opened. This never rejects the code that **select instructions** produces, because it writes no slot between `fst` and `callq`. After the call, only `rax` is set, and no closure is open.

A sequence $s$ has the result type $A$ of its function body, written $s : A$, in its starting state $\psi$ and with the sets $\vec{I}$ of initialized slots of the blocks it may jump to. `retq` requires that `rax` holds an $A$, and `jmp` need not constrain $A$, because the block it jumps to has the same result type $A$. As in CFG, `case` $x\ y\ z\ l_1\ l_2$ writes the payload of the injection in slot $x$ to slot $y$ and jumps to block $l_1$, or to slot $z$ and jumps to block $l_2$. A label $l$ is an index into $\vec{I}$, and $I_l$ is the set of the slots that are initialized where block $l$ starts. Because `jmp` does not change the initialized slots, it requires them to be $I_l$. Because `case` initializes slot $y$ or slot $z$, it requires blocks $l_1$ and $l_2$ to start with $I \cup \{y\}$ and $I \cup \{z\}$. Only the initialized slots pass from one block to the next. The registers, the open closure, and the flag do not, because every block starts with them unset.

**Typing of sequences** $\boxed{\Sigma; \Phi; \psi; \vec{I} \vdash s : A}$

$$\frac{\Sigma;\Phi \vdash i : \psi \Rightarrow \psi' \quad \Sigma;\Phi;\psi';\vec{I} \vdash s : A}{\Sigma;\Phi;\psi;\vec{I} \vdash i;\ s : A} \qquad \frac{\Theta(\texttt{rax}) = A}{\Sigma;\Phi;\langle\Theta;I;O;F\rangle;\vec{I} \vdash \texttt{retq} : A}$$

$$\frac{I_l = I}{\Sigma;\Phi;\langle\Theta;I;O;F\rangle;\vec{I} \vdash \texttt{jmp}\ l : A} \qquad \frac{x : A_1 + A_2 \in \Phi \quad x \in I \quad y : A_1 \in \Phi \quad z : A_2 \in \Phi \quad I_{l_1} = I \cup \{y\} \quad I_{l_2} = I \cup \{z\}}{\Sigma;\Phi;\langle\Theta;I;O;F\rangle;\vec{I} \vdash \texttt{case}\ x\ y\ z\ l_1\ l_2 : A}$$

A function body $(\Phi;\vec{s};\ s)$ has result type $A$ in the state $\psi$ that its start sequence $s$ begins in, written $\Sigma;\psi \vdash d : A$. As in CFG, block $s_l$ is typed over the blocks older than it, whose sets are the first $l$ sets $\vec{I}_{<l}$ of $\vec{I}$. A program is typed by the table $\Sigma$ of the signatures $\Sigma_k = (\Delta_k;A_k \to C_k)$ of its definitions, as in GlobalFun, written $\Sigma \vdash P : A$.

**Typing of function bodies and programs** $\boxed{\Sigma;\psi \vdash d : A \qquad \Sigma \vdash P : A}$

$$\frac{\overrightarrow{\Sigma;\Phi;\langle\emptyset;I_l;\cdot;\texttt{unset}\rangle;\vec{I}_{<l} \vdash s_l : A}^{\,l} \quad \Sigma;\Phi;\psi;\vec{I} \vdash s : A}{\Sigma;\psi \vdash (\Phi;\vec{s};\ s) : A}$$

$$\frac{\overrightarrow{\Sigma_k = (\Delta_k;A_k \to C_k) \quad \Sigma;\langle\texttt{rdi}{:}\texttt{tup}(\Delta_k),\texttt{rsi}{:}A_k;\emptyset;\cdot;\texttt{unset}\rangle \vdash d_k : C_k}^{\,k} \quad \Sigma;\langle\emptyset;\emptyset;\cdot;\texttt{unset}\rangle \vdash d : A}{\Sigma \vdash (\vec{d};\ d) : A}$$

The premises under an arrow stand for one copy for each index $l$ of a block $s_l$ of $\vec{s}$, or each index $k$ of a definition $d_k$ of $\vec{d}$. Each block $s_l$ starts with its registers and flag unset, no closure open, and the slots $I_l$ initialized, and may jump only to the blocks older than it, so blocks are acyclic as in CFG. A definition starts with its environment in `rdi`, its argument in `rsi`, and no slot initialized; the main function body starts with nothing set.

The semantics of a sequence is a function of the frame $\sigma$, the registers $R$ (a map from registers to sets), and the flag $F$ (a set); an instruction transforms the meaning $\kappa$ of the rest of the sequence. An unset register and an unset flag hold $\{\langle\rangle\}$. An operand means $\mathcal{O}[\![a]\!]\sigma R = \mathcal{A}[\![a]\!]\sigma$ for an atom, $R(r)$ for $\%r$, and $\mathcal{O}[\![a]\!]\sigma R$ for `fold` $a$ and `unfold` $a$. The meaning $\mathcal{O}[\![con]\!]\sigma R$ of a constructor, which `alloc` allocates, applies pair, $\mathcal{T}$, inl, or inr to the meanings of its operands. A write to a destination, written *put*, updates a register, or binds a slot as an assignment in CFG does.

**Writing a destination** $\boxed{put(dst, D)\,\kappa\,\sigma RF = E}$

$$put(\%r, D)\,\kappa\,\sigma RF = \kappa\,\sigma\,R[r{:=}D]\,F \qquad put(x, D)\,\kappa\,\sigma RF = \Lambda(\boldsymbol{\lambda} X.\,\kappa\,(\sigma[x{:=}X])\,R\,F) \bullet D$$

**Semantics of instructions** $\boxed{[\![i]\!]\kappa\,\sigma RF = D}$

$$\begin{aligned}
[\![\texttt{movq}\ a, dst]\!]\kappa\,\sigma RF &= put(dst, \mathcal{O}[\![a]\!]\sigma R)\,\kappa\,\sigma RF\\
[\![\texttt{subq}\ a, \%r]\!]\kappa\,\sigma RF &= \kappa\,\sigma\,R[r{:=}\mathrm{prim}_{\mathrm{sub}}(R(r), \mathcal{O}[\![a]\!]\sigma R)]\,F\\
[\![\texttt{cmpq}\ a_2, a_1]\!]\kappa\,\sigma RF &= \kappa\,\sigma\,R\,(\mathrm{prim}_{\mathrm{eq}}(\mathcal{O}[\![a_1]\!]\sigma R, \mathcal{O}[\![a_2]\!]\sigma R))\\
[\![\texttt{sete}\ dst]\!]\kappa\,\sigma RF &= put(dst, F)\,\kappa\,\sigma RF\\
[\![\texttt{alloc}\ con, dst]\!]\kappa\,\sigma RF &= put(dst, \mathcal{O}[\![con]\!]\sigma R)\,\kappa\,\sigma RF\\
[\![\texttt{load}\ \texttt{get}_i\ a, dst]\!]\kappa\,\sigma RF &= put(dst, \pi_{\mathrm{tup}_i}(\mathcal{O}[\![a]\!]\sigma R))\,\kappa\,\sigma RF\\
[\![\texttt{fst}\ x, \%r]\!]\kappa\,\sigma RF &= \kappa\,\sigma\,R[r{:=}\mathrm{car}(\sigma(x))]\,F\\
[\![\texttt{snd}\ x, \%r]\!]\kappa\,\sigma RF &= \kappa\,\sigma\,R[r{:=}\mathrm{cdr}(\sigma(x))]\,F\\
[\![\texttt{callq}\ a]\!]\kappa\,\sigma RF &= \mathrm{guard}(C;\ \kappa\,\sigma\,R_C\,\{\langle\rangle\}) \quad \text{where } C = (\mathcal{O}[\![a]\!]\sigma R \bullet R(\texttt{rdi})) \bullet R(\texttt{rsi})
\end{aligned}$$

Here $R_C$ maps `rax` to $C$ and the other registers to $\{\langle\rangle\}$. Only a call can diverge, so only `callq` needs a guard, which makes the rest of the sequence strict in the result $C$.

**Semantics of sequences** $\boxed{[\![s]\!]_B\,\sigma RF = D}$

$$
\begin{aligned}
[\![i;\ s]\!]_B &= [\![i]\!]\ [\![s]\!]_B \\
[\![\texttt{retq}]\!]_B\,\sigma R F &= R(\texttt{rax}) \\
[\![\texttt{jmp}\ l]\!]_B\,\sigma R F &= B(l)(\sigma) \\
[\![\texttt{case}\ x\ y\ z\ l_1\ l_2]\!]_B\,\sigma R F &= C(\sigma(x),\ \boldsymbol{\lambda} X.\, B(l_1)(\sigma[y{:=}X]),\ \boldsymbol{\lambda} X.\, B(l_2)(\sigma[z{:=}X]))
\end{aligned}
$$

where a block means $B(l)(\sigma) = [\![s_l]\!]_{B_l}\,\sigma\,R_0\,\{\langle\rangle\}$, with $R_0$ mapping every register to $\{\langle\rangle\}$ and $B_l$ giving the blocks older than $l$. A definition means $T(k) = \Lambda(\boldsymbol{\lambda} X.\,\Lambda(\boldsymbol{\lambda} Y.\,[\![s]\!]_B\,\rho_0\,R_{X,Y}\,\{\langle\rangle\}))$ for its function body $s$ with blocks $B$, where $R_{X,Y}$ maps `rdi` to $X$, `rsi` to $Y$, and `rax` to $\{\langle\rangle\}$; the tables are as in GlobalFun. A program means its main function body:

$$[\![(\vec{d};\ (\Phi; s_0, \ldots, s_{j-1};\ s))]\!] = [\![s]\!]_B\,\rho_0\,R_0\,\{\langle\rangle\}$$

Its start sequence $s$ runs with the meanings $B$ of its blocks $s_0, \ldots, s_{j-1}$, as above, in the frame $\rho_0$, which maps every slot to $\{\langle\rangle\}$, with the registers $R_0$, which are all unset, and with the flag $\{\langle\rangle\}$, which is unset.

## C The Compiler Passes

Agda: `FPC/Compiler/Pipeline.agda (compile)`

This appendix defines each pass of Section 2 as a recursive function on the syntax of Figure 2. In each pass, the clauses not shown are homomorphic. They translate the subterms and rebuild the same form. No pass inspects a recursive type; `fold` and `unfold` pass through every pass as operations with one operand. The subsections follow the order of the pipeline; Sections 4.1 and 4.2 and Appendix D prove each definition correct. The compiler is the composition of the passes:

$$\begin{gathered}\textbf{compile} = \textbf{select} \circ \textbf{explicate} \circ \textbf{lift} \circ \textbf{rco} \circ \textbf{hoist} \\ \circ\ \textbf{split} \circ \textbf{enclose} \circ \textbf{simplify}\end{gathered}$$

### C.1 simplify (FPC to FPC)

Agda: `FPC/Compiler/Simplify.agda (simplify)`

The pass works with *syntactic values*, the terms

$$V ::= x \mid \lambda x.\, N \mid c \mid \texttt{tuple}\ (V_1, \ldots, V_n) \mid \texttt{inl}\ V \mid \texttt{inr}\ V \mid \texttt{fold}\ V$$

A syntactic value is *small* when it is a variable, a constant, $\langle\rangle$, or `inl` $a$ or `inr` $a$ of one of those; only small values are duplicated. On a term, **simplify** returns a pair $(M', \theta)$ of the simplified term $M'$ and the static information $\theta$ about the term's value:

$$\theta ::= V \mid \langle \lambda x.\, N, E\rangle \mid \bot$$

that is, a syntactic value $V$ that the term is known to equal, a function $\langle \lambda x.\, N, E\rangle$ (its body before simplification, with the environment $E$ where it was defined), or nothing, $\bot$. An environment $E$ maps each variable to such a pair: the term that replaces the variable and the static information about its value. We write $\mathit{self}(V) = (V, V)$, and $\mathit{info}(M) = (M, M)$ if $M$ is a syntactic value and $(M, \bot)$ otherwise. Under a binder $x$ that stays, $x$ is itself and is known only as itself: $E' = E[x{:=}\mathit{self}(x)]$.

The pass carries fuel $\phi$ that limits nested inlining. Every equation passes $\phi$ along unchanged except the one for a call, which inlines with one less. We write the fuel as a superscript, as in $\textbf{simp}^{\phi}_{E}$, and omit it wherever it does not change.

$$
\begin{aligned}
\mathbf{simp}_E(x) &= E(x)\\
\mathbf{simp}_E(\lambda x.\, N) &= (\lambda x.\, N', \langle \lambda x.\, N, E\rangle)\\
&\qquad \text{where } (N', \theta) = \mathbf{simp}_{E[x:=\mathit{self}(x)]}(N)\\
\mathbf{simp}_E(L\ M) &= \mathbf{app}_E(L, \mathbf{simp}_E(M))\\
\mathbf{simp}_E(c) &= \mathit{self}(c)\\
\mathbf{simp}_E(\mathtt{tuple}\ (M_1, \ldots, M_n)) &= \mathit{info}(\mathtt{tuple}\ (M'_1, \ldots, M'_n))\\
&\qquad \text{where } (M'_i, \theta_i) = \mathbf{simp}_E(M_i)\\
\mathbf{simp}_E(\mathtt{get}_i\ L) &= \mathbf{get}_i(\mathbf{simp}_E(L))\\
\mathbf{simp}_E(\mathtt{inl}\ M) &= \mathit{info}(\mathtt{inl}\ M') \quad \text{where } (M', \theta) = \mathbf{simp}_E(M)\\
&\qquad \text{and likewise for } \mathtt{inr} \text{ and } \mathtt{fold}\\
\mathbf{simp}_E(\mathtt{case}\ L\ (x.\, M)\ (y.\, N)) &= \mathbf{case}_E(\mathbf{simp}_E(L), x.\, M, y.\, N)\\
\mathbf{simp}_E(\mathtt{prim}_p\ L\ M) &= \mathbf{prim}_p(\mathbf{simp}_E(L), \mathbf{simp}_E(M))\\
\mathbf{simp}_E(\mathtt{let}\ x = M\ \mathtt{in}\ N) &= \mathbf{let}_E(x.\, N, \mathbf{simp}_E(M))\\
\mathbf{simp}_E(\mathtt{unfold}\ L) &= \mathbf{unfold}(\mathbf{simp}_E(L))
\end{aligned}
$$

An application of a $\lambda$-abstraction is a $\beta$-redex; any other application is a call, which inlines a known function when there is fuel left and its body, simplified at the call, has at most 24 nodes:

$$
\begin{aligned}
\mathbf{app}_E(\lambda x.\, N, I) &= \mathbf{let}_E(x.\, N, I)\\
\mathbf{app}_E(L, I) &= \mathbf{call}(\mathbf{simp}_E(L), I) \quad \text{otherwise}
\end{aligned}
$$

$$
\begin{aligned}
\mathbf{call}^{\phi+1}((L', \langle \lambda x.\, N, E_f\rangle), I) &= P \quad \text{if } |P| \le 24,\ \text{where } P = \mathbf{let}^{\phi}_{E_f}(x.\, N, I)\\
&= (L'\ M', \bot) \quad \text{otherwise, where } I = (M', \theta')\\
\mathbf{call}^{\phi}((L', \theta), (M', \theta')) &= (L'\ M', \bot) \quad \text{in all other cases}
\end{aligned}
$$

Here $|P|$ is the number of nodes in the term of $P$. A body bound to an operand, by `let`, $\beta$, or inlining, substitutes the operand if it is small, and otherwise keeps it bound by a `let`:

$$
\begin{aligned}
\mathbf{let}_E(x.\, N, (M', \theta)) &= \mathbf{simp}_{E[x:=(M',\theta)]}(N) \quad \text{if } M' \text{ is a small syntactic value}\\
&= (\mathtt{let}\ x = M'\ \mathtt{in}\ N', \bot) \quad \text{if } M' \text{ is another syntactic value,}\\
&\qquad \text{where } (N', \theta'') = \mathbf{simp}_{E[x:=(x,\theta)]}(N)\\
&= (\mathtt{let}\ x = M'\ \mathtt{in}\ N', \bot) \quad \text{otherwise,}\\
&\qquad \text{where } (N', \theta'') = \mathbf{simp}_{E[x:=\mathit{self}(x)]}(N)
\end{aligned}
$$

A projection, `unfold`, `case`, or primitive on a known value reduces:

$$
\begin{aligned}
\mathbf{get}_i((L', \mathtt{tuple}\ (\theta_1, \ldots, \theta_n))) &= \mathit{self}(\theta_i) \quad \text{if } \theta_i \text{ is small}\\
&= (\mathtt{get}_i\ L', \theta_i) \quad \text{otherwise}\\
\mathbf{get}_i((L', \theta)) &= (\mathtt{get}_i\ L', \bot) \quad \text{if } \theta \text{ is not a tuple}\\
\mathbf{unfold}((L', \mathtt{fold}\ \theta)) &= \mathit{self}(\theta) \quad \text{if } \theta \text{ is small}\\
&= (\mathtt{unfold}\ L', \theta) \quad \text{otherwise}\\
\mathbf{unfold}((L', \theta)) &= (\mathtt{unfold}\ L', \bot) \quad \text{if } \theta \text{ is not a } \mathtt{fold}\\
\mathbf{case}_E((L', \mathtt{inl}\ \theta), x.\, M, y.\, N) &= \mathbf{simp}_{E[x:=\mathit{self}(\theta)]}(M) \quad \text{if } \theta \text{ is small}\\
&\qquad \text{(and likewise for } \mathtt{inr}\text{)}\\
\mathbf{case}_E((L', \theta), x.\, M, y.\, N) &= (\mathtt{case}\ L'\ (x.\, M')\ (y.\, N'), \bot) \quad \text{otherwise,}\\
&\qquad \text{where } (M', \theta_1) = \mathbf{simp}_{E[x:=\mathit{self}(x)]}(M)\\
&\qquad \text{and } (N', \theta_2) = \mathbf{simp}_{E[y:=\mathit{self}(y)]}(N)\\
\mathbf{prim}_p((L', n), (M', m)) &= \mathit{self}(\mathit{res}(\delta(p, n, m))) \quad \text{if } n \text{ and } m \text{ are integers}\\
\mathbf{prim}_p((L', \theta_1), (M', \theta_2)) &= (\mathtt{prim}_p\ L'\ M', \bot) \quad \text{otherwise}
\end{aligned}
$$

Here $res(k) = k$ for an integer $k$, $res(\text{true}) = \texttt{inl}\ \langle\rangle$, and $res(\text{false}) = \texttt{inr}\ \langle\rangle$. Finally,

$$\textbf{simplify}(M) = M' \quad \text{where } (M', \theta) = \textbf{simp}^4_{E_0}(M) \text{ and } E_0(x) = \mathit{self}(x) \text{ for every } x.$$

**simplify** drops only evaluations whose value is known, which have no effect, and otherwise keeps the order of evaluation. An operand that is not a syntactic value is never substituted but stays bound by a `let`.

### C.2 enclose (FPC to EnvClos)

Agda: `FPC/Compiler/{Capture,Prune,Concretize,Pipeline}.agda (capture, prune-program, concretize-program, enclose)`

**enclose** works in three steps, which go through the intermediate language Clos:

$$\textbf{enclose} = \mathit{concretize}_{\emptyset} \circ \mathit{prune} \circ \mathit{capture}$$

*Capture.* The first step makes each function a closure that captures the variables in scope, in order:

$$\mathit{capture}(\lambda y.\, N) = \texttt{clos}\ (x_1, \ldots, x_k, y.\, \mathit{capture}(N))\ (x_1, \ldots, x_k)$$

where $x_1, \ldots, x_k$ are the variables in scope.

*Prune.* The second step removes the captured values that the code does not use. Closures are strict, so a captured term is dropped only when the code does not use its name and the term is a variable, which is all that *capture* captures:

$$\begin{aligned} &\mathit{prune}(\texttt{clos}\ (x_1, \ldots, x_n, y.\, N)\ (M_1, \ldots, M_n)) = \\ &\quad \texttt{clos}\ (x_{i_1}, \ldots, x_{i_k}, y.\, \mathit{prune}(N))\ (\mathit{prune}(M_{i_1}), \ldots, \mathit{prune}(M_{i_k})) \\ &\quad \text{where } \{i_1 < \cdots < i_k\} = \{i \mid x_i \in FV(N) \text{ or } M_i \text{ is not a variable}\} \end{aligned}$$

*Concretize.* The last step stores the captured values of each closure in one tuple, bound to $z$ in its code, and the code reads the name $x_i$ as $\texttt{get}_i\ z$. The map $\varphi$ sends each Clos variable to a reference, $x$ or $\texttt{get}_i\ z$, in EnvClos:

$$\begin{aligned} &\mathit{concretize}_{\varphi}(x) = \varphi(x) \\ &\mathit{concretize}_{\varphi}(\texttt{clos}\ (x_1, \ldots, x_n, y.\, N)\ (M_1, \ldots, M_n)) = \\ &\quad \texttt{clos}\ (z, y.\, \mathit{concretize}_{\varphi'}(N))\ (\mathit{concretize}_{\varphi}(M_1), \ldots, \mathit{concretize}_{\varphi}(M_n)) \\ &\quad \text{where } \varphi' = \{y \mapsto y,\ x_1 \mapsto \texttt{get}_1\ z,\ \ldots,\ x_n \mapsto \texttt{get}_n\ z\} \end{aligned}$$

Under a `let` or `case` the bound variable maps to itself, and a program starts with the empty map.

### C.3 split closure (EnvClos to PairClos)

Agda: `FPC/Compiler/Split.agda (split-program)`

A closure becomes a pair of its code and the tuple of its captured values, and an application opens the closure and calls its code:

$$\textbf{split}(\texttt{clos}\ (z, y.\, N)\ \vec{M}) = \texttt{pair}\ (\texttt{code}\ (z, y.\, \textbf{split}(N)))\ (\texttt{tuple}\ \textbf{split}(\vec{M}))$$

$$\textbf{split}(L\ M) = \texttt{callc}\ \textbf{split}(L)\ \textbf{split}(M)$$

### C.4 hoist (PairClos to GlobalFun)

Agda: `FPC/Compiler/Hoist.agda (hoist)`

The code of every function becomes a global definition, and the code itself a reference to it. The pass threads the list $\vec{d}$ of the definitions made so far, from left to right. For code, it first translates the body, so the functions nested in it become definitions first. For any other form $op$ with subterms $A_1, \ldots, A_n$ (each under its binders $\bar{x}_i$):

$$
\begin{aligned}
\mathbf{glob}(x, \vec{d}) &= (x, \vec{d}) \\
\mathbf{glob}(\texttt{code}\ (z, y.\, N), \vec{d}) &= (f_{|\vec{d}_1|},\ (\vec{d}_1, \texttt{code}\ (z, y.\, N'))) \\
&\qquad \text{where } (N', \vec{d}_1) = \mathbf{glob}(N, \vec{d}) \\
\mathbf{glob}(op(\bar{x}_1.\, A_1, \ldots, \bar{x}_n.\, A_n), \vec{d}_0) &= (op(\bar{x}_1.\, A'_1, \ldots, \bar{x}_n.\, A'_n), \vec{d}_n) \\
&\qquad \text{where } (A'_i, \vec{d}_i) = \mathbf{glob}(A_i, \vec{d}_{i-1}) \\
\mathbf{hoist}(M) &= (\vec{d};\ M') \quad \text{where } (M', \vec{d}) = \mathbf{glob}(M, \epsilon)
\end{aligned}
$$

Each definition refers only to the definitions made before it.

### C.5 remove complex operands (GlobalFun to MNF)

Agda: `FPC/Compiler/RCO.agda (rco-program)`

Every complex operand is bound to a fresh variable by a `let`, from left to right, so the operands of every operation become atoms. A `case` or `let` is complex; its branches and body are translated in place, so nothing is duplicated. An operand that is an atom stays in place; any other operand is translated and bound to a fresh variable:

$$
\begin{aligned}
\mathbf{atomize}(a) &= (\epsilon;\ a) \\
\mathbf{atomize}(M) &= (x{=}\mathbf{rco}(M);\ x) \quad \text{otherwise } (x \text{ fresh}) \\
\mathbf{atomize}(M_1, \ldots, M_n) &= (\vec{b}_1 \cdots \vec{b}_n;\ a_1, \ldots, a_n) \quad \text{where } (\vec{b}_i; a_i) = \mathbf{atomize}(M_i) \\
\mathbf{lets}(x_1{=}N_1 \cdots x_k{=}N_k,\ N) &= \texttt{let}\ x_1 = N_1\ \texttt{in}\ \cdots\ \texttt{let}\ x_k = N_k\ \texttt{in}\ N \\
\mathbf{rco}(a) &= a \\
\mathbf{rco}(o\ (M_1, \ldots, M_n)) &= \mathbf{lets}(\vec{b},\ o\ (a_1, \ldots, a_n)) \\
&\qquad \text{where } (\vec{b}; a_1, \ldots, a_n) = \mathbf{atomize}(M_1, \ldots, M_n) \\
\mathbf{rco}(\texttt{case}\ L\ (x.\, M)\ (y.\, N)) &= \mathbf{lets}(\vec{b},\ \texttt{case}\ a\ (x.\, \mathbf{rco}(M))\ (y.\, \mathbf{rco}(N))) \\
&\qquad \text{where } (\vec{b}; a) = \mathbf{atomize}(L) \\
\mathbf{rco}(\texttt{let}\ x = M\ \texttt{in}\ N) &= \texttt{let}\ x = \mathbf{rco}(M)\ \texttt{in}\ \mathbf{rco}(N) \\
\mathbf{rco}(\texttt{code}\ (z, y.\, N)) &= \texttt{code}\ (z, y.\, \mathbf{rco}(N)) \\
\mathbf{rco}((d_0, \ldots, d_{m-1};\ M)) &= (\mathbf{rco}(d_0), \ldots, \mathbf{rco}(d_{m-1});\ \mathbf{rco}(M))
\end{aligned}
$$

where $o$ ranges over the operations of MNF (Figure 2).

### C.6 lift locals (MNF to Imp)

Agda: `FPC/Compiler/LiftLocals.agda (lift-program)`

Every variable that a function body binds gets a slot of its own. The translation $\mathbf{lift}_{\Gamma,n}(e)$ takes the map $\Gamma$ from the variables in scope to their slots and the first unused slot $n$, and returns the translation of $e$ and the first slot that it leaves unused. A variable $x$ becomes its slot $\Gamma(x)$. A `let` gives its variable the slot after those of the variables bound in its right-hand side, and a `case` gives a slot to the variable of each branch, before the variables bound in that branch. A definition `code` $(z, y.\, N)$ starts with $z$ in slot 0 and $y$ in slot 1, and the main expression starts with no variables in scope.

$$
\begin{aligned}
\textbf{lift}_{\Gamma,n}(x) &= (\Gamma(x), n)\\
\textbf{lift}_{\Gamma,n}(\texttt{let } x = M \texttt{ in } N) &= (s := e_1;\ e_2,\ n_2)\\
&\qquad \text{where } (e_1, s) = \textbf{lift}_{\Gamma,n}(M) \text{ and } (e_2, n_2) = \textbf{lift}_{\Gamma[x:=s],\, s+1}(N)\\
\textbf{lift}_{\Gamma,n}(\texttt{case } a\ (x.\,M)\ (y.\,N)) &= (\texttt{case } \textbf{lift}_{\Gamma}(a)\ n\ n_1\ e_1\ e_2,\ n_2)\\
&\qquad \text{where } (e_1, n_1) = \textbf{lift}_{\Gamma[x:=n],\, n+1}(M) \text{ and } (e_2, n_2) = \textbf{lift}_{\Gamma[y:=n_1],\, n_1+1}(N)\\
\textbf{lift}(\texttt{code } (z, y.\,N)) &= (\Phi; e) \quad \text{where } (e, n) = \textbf{lift}_{[z:=0,\ y:=1],\, 2}(N)\\
\textbf{lift}((d_0, \ldots, d_{m-1};\ M)) &= (\textbf{lift}(d_0), \ldots, \textbf{lift}(d_{m-1});\ (\Phi; e)) \quad \text{where } (e, n) = \textbf{lift}_{\emptyset,0}(M)
\end{aligned}
$$

Every other form is a simple expression, with $\textbf{lift}_\Gamma$ applied to its atoms, and leaves $n$ unchanged. The frame $\Phi$ of a function body lists the types of its $n$ slots. When a variable is bound, its slot is uninitialized, as Imp's assignment and `case` rules require. The initialized slots are those of the variables in scope, which are all below $n$, and the new variable's slot is $n$ or above. For example, **lift locals** translates `let` $y$ = (`let` $z$ = 5 `in let` $w = z - 1$ `in` $w - 1$) `in` $y - 2$ into the following expression, in which, as in AT&T syntax, $\$n$ is the integer constant $n$ and a bare number is a slot:

$$2 := (0 := \$5;\ 1 := \texttt{prim}_{\texttt{sub}}\ 0\ \$1;\ \texttt{prim}_{\texttt{sub}}\ 1\ \$1);\ \texttt{prim}_{\texttt{sub}}\ 2\ \$2.$$

The variables $z$ and $w$ of the right-hand side get slots 0 and 1, and $y$ gets slot 2, which is the first slot that the right-hand side leaves unused. Each assignment writes an uninitialized slot. The assignment to $z$ writes slot 0 where no other slot is initialized, the one to $w$ writes slot 1 where only slot 0 is, and the one to $y$ writes slot 2 where again no slot is initialized. The right-hand side writes slots 0 and 1, but they count as initialized only in the scopes of $z$ and $w$, which end with the right-hand side. The Agda development lists the frame from the newest slot to the oldest, and a slot is a de Bruijn index into that list, so slot $k$ here is index $|\Phi| - 1 - k$ there. In the example, $y$, $w$, and $z$ are the indices 0, 1, and 2.

## C.7 explicate control (Imp to CFG)

Agda: `FPC/Compiler/Explicate.agda (explicate)`

**tail**($e$) translates $e$ in tail position, and **assign**($e, x, k$) translates "write $e$ to slot $x$, then run the tail $k$" in continuation-passing style. A `case` on the right of an assignment adds $k$ as a block, the join point, to which both branches jump. **add**($t$) adds the block $t$ and returns its label. Both functions thread the blocks made so far, each typed over the older ones, so a label made earlier stays valid as more blocks are added. Blocks are added in the order in which the equations compute them, inner calls first and from left to right: **tail**($e_2$) before the assignment of $e_1$, $l$ before $l_1$, and $l_1$ before $l_2$.

$$
\begin{aligned}
\textbf{tail}(\mathit{rhs}) &= \texttt{return } \mathit{rhs}\\
\textbf{tail}(x := e_1;\ e_2) &= \textbf{assign}(e_1, x, \textbf{tail}(e_2))\\
\textbf{tail}(\texttt{case } a\ x\ y\ e_1\ e_2) &= \texttt{case } a\ x\ y\ l_1\ l_2 \quad \text{where } l_i = \textbf{add}(\textbf{tail}(e_i))
\end{aligned}
$$

$$
\begin{aligned}
\textbf{assign}(\mathit{rhs}, x, k) &= x := \mathit{rhs};\ k\\
\textbf{assign}(x' := e_1;\ e_2, x, k) &= \textbf{assign}(e_1, x', \textbf{assign}(e_2, x, k))\\
\textbf{assign}(\texttt{case } a\ y\ z\ e_1\ e_2, x, k) &= \texttt{case } a\ y\ z\ l_1\ l_2\\
&\qquad \text{where } l = \textbf{add}(k) \text{ and } l_i = \textbf{add}(\textbf{assign}(e_i, x, \texttt{goto } l))
\end{aligned}
$$

$$\textbf{explicate}(e) = (\vec{t};\ t) \quad \text{where } t = \textbf{tail}(e) \text{ and } \vec{t} \text{ are the blocks it adds}$$

and **explicate** maps over the definitions and main expression of a program, keeping the frame of each function body. The second equation for **assign** writes $x'$ and the slots that $e_2$ writes before it writes $x$. The tail $k$ starts with the slots initialized where the assignment starts, plus $x$, so the typing of CFG lets it forget the others. For the example of §C.6, **explicate control** gives the tail

$$0 := \$5;\ 1 := \texttt{prim}_{\texttt{sub}}\ 0\ \$1;\ 2 := \texttt{prim}_{\texttt{sub}}\ 1\ \$1;\ \texttt{return prim}_{\texttt{sub}}\ 2\ \$2,$$

whose last tail starts with only slot 2 initialized.

### C.8 select instructions (CFG to $\text{Instr}_\mu$)

Agda: `FPC/Compiler/Select.agda (select)`

The code of a simple expression *rhs* is selected in continuation-passing style. The code $\textbf{sel}(rhs, \omega, k)$ sends the value of *rhs* to $\omega$, which is either `rax` or a slot $x$,

$$\omega ::= \texttt{ret} \mid \texttt{to}\ x$$

and then the instructions $k$ run. We write $\bar{\omega}$ for the destination of $\omega$, where $\overline{\texttt{ret}} = \texttt{\%rax}$ and $\overline{\texttt{to}\ x} = x$. An atom is a slot, a constant, or a function reference, and a constant has a base type and a function reference a code type. So an atom of a function type or a sum type is a slot $x$, and therefore `callc` and `case` take a slot.

$$\begin{aligned}
\textbf{sel}(a, \omega, k) &= \texttt{movq}\ a, \bar{\omega};\ k \\
\textbf{sel}(\texttt{pair}\ a_1\ a_2, \omega, k) &= \texttt{alloc pair}(a_1, a_2), \bar{\omega};\ k \\
\textbf{sel}(\texttt{callc}\ x\ a, \omega, k) &= \texttt{fst}\ x, \texttt{\%rax};\ \texttt{snd}\ x, \texttt{\%rdi};\ \texttt{movq}\ a, \texttt{\%rsi}; \\
&\phantom{=}\ \texttt{callq \%rax};\ \texttt{movq \%rax}, \bar{\omega};\ k \\
\textbf{sel}(\texttt{tuple}\ (a_1, \ldots, a_n), \omega, k) &= \texttt{alloc tuple}(a_1, \ldots, a_n), \bar{\omega};\ k \\
\textbf{sel}(\texttt{get}_i\ a, \omega, k) &= \texttt{load get}_i\ a, \bar{\omega};\ k \\
\textbf{sel}(\texttt{inl}\ a, \omega, k) &= \texttt{alloc inl}(a), \bar{\omega};\ k \qquad \text{and likewise for } \texttt{inr} \\
\textbf{sel}(\texttt{prim}_{\texttt{sub}}\ a_1\ a_2, \omega, k) &= \texttt{movq}\ a_1, \texttt{\%rax};\ \texttt{subq}\ a_2, \texttt{\%rax};\ \texttt{movq \%rax}, \bar{\omega};\ k \\
\textbf{sel}(\texttt{prim}_{\texttt{eq}}\ a_1\ a_2, \omega, k) &= \texttt{cmpq}\ a_2, a_1;\ \texttt{sete}\ \bar{\omega};\ k \\
\textbf{sel}(\texttt{fold}\ a, \omega, k) &= \texttt{movq fold}\ a, \bar{\omega};\ k \\
\textbf{sel}(\texttt{unfold}\ a, \omega, k) &= \texttt{movq unfold}\ a, \bar{\omega};\ k
\end{aligned}$$

$$\begin{aligned}
\textbf{tail}(\texttt{return}\ rhs) &= \textbf{sel}(rhs, \texttt{ret}, \texttt{retq}) \\
\textbf{tail}(x := rhs;\ t) &= \textbf{sel}(rhs, \texttt{to}\ x, \textbf{tail}(t)) \\
\textbf{tail}(\texttt{goto}\ l) &= \texttt{jmp}\ l \\
\textbf{tail}(\texttt{case}\ x\ y\ z\ l_1\ l_2) &= \texttt{case}\ x\ y\ z\ l_1\ l_2
\end{aligned}$$

$$\begin{aligned}
\textbf{select}((\Phi; t_0, \ldots, t_{j-1};\ t)) &= (\Phi;\ \textbf{tail}(t_0), \ldots, \textbf{tail}(t_{j-1});\ \textbf{tail}(t)) \\
\textbf{select}^+((\Phi; t_0, \ldots, t_{j-1};\ t)) &= (\Phi;\ \textbf{tail}(t_0), \ldots, \textbf{tail}(t_{j-1}); \\
&\phantom{=}\quad \texttt{movq \%rdi}, 0;\ \texttt{movq \%rsi}, 1;\ \textbf{tail}(t))
\end{aligned}$$

Each function body keeps its frame $\Phi$, and an assignment to a slot becomes a write to that slot in place. The main function body uses **select**, and each definition uses $\textbf{select}^+$, whose prologue moves the environment and argument from `rdi` and `rsi` to slots 0 and 1: $\textbf{select}((d_0, \ldots, d_{m-1};\ d)) = (\textbf{select}^+(d_0), \ldots, \textbf{select}^+(d_{m-1});\ \textbf{select}(d))$.

## D Calculations for the Other Equational Passes

Agda: `FPC/proof/Correctness/GraphModel/, one file per pass or step (CaptureCorrect.agda, ...)`

This appendix gives the key case of the proofs of the first two steps of **enclose** and of the five equational passes that Section 4.1 does not show, in the order of the pipeline. As there, each proof generalizes the equation to open terms under related environments, the other cases are congruences, and where a step needs a set to be nonempty we show the nonempty case.

*Enclose: capture.* The environments agree on the variables $\vec{x}$ in scope, and the target environment is nonempty on them. A $\lambda$-abstraction becomes a closure that captures all of $\vec{x}$. With $N' = \mathit{capture}(N)$ and $F = \boldsymbol{\lambda} Y.\, [\![N']\!]\rho_0[\vec{x}{:=}\rho'(\vec{x}), y{:=}Y]$:

$$\begin{aligned}
[\![\lambda y.\, N]\!]\rho &= \Lambda(\boldsymbol{\lambda} Y.\, [\![N]\!]\rho[y{:=}Y]) && \text{(meaning of } \lambda\text{)}\\
&= \Lambda(F) && \text{(induction hypothesis for } N\text{)}\\
&= \mathtt{guard}(\rho'(x_1), \ldots, \rho'(x_k);\ \Lambda(F)) && (\rho' \text{ is nonempty on } \vec{x})\\
&= [\![\mathtt{clos}\ (\vec{x}, y.\, N')\ \vec{x}]\!]\rho' && \text{(meaning of a closure)}
\end{aligned}$$

*Enclose: prune.* The environments agree on the variables that a term uses, and the source environment is nonempty. Let $D_i = [\![M_i]\!]\rho$ and $D'_i = [\![\mathit{prune}(M_i)]\!]\rho'$, let $I$ be the indices of the kept captured values, and let $N' = \mathit{prune}(N)$:

$$\begin{aligned}
[\![\mathtt{clos}\ (\vec{x}, y.\, N)\ \vec{M}]\!]\rho &= \mathtt{guard}(\vec{D};\ \Lambda(\boldsymbol{\lambda} Y.\, [\![N]\!]\rho_0[\vec{x}{:=}\vec{D}, y{:=}Y])) && \text{(meaning of a closure)}\\
&= \mathtt{guard}(\vec{D}'_I;\ \Lambda(\boldsymbol{\lambda} Y.\, [\![N]\!]\rho_0[\vec{x}{:=}\vec{D}, y{:=}Y])) && \text{(the guards hold together)}\\
&= \mathtt{guard}(\vec{D}'_I;\ \Lambda(\boldsymbol{\lambda} Y.\, [\![N']\!]\rho_0[\vec{x}_I{:=}\vec{D}'_I, y{:=}Y])) && \text{(induction hypothesis for } N\text{)}\\
&= [\![\mathtt{clos}\ (\vec{x}_I, y.\, N')\ \vec{M}'_I]\!]\rho' && \text{(meaning of a closure)}
\end{aligned}$$

The guards hold together because a dropped captured value is a variable, which the environment makes nonempty. The induction hypothesis applies because $N$ uses only $\vec{x}_I$ and $y$.

*Hoist.* The induction hypothesis says that a term means the same as its translation under the table $T'$ of the definitions made so far. For code, **hoist** first translates the body $N$ to $N'$ and then adds the definition $d_k = \mathtt{code}\ (z, y.\, N')$:

$$\begin{aligned}
[\![\mathtt{code}\ (z, y.\, N)]\!]\rho &= \Lambda(\boldsymbol{\lambda} X.\, \Lambda(\boldsymbol{\lambda} Y.\, [\![N]\!]\rho_0[z{:=}X, y{:=}Y])) && \text{(meaning of code)}\\
&= \Lambda(\boldsymbol{\lambda} X.\, \Lambda(\boldsymbol{\lambda} Y.\, [\![N']\!]_{T_k}\rho_0[z{:=}X, y{:=}Y])) && \text{(induction hypothesis for } N\text{)}\\
&= T(k) && \text{(meaning of definition } d_k\text{)}\\
&= [\![f_k]\!]_T\, \rho && \text{(meaning of } f_k\text{)}
\end{aligned}$$

The other cases also use a weakening lemma, which says that a term means the same under a larger table that agrees on the definitions the term refers to.

*Remove complex operands.* When an operand $M$ is not an atom, **remove complex operands** binds it to a fresh variable $x$ with a `let`. For a projection, with $D = [\![\mathbf{rco}(M)]\!]_{T'}\rho'$:

$$\begin{aligned}
[\![\mathtt{get}_i\ M]\!]_T\, \rho &= \pi_{\mathrm{tup}_i}([\![M]\!]_T\, \rho) && \text{(meaning of } \mathtt{get}_i\text{)}\\
&= \pi_{\mathrm{tup}_i}(D) && \text{(induction hypothesis for } M\text{)}\\
&= [\![\mathtt{get}_i\ x]\!]_{T'}\, \rho'[x{:=}D] && \text{(meaning of } \mathtt{get}_i\text{)}\\
&= \Lambda(\boldsymbol{\lambda} X.\, [\![\mathtt{get}_i\ x]\!]_{T'}\, \rho'[x{:=}X]) \bullet D && \text{(Lemma 3.1, part 1)}\\
&= [\![\mathtt{let}\ x = \mathbf{rco}(M)\ \mathtt{in}\ \mathtt{get}_i\ x]\!]_{T'}\, \rho' && \text{(meaning of } \mathtt{let}\text{)}
\end{aligned}$$

The Agda development proves this once for every strict operation and every continuation, and chains it over several operands.

*Lift locals.* The semantics of Imp is in continuation-passing style, so the induction hypothesis relates an MNF term $e$ and its translation under a continuation $\kappa$: $\kappa([\![e]\!]_T\, \rho) = [\![\mathbf{lift}_\Gamma(e)]\!]_{T'}\, \sigma\, \kappa$, for an environment $\rho$ and a frame $\sigma$ that holds the value of each variable $v$ in scope in its slot, $\rho(v) = \sigma(\Gamma(v))$. A `let` becomes an assignment to the slot $s$ of its variable $x$. Because $s$ is not initialized, it is not the slot of a variable in scope, so $\sigma[s{:=}X]$ holds $\rho[x{:=}X]$ in the slots of $\Gamma, x{:}s$. With $\kappa' = \boldsymbol{\lambda} D.\, \Lambda(\boldsymbol{\lambda} X.\, [\![\mathbf{lift}_{\Gamma,x:s}(N)]\!]_{T'}\, \sigma[s{:=}X]\, \kappa) \bullet D$:

$$\begin{aligned}
\kappa([\![\mathtt{let}\ x = M\ \mathtt{in}\ N]\!]_T\, \rho) &= \kappa(\Lambda(\boldsymbol{\lambda} X.\, [\![N]\!]_T\, \rho[x{:=}X]) \bullet [\![M]\!]_T\, \rho) && \text{(meaning of } \mathtt{let}\text{)}\\
&= \Lambda(\boldsymbol{\lambda} X.\, \kappa([\![N]\!]_T\, \rho[x{:=}X])) \bullet [\![M]\!]_T\, \rho && \text{(a } \mathtt{let} \text{ commutes with } \kappa\text{)}\\
&= \Lambda(\boldsymbol{\lambda} X.\, [\![\mathbf{lift}_{\Gamma,x:s}(N)]\!]_{T'}\, \sigma[s{:=}X]\, \kappa) \bullet [\![M]\!]_T\, \rho && \text{(induction hypothesis for } N\text{)}\\
&= [\![\mathbf{lift}_\Gamma(M)]\!]_{T'}\, \sigma\, \kappa' && \text{(induction hypothesis for } M\text{)}\\
&= [\![s := \mathbf{lift}_\Gamma(M);\ \mathbf{lift}_{\Gamma,x:s}(N)]\!]_{T'}\, \sigma\, \kappa && \text{(meaning of assignment)}
\end{aligned}$$

The second step, a commuting conversion, needs $[\![M]\!]_T\,\rho$ to be consistent and the body to be monotone. The `case` case follows the same pattern.

*Explicate control.* The induction hypothesis says that an Imp expression $e$, run with the continuation $\kappa_{x,k}$ that writes slot $x$ and then runs the tail $k$, means the same as the tail $\mathbf{assign}(e, x, k)$, given the meanings $B$ of the blocks. The key case is an assignment whose right-hand side is a `case`. The pass adds $k$ as a block $l$, the join point, so $B(l)$ is the meaning of $k$, and each branch $e_i$ becomes a block $l_i = \mathbf{add}(\mathbf{assign}(e_i, x, \texttt{goto}\ l))$. The continuation $\kappa_{x,k}$ writes slot $x$ of the frame $\sigma$ in which the assignment starts, while the tail writes slot $x$ of the frame in which $e$ ends, which may also hold the slots that $e$ writes. The tail $k$ does not read those slots, because they are not initialized where the assignment starts, so $k$ means the same in both frames, because the meaning of a tail depends only on the slots it starts with. We show the first branch; the second is the same:

$$
\begin{aligned}
[\![\texttt{case}\ a\ y\ z\ e_1\ e_2]\!]_T\,\sigma\,\kappa_{x,k} &= C(\mathcal{A}[\![a]\!]\sigma,\ \boldsymbol{\lambda} X.\,[\![e_1]\!]_T\,\sigma[y{:=}X]\,\kappa_{x,k},\ \ldots) && \text{(Imp)}\\
&= C(\mathcal{A}[\![a]\!]\sigma,\ \boldsymbol{\lambda} X.\,[\![e_1]\!]_T\,\sigma[y{:=}X]\,\kappa_{x,\texttt{goto}\,l},\ \ldots) && \text{(join point } l\text{; } k \text{ does not read } y)\\
&= C(\mathcal{A}[\![a]\!]\sigma,\ \boldsymbol{\lambda} X.\,B(l_1)(\sigma[y{:=}X]),\ \ldots) && \text{(ind. hyp.)}\\
&= [\![\texttt{case}\ a\ y\ z\ l_1\ l_2]\!]_{T',B}\,\sigma && \text{(CFG)}
\end{aligned}
$$

A further lemma says that blocks added later do not change the meaning of a label made earlier.

*Select instructions.* The induction hypothesis says that a CFG tail $t$ means the same as its instructions $\mathbf{tail}(t)$, for any registers $R$ and flag $F$. A call $x$ := `callc` $y$ $a$; $t$ writes the result $C = (\mathsf{car}(\sigma(y)) \bullet \mathsf{cdr}(\sigma(y))) \bullet \mathcal{A}[\![a]\!]\sigma$ to slot $x$. Let $s$ be the sequence `movq %rax,` $x$; $\mathbf{tail}(t)$, let $R_C$ hold $C$ in `rax`, and let $R_1$ be $R$ with the curried function of the closure in `rax`, its environment in `rdi`, and the argument in `rsi`:

$$
\begin{aligned}
[\![x := \texttt{callc}\ y\ a;\ t]\!]_{T,B}\,\sigma &= \Lambda(\boldsymbol{\lambda} X.\,[\![t]\!]_{T,B}(\sigma[x{:=}X])) \bullet C && \text{(meaning of assignment)}\\
&= \Lambda(\boldsymbol{\lambda} X.\,[\![\mathbf{tail}(t)]\!](\sigma[x{:=}X])\,R_C\,\{\langle\rangle\}) \bullet C && \text{(induction hypothesis for } t)\\
&= [\![s]\!]\,\sigma\,R_C\,\{\langle\rangle\} && \text{(meaning of a write to } x)\\
&= \mathsf{guard}(C;\ [\![s]\!]\,\sigma\,R_C\,\{\langle\rangle\}) && \text{(an application to } \emptyset \text{ is } \emptyset)\\
&= [\![\texttt{callq \%rax};\ s]\!]\,\sigma\,R_1\,F && \text{(meaning of callq)}\\
&= [\![\texttt{fst}\ y,\texttt{\%rax};\ \texttt{snd}\ y,\texttt{\%rdi};\ \texttt{movq}\ a,\texttt{\%rsi};\ s]\!]\,\sigma\,R\,F && \text{(meaning of the loads)}
\end{aligned}
$$

The step for `callq` holds because `callq` applies the curried function $\mathsf{car}(\sigma(y))$ in `rax` to the environment $\mathsf{cdr}(\sigma(y))$ in `rdi` and then to the argument $\mathcal{A}[\![a]\!]\sigma$ in `rsi`, which gives $C$.

## E An Example of the Approximant Logical Relation

Agda: `FPC/proof/Correctness/GraphModel/ApproxLR.agda (the definitions; the example itself is not mechanized)`

This appendix works clause 3 of the approximant logical relation (Figure 5) through the example of Section 4.2: $\lambda y.\,x + y$ with $x$ bound to 9, whose source meaning $D$ contains $[1] \mapsto 10$. Take the one-element list $V = [\langle\!|u, 9|\!\rangle]$ with $u = [9] \mapsto ([1] \mapsto 10)$. The environment $[9]$ of the curried entry is in $\mathsf{cdr}(V)$, so $[1] \mapsto 10 \mathrel{\dot\in} V$ is a usable entry, and $V$ maps $[1]$ to $[10]$. Clause 3 then asks that $D \bullet E \trianglerighteq [10]$ for every consistent $E$ related to $[1]$, that is, every $E$ that contains 1. It holds because $D$ has the entry $[1] \mapsto 10$ and $[1] \subseteq E$, so $10 \in D \bullet E$, which is clause 1 for $[10]$. Now take $V' = [\langle\!|u', 9|\!\rangle]$ with the junk entry $u' = [2] \mapsto ([1] \mapsto 3)$. Its environment $[2]$ is not in $\mathsf{cdr}(V') = \{9\}$, so $V'$ has no usable entry, clause 3 asks nothing, and $D \trianglerighteq V'$ holds although $3 \notin D \bullet E$. Both $V$ and $V'$ are lists of elements of the target closure, which pairs every entry of its curried function with the environment 9.

*The shapes that fail.* Section 4.2 rules out all but one of the nine combinations of a source side and a target side. The details follow.

*A single target element.* A clause for one pair that accepted environments from anywhere would have to account for junk entries such as the curried entry for $[2]$ in the example above. We tried two variants in Agda: a single element against a single element, and one target element inside its whole target set. In the second, the function case must relate the whole argument set, which cannot shrink to match the index.

*A single source element or a finite source list.* A target closure that shows no usable curried entry is related to every nonempty source set, such as the singleton of $[n] \mapsto n$ for each $n$, and no finite source side for $\lambda f.\, f$ has an entry for all of them; we check this case in Agda.

*Whole denotations on both sides.* A later modality would need a step to discharge, which a denotational semantics does not have; the Scott-domain proof instead builds its relation by stages along the bilimit (Appendix G).

*A target set of bounded depth.* The argument for a call succeeds for finitely many outputs at a time. We show in Agda, by induction on the bound, that the relation holds on a bounded set exactly when it holds on each finite sublist.

## F The Other Four Semantics and Their Correctness Results

Agda: `FPC/Theorems/CorrectWrtClockedInterpreters.agda, CorrectWrtAbstractMachines.agda, CorrectWrtReductionSemantics.agda, CorrectWrtScottDomains.agda`

Each of the five developments of Section 6 proves the whole compiler correct with respect to its own semantics of the languages. For the graph model, the semantics are those of Appendix B and the theorem is Theorem 4.1. This appendix gives, for each of the other four styles, the semantics of the source language FPC and the target language $\text{Instr}_\mu$. For the operational styles, it also states the correctness result for the whole compiler, so that each can be read on its own. The theorem for Scott domains is Theorem 4.2. The semantics of the intermediate languages follow the same pattern and are in the artifact.

### F.1 Values and Observations

Agda: `Semantics/Observation.agda, Framework/Smallstep.agda, Framework/Clocked.agda`

The three operational styles share their values and their observations. A value of type $A$ is one of

$$v, w ::= c \mid \mathtt{code}\, f \mid \mathtt{pair}(v, w) \mid \mathtt{tup}(v_1, \dots, v_n) \mid \mathtt{inl}\, v \mid \mathtt{inr}\, v \mid \mathtt{fold}\, v$$

where a closure $\mathtt{pair}(v, w)$ pairs a code value $v$ with the tuple $w$ of its environment. In FPC, the code of a closure is the body $N$ of a $\lambda$-abstraction and its environment is the environment $\rho$ of the $\lambda$-abstraction, a list of values; we write $\langle \lambda x.\, N, \rho \rangle$ for $\mathtt{pair}(\mathtt{code}\, N, \mathtt{tup}(\rho))$. In $\text{Instr}_\mu$, the code of a closure is the name $f$ of a definition. A run is observed through its final value, forgetting the code and environment inside closures. An observation is one of

$$o ::= c \mid \mathit{fun} \mid \mathit{tuple}(o_1, \dots, o_n) \mid \mathit{left}(o) \mid \mathit{right}(o) \mid \mathit{error}$$

and the observation of a value is

$$\begin{aligned} \mathit{obs}(c) &= c & \mathit{obs}(\mathtt{pair}(v, w)) &= \mathit{fun} \\ \mathit{obs}(\mathtt{tup}(v_1, \dots, v_n)) &= \mathit{tuple}(\mathit{obs}(v_1), \dots, \mathit{obs}(v_n)) & \mathit{obs}(\mathtt{fold}\, v) &= \mathit{obs}(v) \\ \mathit{obs}(\mathtt{inl}\, v) &= \mathit{left}(\mathit{obs}(v)) & \mathit{obs}(\mathtt{inr}\, v) &= \mathit{right}(\mathit{obs}(v)) \end{aligned}$$

and $\mathit{obs}(\mathtt{code}\, f) = \mathit{fun}$. A run stops with the observation *error* when it gets stuck, which no well-typed program reaches. The primitives compute $\delta_{\mathsf{sub}}(n, m) = n - m$ and $\delta_{\mathsf{eq}}(n, m) = \mathtt{inl}\, \mathtt{tup}()$ if $n = m$ and $\mathtt{inr}\, \mathtt{tup}()$ otherwise.

*Machines.* A machine is a set of states, a start state, and a step function that maps each state $\varsigma$ either to a next state, written $\varsigma \longrightarrow \varsigma'$, or to a final observation $o$, written $\varsigma \longrightarrow \mathrm{stop}(o)$. A machine *terminates with* $o$ when a sequence of steps leads from the start state to a state $\varsigma$ with $\varsigma \longrightarrow \mathrm{stop}(o)$, and it *diverges* when for every $n$ there is a sequence of $n$ steps from the start state. We use the same notation for each machine; the CEK machines of FPC and $\text{Instr}_\mu$ (Appendix F.3) and the reduction semantics of FPC (Appendix F.4) each define their own $\longrightarrow$ and $\longrightarrow$ stop.

*Clocked interpreters.* A clocked interpreter maps a clock $k \in \mathbb{N}$ to a *result*, ranged over by $e$, which times out, reports an error, or returns a value $v$:

$$e ::= \mathsf{timeout} \mid \mathsf{error} \mid \mathsf{ret}\, v$$

We write $v \leftarrow e;\ e'$ for the result $e'$ which depends on $v$ when $e$ is $\mathsf{ret}\,v$, and for $e$ itself when $e$ is timeout or error. An interpreter *terminates with* $o$ when for some clock $k$ its result is $\mathsf{ret}\,v$ with $obs(v) = o$, or error with $o = error$, and it *diverges* when its result is timeout for every clock.

*Equivalence.* Two machines, or two interpreters, are *equivalent* when, for every observation $o$, one terminates with $o$ if and only if the other does, and one diverges if and only if the other does. The compiler correctness proofs for the three operational semantics state that a program and its compilation are equivalent.

## F.2 Clocked Interpreters

Agda: `FPC/Semantics/ClockedInterpreters/FPC.agda`,
`FPC/Semantics/ClockedInterpreters/X86.agda`

*Interpreter for* FPC. The interpreter $eval_k(\rho, M)$ evaluates a term $M$ in an environment $\rho$ of values at clock $k$. The clock ticks at each call of the interpreted program.

**The interpreter of** FPC $\quad$ $\boxed{eval_k(\rho, M) = e \qquad call_k(f, w) = e}$

$$\begin{aligned}
eval_k(\rho, x) &= \mathsf{ret}\,\rho(x)\\
eval_k(\rho, \lambda x.\, N) &= \mathsf{ret}\,\langle \lambda x.\, N, \rho\rangle\\
eval_k(\rho, L\ M) &= f \leftarrow eval_k(\rho, L);\ w \leftarrow eval_k(\rho, M);\ call_k(f, w)\\
eval_k(\rho, c) &= \mathsf{ret}\, c\\
eval_k(\rho, \mathtt{tuple}\,(M_1, \ldots, M_n)) &= v_1 \leftarrow eval_k(\rho, M_1);\ \ldots;\ v_n \leftarrow eval_k(\rho, M_n);\ \mathsf{ret}\,\mathtt{tup}(v_1, \ldots, v_n)\\
eval_k(\rho, \mathtt{get}_i\, M) &= \mathtt{tup}(v_1, \ldots, v_n) \leftarrow eval_k(\rho, M);\ \mathsf{ret}\, v_i\\
eval_k(\rho, \mathtt{inl}\, M) &= v \leftarrow eval_k(\rho, M);\ \mathsf{ret}\,(\mathtt{inl}\, v) \qquad \text{(likewise } \mathtt{inr} \text{ and } \mathtt{fold}\text{)}\\
eval_k(\rho, \mathtt{case}\ L\ (x.\, M)\ (y.\, N)) &= u \leftarrow eval_k(\rho, L);\ \begin{cases} eval_k(\rho[x{:=}v], M) & \text{if } u = \mathtt{inl}\, v\\ eval_k(\rho[y{:=}v], N) & \text{if } u = \mathtt{inr}\, v\end{cases}\\
eval_k(\rho, \mathtt{prim}_p\ L\ M) &= n \leftarrow eval_k(\rho, L);\ m \leftarrow eval_k(\rho, M);\ \mathsf{ret}\,\delta_p(n, m)\\
eval_k(\rho, \mathtt{let}\ x = M\ \mathtt{in}\ N) &= v \leftarrow eval_k(\rho, M);\ eval_k(\rho[x{:=}v], N)\\
eval_k(\rho, \mathtt{unfold}\ M) &= \mathtt{fold}\, v \leftarrow eval_k(\rho, M);\ \mathsf{ret}\, v
\end{aligned}$$

$$\begin{aligned}
call_0(f, w) &= \mathsf{timeout}\\
call_{k+1}(\langle \lambda x.\, N, \rho\rangle, w) &= eval_k(\rho[x{:=}w], N)
\end{aligned}$$

A pattern on the left of $\leftarrow$ always matches, because the interpreter is defined on well-typed terms and values. The interpreter of a closed program $M$ is $k \mapsto eval_k(\emptyset, M)$.

*Interpreter for* $\text{Instr}_\mu$. A machine state $m = \langle R, \sigma, u, b\rangle$ of $\text{Instr}_\mu$ has the registers $R$, which map each of `rax`, `rdi`, and `rsi` to a value or unset; the frame $\sigma$, which holds a value in each initialized slot; the environment $u$ of the open closure, if any; and the flag $b$, which is a Boolean or unset. We write $m(r)$ for $R(r)$. $arg(a, m)$ converts an operand $a$ into a value, $obj(con, m)$ converts a constructor $con$ to be allocated into a value, and $put(dst, v, m)$ stores $v$ in a destination $dst$.

**Auxiliary definitions** $\boxed{arg(a, m) = v \qquad obj(con, m) = v \qquad put(dst, v, m) = m'}$

$$\begin{aligned}
arg(x,\ \langle R, \sigma, u, b\rangle) &= \sigma(x)\\
arg(c,\ m) &= c\\
arg(f,\ m) &= \mathsf{code}\, f\\
arg(\%r,\ \langle R, \sigma, u, b\rangle) &= R(r)\\
arg(\mathtt{fold}\, a,\ m) &= \mathtt{fold}\,(arg(a, m))\\
arg(\mathtt{unfold}\, a,\ m) &= \mathtt{fold}\, v \leftarrow arg(a, m);\ v
\end{aligned}$$

$$\begin{aligned}
obj(\mathtt{pair}(a_1, a_2),\ m) &= \mathtt{pair}(arg(a_1, m), arg(a_2, m))\\
obj(\mathtt{tuple}(a_1, \ldots, a_n),\ m) &= \mathtt{tup}(arg(a_1, m), \ldots, arg(a_n, m))\\
obj(\mathtt{inl}(a),\ m) &= \mathtt{inl}\,(arg(a, m))\\
obj(\mathtt{inr}(a),\ m) &= \mathtt{inr}\,(arg(a, m))
\end{aligned}$$

$$\begin{aligned}
put(\%r, v,\ \langle R, \sigma, u, b\rangle) &= \langle R[r{:=}v], \sigma, u, b\rangle\\
put(x, v,\ \langle R, \sigma, u, b\rangle) &= \langle R, \sigma[x{:=}v]|_{I'}, \mathsf{none}, b\rangle
\end{aligned}$$

Here $\sigma[x{:=}v]$ updates slot $x$ in place, and $|_{I'}$ keeps the slots $I'$ that the rest of the sequence starts with (DST-SLOT, Appendix B.7), which forgets the values of variables whose scope has ended. Writing a slot also closes the open closure. DST-SLOT resets the open closure to $\cdot$, and $put$ closes it by setting it to none. The instructions other than `callq` update the state as $instr$ specifies.

**The effect of an instruction** $\boxed{instr(i, m) = m'}$

$$\begin{aligned}
instr(\mathsf{movq}\ a, dst,\ m) &= put(dst, arg(a, m), m)\\
instr(\mathsf{subq}\ a, \%r,\ m) &= put(\%r, \delta_{\mathsf{sub}}(m(r), arg(a, m)), m)\\
instr(\mathsf{cmpq}\ a_2, a_1,\ \langle R, \sigma, u, b\rangle) &= \langle R, \sigma, u, (arg(a_1, m) = arg(a_2, m))\rangle\\
instr(\mathsf{sete}\ dst,\ \langle R, \sigma, u, b\rangle) &= \begin{cases} put(dst, \mathtt{inl}\ \mathtt{tup}(), m) & \text{if } b = \mathsf{true}\\ put(dst, \mathtt{inr}\ \mathtt{tup}(), m) & \text{if } b = \mathsf{false}\end{cases}\\
instr(\mathtt{alloc}\ con, dst,\ m) &= put(dst, obj(con, m), m)\\
instr(\mathtt{load}\ \mathtt{get}_i\ a, dst,\ m) &= \mathtt{tup}(v_1, \ldots, v_n) \leftarrow arg(a, m);\ put(dst, v_i, m)\\
instr(\mathtt{fst}\ x, \%r,\ \langle R, \sigma, u, b\rangle) &= \mathtt{pair}(c, u') \leftarrow \sigma(x);\ \langle R^\circ[r{:=}c], \sigma, u', b\rangle\\
instr(\mathtt{snd}\ x, \%r,\ \langle R, \sigma, u, b\rangle) &= \langle R[r{:=}u], \sigma, u, b\rangle
\end{aligned}$$

Here $m$ on the right is the state on the left, and $R^\circ$ is the result of wiping $R$, which unsets the registers that held the code or the environment of an earlier open closure, as $wipe(\Theta)$ unsets their types in INS-FST (Appendix B.7). As in the interpreter of FPC, a pattern on the left of $\leftarrow$ always matches, because instructions are typed by the state they start in. The interpreter $run_k(B, m, s)$ runs a sequence $s$ in the state $m$, where $B = s_0, \ldots, s_{n-1}$ lists the blocks that $s$ may jump to. A label $l$ names the block $s_l$ of $B$. Because control flow within a function is acyclic, a block may jump only to the blocks created before it, which run after it, so jumping to $l$ runs $s_l$ with the blocks $B_l = s_0, \ldots, s_{l-1}$ and pops the newer ones, which $s_l$ cannot reach. A `callq` to the code $\mathsf{code}\, f$ runs $call_k(\texttt{code}\, f, u, w)$, which does not use $B$. The program has a global list of definitions, fixed for the whole run, and $call_k$ looks up definition $f$ in it, whose function body has the start $s_f$ and the blocks $B_f$; when $call_k$ returns, the caller resumes with its own $B$.

**The interpreter of** $\mathrm{Instr}_\mu$ $\boxed{run_k(B, m, s) = e \qquad jump_k(B, l, \sigma) = e \qquad call_k(c, u, w) = e}$

$$
\begin{aligned}
run_k(B, m, i;\ s) &= run_k(B, instr(i, m), s) \qquad \text{for } i \text{ not } \texttt{callq} \\
run_k(B, m, \texttt{callq}\ a;\ s) &= v \leftarrow call_k(arg(a, m), m(\texttt{rdi}), m(\texttt{rsi}));\ run_k(B, m_v, s) \\
run_k(B, m, \texttt{retq}) &= \mathrm{ret}\ m(\texttt{rax}) \\
run_k(B, \langle R, \sigma, u, b\rangle, \texttt{jmp}\ l) &= jump_k(B, l, \sigma) \\
run_k(B, \langle R, \sigma, u, b\rangle, \texttt{case}\ x\ y\ z\ l_1\ l_2) &= \begin{cases} jump_k(B, l_1, \sigma[y{:=}v]) & \text{if } \sigma(x) = \texttt{inl}\, v \\ jump_k(B, l_2, \sigma[z{:=}v]) & \text{if } \sigma(x) = \texttt{inr}\, v \end{cases} \\
jump_k(B, l, \sigma) &= run_k(B_l, m_\sigma, s_l) \\
call_0(\texttt{code}\, f, u, w) &= \mathrm{timeout} \\
call_{k+1}(\texttt{code}\, f, u, w) &= run_k(B_f, m_{u,w}, s_f)
\end{aligned}
$$

Here $m_v$ has the frame of $m$, `rax` set to $v$, and nothing else set; $m_\sigma$ has the frame $\sigma$ and nothing else set; and $m_{u,w}$ has $u$ in `rdi`, $w$ in `rsi`, a frame with no slot initialized, and nothing else set. The interpreter of a program is $k \mapsto run_k(B, m_0, s)$ for the main function body $s$ with blocks $B$, where $m_0$ has no slot initialized and nothing set.

Proposition F.1 (Correctness, clocked interpreters). *Let $M$ be a closed term of* FPC *of type* `Int`. *For every integer $n$, the interpreter of* FPC *on $M$ terminates with $n$ if and only if the interpreter of* $\mathrm{Instr}_\mu$ *on* ***compile***$(M)$ *does, and one diverges if and only if the other does. More generally, for every closed term $M$, the two interpreters are equivalent.*

## F.3 Abstract Machines

Agda: `FPC/Semantics/AbstractMachines/FPC.agda, FPC/Semantics/AbstractMachines/X86.agda`

*Machine for* FPC. The machine is the CEK machine of Felleisen and Friedman [1986], extended from the $\lambda$-calculus to the constructs of FPC.

**States and continuations** $\boxed{\varsigma \text{ and } \kappa}$

$$
\begin{aligned}
\varsigma &::= eval(M, \rho, \kappa) \mid ret(v, \kappa) \\
\kappa &::= halt \mid appL(M, \rho, \kappa) \mid appR(f, \kappa) \mid tupK(\vec{v}, \vec{M}, \rho, \kappa) \mid getK_i(\kappa) \mid inlK(\kappa) \mid inrK(\kappa) \\
&\quad \mid caseK(x.\, M, y.\, N, \rho, \kappa) \mid primL_p(M, \rho, \kappa) \mid primR_p(n, \kappa) \\
&\quad \mid letK(x.\, N, \rho, \kappa) \mid foldK(\kappa) \mid unfoldK(\kappa)
\end{aligned}
$$

A state either evaluates a term $M$ in an environment $\rho$ with a continuation $\kappa$, $eval(M, \rho, \kappa)$, or returns a value $v$ to a continuation, $ret(v, \kappa)$. The first is the CEK triple $\langle M, \rho, \kappa\rangle$. The second is the return state of Felleisen and Friedman [1986], $\langle \ddagger, \emptyset, (\kappa\ ret\ v)\rangle$, whose environment is always empty, so we drop it. A continuation is a stack of frames, and each frame holds the work that remains after a subterm returns its value. For example, *halt* is the empty continuation (their *stop*), $appL(M, \rho, \kappa)$ holds the argument $M$ that the machine evaluates next (their *arg*), and $appR(f, \kappa)$ holds the function $f$ that awaits the argument's value (their *fun*). The other frames are for primitives, `let`, tuples, sums and recursive types. An environment is a list of values, and the environment inside a closure is stored as a tuple.

**Steps of the CEK machine of** FPC $\boxed{\varsigma \longrightarrow \varsigma'}$

$$
\begin{aligned}
\mathit{eval}(x, \rho, \kappa) &\longrightarrow \mathit{ret}(\rho(x), \kappa)\\
\mathit{eval}(\lambda x.\, N, \rho, \kappa) &\longrightarrow \mathit{ret}(\langle \lambda x.\, N, \rho\rangle, \kappa)\\
\mathit{eval}(L\ M, \rho, \kappa) &\longrightarrow \mathit{eval}(L, \rho, \mathit{appL}(M, \rho, \kappa))\\
\mathit{eval}(c, \rho, \kappa) &\longrightarrow \mathit{ret}(c, \kappa)\\
\mathit{eval}(\texttt{tuple}\ (), \rho, \kappa) &\longrightarrow \mathit{ret}(\texttt{tup}(), \kappa)\\
\mathit{eval}(\texttt{tuple}\ (M_1, \ldots, M_n), \rho, \kappa) &\longrightarrow \mathit{eval}(M_1, \rho, \mathit{tupK}((), (M_2, \ldots, M_n), \rho, \kappa))\\
\mathit{eval}(\texttt{get}_i\ M, \rho, \kappa) &\longrightarrow \mathit{eval}(M, \rho, \mathit{getK}_i(\kappa)) \qquad \text{(likewise } \texttt{inl}, \texttt{inr}, \texttt{fold}, \texttt{unfold})\\
\mathit{eval}(\texttt{case}\ L\ (x.\, M)\ (y.\, N), \rho, \kappa) &\longrightarrow \mathit{eval}(L, \rho, \mathit{caseK}(x.\, M, y.\, N, \rho, \kappa))\\
\mathit{eval}(\texttt{prim}_p\ L\ M, \rho, \kappa) &\longrightarrow \mathit{eval}(L, \rho, \mathit{primL}_p(M, \rho, \kappa))\\
\mathit{eval}(\texttt{let}\ x = M\ \texttt{in}\ N, \rho, \kappa) &\longrightarrow \mathit{eval}(M, \rho, \mathit{letK}(x.\, N, \rho, \kappa))
\end{aligned}
$$

$$
\begin{aligned}
\mathit{ret}(f, \mathit{appL}(M, \rho, \kappa)) &\longrightarrow \mathit{eval}(M, \rho, \mathit{appR}(f, \kappa))\\
\mathit{ret}(w, \mathit{appR}(\langle \lambda x.\, N, \rho\rangle, \kappa)) &\longrightarrow \mathit{eval}(N, \rho[x{:=}w], \kappa)\\
\mathit{ret}(v, \mathit{tupK}(\vec{v}, (), \rho, \kappa)) &\longrightarrow \mathit{ret}(\texttt{tup}(\vec{v}, v), \kappa)\\
\mathit{ret}(v, \mathit{tupK}(\vec{v}, (M, \vec{M}), \rho, \kappa)) &\longrightarrow \mathit{eval}(M, \rho, \mathit{tupK}((\vec{v}, v), \vec{M}, \rho, \kappa))\\
\mathit{ret}(\texttt{tup}(v_1, \ldots, v_n), \mathit{getK}_i(\kappa)) &\longrightarrow \mathit{ret}(v_i, \kappa)\\
\mathit{ret}(v, \mathit{inlK}(\kappa)) &\longrightarrow \mathit{ret}(\texttt{inl}\, v, \kappa) \qquad \text{(likewise } \texttt{inr}, \texttt{fold})\\
\mathit{ret}(\texttt{inl}\, v, \mathit{caseK}(x.\, M, y.\, N, \rho, \kappa)) &\longrightarrow \mathit{eval}(M, \rho[x{:=}v], \kappa)\\
\mathit{ret}(\texttt{inr}\, v, \mathit{caseK}(x.\, M, y.\, N, \rho, \kappa)) &\longrightarrow \mathit{eval}(N, \rho[y{:=}v], \kappa)\\
\mathit{ret}(n, \mathit{primL}_p(M, \rho, \kappa)) &\longrightarrow \mathit{eval}(M, \rho, \mathit{primR}_p(n, \kappa))\\
\mathit{ret}(m, \mathit{primR}_p(n, \kappa)) &\longrightarrow \mathit{ret}(\delta_p(n, m), \kappa)\\
\mathit{ret}(v, \mathit{letK}(x.\, N, \rho, \kappa)) &\longrightarrow \mathit{eval}(N, \rho[x{:=}v], \kappa)\\
\mathit{ret}(\texttt{fold}\, v, \mathit{unfoldK}(\kappa)) &\longrightarrow \mathit{ret}(v, \kappa)
\end{aligned}
$$

The machine of a closed program $M$ starts in $\mathit{eval}(M, \emptyset, \mathit{halt})$. The terminal state is $\mathit{ret}(v, \mathit{halt})$. We define the final observation as follows.

**Final observation of the CEK machine of** FPC $\boxed{\varsigma \longrightarrow \mathsf{stop}(o)}$

$$\mathit{ret}(v, \mathit{halt}) \longrightarrow \mathsf{stop}(\mathit{obs}(v))$$

In the Agda development, terms, environments and continuations are intrinsically typed, and a variable is a de Bruijn index into the environment, so a state is well typed by construction. The CEK transitions are defined in the step function, `step : State → Next State`, which maps each state to a next state or to a final observation.

*Machine for* $\mathrm{Instr}_\mu$. In the notation of Appendix F.2, a state of the machine of $\mathrm{Instr}_\mu$, $\mathit{run}(s, B, \Sigma, m)$, runs a sequence $s$ with blocks $B$, a stack $\Sigma$ of the callers' return points, and registers, frame, open closure and flag $m$. A return point $\mathit{retK}(s, B, \sigma)$ holds the rest $s$ of the caller's sequence, its blocks, and its frame, which the callee's `retq` restores.

**States and stacks** $\boxed{\varsigma \quad \Sigma}$

$$\varsigma ::= run(s, B, \Sigma, m)$$
$$\Sigma ::= halt \mid retK(s, B, \sigma) :: \Sigma$$

**Steps of the machine of** $\mathrm{Instr}_\mu$ $\boxed{\varsigma \longrightarrow \varsigma'}$

$$
\begin{aligned}
run(i;\ s, B, \Sigma, m) &\longrightarrow run(s, B, \Sigma, instr(i, m)) \qquad \text{for } i \text{ not } \texttt{callq} \\
run(\texttt{callq}\ a;\ s, B, \Sigma, m) &\longrightarrow run(s_f, B_f, retK(s, B, \sigma) :: \Sigma, m_{u,w}) \\
&\quad \text{where } arg(a, m) = \mathsf{code}\, f,\ u = m(\texttt{rdi}),\ w = m(\texttt{rsi}) \\
run(\texttt{retq}, B, retK(s, B', \sigma') :: \Sigma, m) &\longrightarrow run(s, B', \Sigma, m') \\
&\quad \text{where } m' \text{ has the frame } \sigma' \text{ and only } \texttt{rax} \text{ set, to } m(\texttt{rax}) \\
run(\texttt{jmp}\ l, B, \Sigma, m) &\longrightarrow run(s_l, B_l, \Sigma, m_\sigma) \\
run(\texttt{case}\ x\ y\ z\ l_1\ l_2, B, \Sigma, m) &\longrightarrow run(s_{l_1}, B_{l_1}, \Sigma, m_{\sigma[y:=v]}) \\
&\quad \text{if } \sigma(x) = \texttt{inl}\, v \\
&\text{(and likewise for } \texttt{inr}, z, \text{ and } l_2)
\end{aligned}
$$

**Final observation of the machine of** $\mathrm{Instr}_\mu$ $\boxed{\varsigma \longrightarrow \mathsf{stop}(o)}$

$$run(\texttt{retq}, B, halt, m) \longrightarrow \mathsf{stop}(obs(m(\texttt{rax})))$$

The machine of a program starts in $run(s, B, halt, m_0)$ for the main function body $s$ with blocks $B$.

Proposition F.2 (Correctness, abstract machines). *Let $M$ be a closed term of* FPC *of type* `Int`. *For every integer $n$, the machine of* FPC *on $M$ terminates with $n$ if and only if the machine of* $\mathrm{Instr}_\mu$ *on* $\textbf{compile}(M)$ *does, and one diverges if and only if the other does. More generally, for every closed term $M$, the two machines are equivalent.*

## F.4 Reduction Semantics

Agda: `FPC/Semantics/ReductionSemantics/FPC.agda, Framework/Reduction.agda`

The reduction semantics of FPC rewrites closed terms by substitution. The evaluation order is call by value and left to right. We define values as follows.

$$V, W ::= \lambda x.\, N \mid c \mid \texttt{tuple}\,(V_1, \ldots, V_n) \mid \texttt{inl}\, V \mid \texttt{inr}\, V \mid \texttt{fold}\, V$$

The Agda development defines the reduction semantics as a function (instead of a relation). The function *red* maps a closed term $M$ to a *result*, ranged over by *res*, which says whether $M$ is a value (val), reduces to a term $M'$ (to $M'$), or is stuck (error). The stuck case is there only to make *red* total. A closed well-typed term does not get stuck, because FPC is type safe.

$$res ::= \mathsf{val} \mid \mathsf{to}\, M' \mid \mathsf{error}$$

The definition of *red* below is the core of the reduction semantics. The function *red* is deterministic and total by construction. In each case, the subterms are reduced left to right. A subterm that reduces makes the whole term reduce in the same position, and once the subterms are values the term takes its own step.

**Reduction of a closed term** $\boxed{red(M) = res}$

$$
\begin{aligned}
red(V) &= \mathsf{val} \\
red(L\ M) &= \mathsf{to}\,(L'\ M) && \text{if } red(L) = \mathsf{to}\,L' \\
&= \mathsf{to}\,(V\ M') && \text{if } L = V,\ red(M) = \mathsf{to}\,M' \\
&= \mathsf{to}\,(N[W/x]) && \text{if } L = \lambda x.\,N,\ M = W \\
red(\mathtt{tuple}\,(\vec{V}, M, \vec{M})) &= \mathsf{to}\,(\mathtt{tuple}\,(\vec{V}, M', \vec{M})) && \text{if } red(M) = \mathsf{to}\,M' \\
red(\mathtt{get}_i\,M) &= \mathsf{to}\,(\mathtt{get}_i\,M') && \text{if } red(M) = \mathsf{to}\,M' \\
&= \mathsf{to}\,V_i && \text{if } M = \mathtt{tuple}\,(V_1, \ldots, V_n) \\
red(\mathtt{inl}\,M) &= \mathsf{to}\,(\mathtt{inl}\,M') && \text{if } red(M) = \mathsf{to}\,M' \\
& && \text{(likewise } \mathtt{inr} \text{ and } \mathtt{fold}\text{)} \\
red(\mathtt{case}\ L\ (x.\,M)\ (y.\,N)) &= \mathsf{to}\,(\mathtt{case}\ L'\ (x.\,M)\ (y.\,N)) && \text{if } red(L) = \mathsf{to}\,L' \\
&= \mathsf{to}\,(M[V/x]) && \text{if } L = \mathtt{inl}\,V \\
&= \mathsf{to}\,(N[V/y]) && \text{if } L = \mathtt{inr}\,V \\
red(\mathtt{prim}_p\ L\ M) &= \mathsf{to}\,(\mathtt{prim}_p\ L'\ M) && \text{if } red(L) = \mathsf{to}\,L' \\
&= \mathsf{to}\,(\mathtt{prim}_p\ V\ M') && \text{if } L = V,\ red(M) = \mathsf{to}\,M' \\
&= \mathsf{to}\,\delta_p(n, m) && \text{if } L = n,\ M = m \\
red(\mathtt{let}\ x{=}M\ \mathtt{in}\ N) &= \mathsf{to}\,(\mathtt{let}\ x{=}M'\ \mathtt{in}\ N) && \text{if } red(M) = \mathsf{to}\,M' \\
&= \mathsf{to}\,(N[V/x]) && \text{if } M = V \\
red(\mathtt{unfold}\ M) &= \mathsf{to}\,(\mathtt{unfold}\ M') && \text{if } red(M) = \mathsf{to}\,M' \\
&= \mathsf{to}\,V && \text{if } M = \mathtt{fold}\,V \\
red(M) &= \mathsf{error} && \text{otherwise}
\end{aligned}
$$

Here $\delta_p(n, m)$ is the result of a primitive operator, which is a value, read as a term. The observation $obs(V)$ of a value is as in Appendix F.1, with *fun* for $\lambda x.\,N$.

In the Agda development, the cases that reduce a subterm are written with one combinator, *step-in*$(f, res, k)$. If the subterm reduces, by *res*, then so does the whole term, in the evaluation context $f$; if the subterm is a value, the reduction continues with $k$; and if the subterm is stuck, so is the whole term. The equations above are the Agda definition with the combinator unfolded.

With the function *red* defined, we define single-step reduction and final observation as follows. A closed program $M$ runs by these steps, so it terminates and diverges in the sense of Appendix F.1.

**Steps and final observation of the reduction semantics** $\boxed{M \longrightarrow M' \qquad M \longrightarrow \mathsf{stop}(o)}$

$$M \longrightarrow M' \ \text{ if } red(M) = \mathsf{to}\,M' \qquad M \longrightarrow \mathsf{stop}(obs(M)) \ \text{ if } red(M) = \mathsf{val}$$
$$M \longrightarrow \mathsf{stop}(error) \ \text{ if } red(M) = \mathsf{error}$$

The languages Imp, CFG, and $\mathrm{Instr}_\mu$ are first order, so their abstract machines of Appendix F.3 serve as their reduction semantics.

Proposition F.3 (Correctness, reduction semantics). *Let $M$ be a closed term of* FPC *of type* `Int`. *For every integer $n$, the reduction semantics of* FPC *on $M$ terminates with $n$ if and only if the machine of* $\mathrm{Instr}_\mu$ *on* ***compile***$(M)$ *does, and one diverges if and only if the other does. More generally, for every closed term $M$, the two are equivalent.*

### F.5 Scott Domains

Agda: `Framework/Scott/Lift.agda`, `Framework/Scott/Domain.agda`, `Framework/Scott/Ops.agda`, `FPC/Semantics/ScottDomains/FPC.agda`, `FPC/Semantics/ScottDomains/X86.agda`

Our Agda development builds the domain of values with the constructive domain theory of TypeTopology [de Jong 2023], in which domains are directed-complete partial orders, as the bilimit of Appendix G. The domain of values is the solution of

$$\mathbb{D} \cong \mathcal{L}(\Sigma_{t \in Tag}\, Payload_t(\mathbb{D}))$$

where $\mathcal{L}$ is lifting, the standard construction that adds a least element $\bot$ below a domain [Escardó and Knapp 2017; Winskel 1993a]. The tag $t$ of a value distinguishes constants, tuples of each arity, injections, functions, and pairs, and the payload holds its contents:

$$t ::= c \mid tup_n \mid left \mid right \mid fun \mid pr$$

$$\begin{aligned} Payload_c(\mathbb{D}) &= \mathbf{1} & Payload_{fun}(\mathbb{D}) &= [\mathbb{D} \to \mathbb{D}] \\ Payload_{tup_n}(\mathbb{D}) &= \mathbb{D}^n & Payload_{pr}(\mathbb{D}) &= \mathbb{D}^2 \\ Payload_{left}(\mathbb{D}) &= \mathbb{D} & Payload_{right}(\mathbb{D}) &= \mathbb{D} \end{aligned}$$

where $\mathbf{1}$ is the one-point domain and $[\mathbb{D} \to \mathbb{D}]$ the domain of continuous functions. The least element $\bot$ of $\mathbb{D}$ means divergence; a value $d$ is *defined* when $d \neq \bot$, that is, when $d = \lfloor(t, x)\rfloor$ for a tag $t$ and a payload $x$ in $Payload_t(\mathbb{D})$, where $\lfloor\cdot\rfloor$ is the injection of lifting. The empty tuple is $\langle\rangle = \lfloor(tup_0, ())\rfloor$, where $()$ is the element of $\mathbf{1}$ and of $\mathbb{D}^0$. As in the graph model, the operators are strict, and each of them is also continuous. We define them as follows, where $g'$ is the strict function that agrees with $g$ on defined values and maps $\bot$ to $\bot$.

$$\begin{aligned}
\Lambda(g) &= \lfloor(fun, g')\rfloor \\
\mathsf{app}(d, e) &= h(e) && \text{if } d = \lfloor(fun, h)\rfloor \text{ and } e \text{ is defined} \\
\mathcal{T}(d_1, \ldots, d_n) &= \lfloor(tup_n, (d_1, \ldots, d_n))\rfloor && \text{if every } d_i \text{ is defined} \\
\pi_{tup_i}(d) &= d_i && \text{if } d = \lfloor(tup_n, (d_1, \ldots, d_n))\rfloor \\
\mathsf{pair}(d, e) &= \lfloor(pr, (d, e))\rfloor && \text{if } d \text{ and } e \text{ are defined} \\
\mathsf{car}(d) &= d_1 && \text{if } d = \lfloor(pr, (d_1, d_2))\rfloor \\
\mathsf{cdr}(d) &= d_2 && \text{if } d = \lfloor(pr, (d_1, d_2))\rfloor \\
\mathsf{inl}(d) &= \lfloor(left, d)\rfloor && \text{if } d \text{ is defined} \\
\mathsf{inr}(d) &= \lfloor(right, d)\rfloor && \text{if } d \text{ is defined} \\
\mathsf{cases}(d, f, g) &= f(e) && \text{if } d = \lfloor(left, e)\rfloor \\
&= g(e) && \text{if } d = \lfloor(right, e)\rfloor \\
\mathsf{prim}_p(d, e) &= out(\delta(p, n, m)) && \text{if } d = \lfloor(n, ())\rfloor \text{ and } e = \lfloor(m, ())\rfloor \text{ for integers } n, m \\
\mathsf{seq}(d, e) &= e && \text{if } d \text{ is defined}
\end{aligned}$$

In every case that the equations do not cover, the result is $\bot$. The function *out* turns the result of a primitive into a value, with booleans as injections, as in the graph model (Figure 3): $out(k) = \lfloor(k, ())\rfloor$ for an integer $k$, $out(\text{true}) = \lfloor(left, \langle\rangle)\rfloor$, and $out(\text{false}) = \lfloor(right, \langle\rangle)\rfloor$. An environment maps variables to values, and the environment $\rho_0$ of a closed program maps every variable to the empty tuple $\langle\rangle$.

*Semantics of* FPC. The semantics of FPC has the equations of the graph-model semantics (Appendix B), with the operators of $\mathbb{D}$:

$$
\begin{aligned}
[\![x]\!]\rho &= \rho(x) & [\![c]\!]\rho &= \lfloor (c, ()) \rfloor \\
[\![\lambda x.\, N]\!]\rho &= \Lambda(\boldsymbol{\lambda} d.\, [\![N]\!]\rho[x{:=}d]) & [\![L\ M]\!]\rho &= \mathsf{app}([\![L]\!]\rho, [\![M]\!]\rho) \\
[\![\mathtt{tuple}\ (M_1, \ldots, M_n)]\!]\rho &= \mathcal{T}([\![M_1]\!]\rho, \ldots, [\![M_n]\!]\rho) & [\![\mathtt{get}_i\ M]\!]\rho &= \pi_{tup_i}([\![M]\!]\rho) \\
[\![\mathtt{inl}\ M]\!]\rho &= \mathsf{inl}([\![M]\!]\rho) & [\![\mathtt{inr}\ M]\!]\rho &= \mathsf{inr}([\![M]\!]\rho) \\
[\![\mathtt{prim}_p\ L\ M]\!]\rho &= \mathsf{prim}_p([\![L]\!]\rho, [\![M]\!]\rho) & [\![\mathtt{fold}\ M]\!]\rho &= [\![M]\!]\rho \\
[\![\mathtt{let}\ x = M\ \mathtt{in}\ N]\!]\rho &= \mathsf{app}(\Lambda(\boldsymbol{\lambda} d.\, [\![N]\!]\rho[x{:=}d]), [\![M]\!]\rho) & [\![\mathtt{unfold}\ M]\!]\rho &= [\![M]\!]\rho
\end{aligned}
$$

$$
[\![\mathtt{case}\ L\ (x.\,M)\ (y.\,N)]\!]\rho = \mathsf{cases}([\![L]\!]\rho,\ (\boldsymbol{\lambda} d.\, [\![M]\!]\rho[x{:=}d])',\ (\boldsymbol{\lambda} d.\, [\![N]\!]\rho[y{:=}d])')
$$

where $g'$ is the strict function that agrees with $g$ on defined values, as in $\Lambda$.

*Semantics of* $\mathrm{Instr}_\mu$. The semantics of $\mathrm{Instr}_\mu$ has the equations of the graph-model semantics of $\mathrm{Instr}_\mu$ (Appendix B), with the operators of $\mathbb{D}$ in place of those of the graph model, seq in place of guard in the meaning of `callq`, and the empty tuple $\langle\rangle$ in place of $\{\langle\rangle\}$ as the content of an unset register and an unset flag. A sequence means a continuous function of the frame, the registers, and the flag.

Theorem 4.2 (Section 4) states the correctness of the compiler with respect to these semantics.

## G The Scott-Domain Proof of split closure

Agda: `FPC/proof/Correctness/ScottDomains/SplitRel.agda, SplitPreserve.agda, SplitReflect.agda, SplitTerms.agda, SplitCorrect.agda; proof/Framework/Scott/Stages.agda, StageCall.agda`

As in the graph model (Section 4.2), **split closure** changes the representation of closures, so the Scott-domain proof relates the meaning of a program to the meaning of its translation by a logical relation, one for each direction. The universal domain $\mathbb{D}$ (Appendix F.5) rules out the two usual ways of defining such a relation. Recursion on types does not terminate, because the types include recursive types, and recursion on values is not well founded, because values are infinite. We instead define the relations by recursion on the stages of the bilimit that constructs $\mathbb{D}$, as Reynolds [1974] did. Each relation relates an approximation of one value at a stage to the whole other value, so, like the approximant logical relation, it approximates only one side. The preservation relation approximates the source, and the reflection relation approximates the target.

*Stages.* The stages of $\mathbb{D}$ are the domains

$$\mathbb{D}_0 = \mathbf{1} \qquad \mathbb{D}_{n+1} = \mathcal{L}(\Sigma_{t \in Tag}\, Payload_t(\mathbb{D}_n))$$

with the maps $\pi_n : \mathbb{D}_{n+1} \to \mathbb{D}_n$ and $\varepsilon_n : \mathbb{D}_n \to \mathbb{D}_{n+1}$ between consecutive stages, and the maps $\pi_n^\infty : \mathbb{D} \to \mathbb{D}_n$ and $\varepsilon_n^\infty : \mathbb{D}_n \to \mathbb{D}$ between a stage and the limit. Every value $d$ is the supremum of its approximations $\varepsilon_n^\infty(\pi_n^\infty(d))$. At stage $n+1$, the payload of a function is in $[\mathbb{D}_n \to \mathbb{D}_n]$ and the payload of a pair is in $\mathbb{D}_n^2$.

A target closure is a pair $(c, e)$ of a curried function and an environment, and the meaning of `callc` calls it on an argument $a$ by $call(c, e, a) = \mathsf{app}(\mathsf{app}(c, e), a)$. The call applies the curried function twice, and applying a function of stage $k+1$ gives a result of stage $k$. A call of a pair of stage $m+2$ therefore gives its result at stage $m$:

$$call_m((c, e), v) = \pi_m^\infty(call(\varepsilon_{m+2}^\infty(c),\ \varepsilon_{m+1}^\infty(\pi_{m+1}(e)),\ \varepsilon_m^\infty(v))) \qquad \text{for } c, e \in \mathbb{D}_{m+2} \text{ and } v \in \mathbb{D}_m.$$

**Related tags** $\boxed{t \asymp^+ t' \qquad t \asymp^- t'}$

$$t \asymp^\pm t \text{ for } t \notin \{\mathit{fun}, \mathit{pr}\} \qquad \mathit{fun} \asymp^+ \mathit{pr} \qquad \mathit{pr} \asymp^- \mathit{fun}$$

**The stage relations** $\boxed{s\ \mathcal{R}^\pm_n\ w \quad (s \in \mathbb{D}_n,\ w \in \mathbb{D})}$

$$\begin{aligned} s\ \mathcal{R}^\pm_0\ w &\iff \text{true} \\ s\ \mathcal{R}^\pm_{n+1}\ w &\iff \forall t, x.\ s = \lfloor (t,x) \rfloor \Rightarrow \exists t', y.\ w = \lfloor (t', y) \rfloor \wedge t \asymp^\pm t' \wedge (t,x)\ \mathcal{M}^\pm_n\ (t', y) \end{aligned}$$

**Related payloads** $\boxed{(t,x)\ \mathcal{M}^\pm_n\ (t',y) \quad (x \in \mathit{Payload}_t(\mathbb{D}_n),\ y \in \mathit{Payload}_{t'}(\mathbb{D}))}$

$$\begin{aligned} (c, ())\ \mathcal{M}^\pm_n\ (c, ()) &\iff \text{true} \\ (\mathit{tup}_k, (x_1, \ldots, x_k))\ \mathcal{M}^\pm_n\ (\mathit{tup}_k, (y_1, \ldots, y_k)) &\iff \forall i.\ x_i\ \mathcal{R}^\pm_n\ y_i \\ (\mathit{left}, x)\ \mathcal{M}^\pm_n\ (\mathit{left}, y) &\iff x\ \mathcal{R}^\pm_n\ y \qquad \text{(likewise } \mathit{right}\text{)} \\ (\mathit{fun}, h)\ \mathcal{M}^+_n\ (\mathit{pr}, (c, e)) &\iff \forall v \in \mathbb{D}_n,\ a \in \mathbb{D}.\ v\ \mathcal{R}^+_n\ a \Rightarrow h(v)\ \mathcal{R}^+_n\ \mathit{call}(c, e, a) \\ (\mathit{pr}, (c, e))\ \mathcal{M}^-_{m+2}\ (\mathit{fun}, g) &\iff \forall v \in \mathbb{D}_m,\ a \in \mathbb{D}.\ v\ \mathcal{R}^-_m\ a \Rightarrow \mathit{call}_m((c,e), v)\ \mathcal{R}^-_m\ g'(a) \\ (\mathit{pr}, (c, e))\ \mathcal{M}^-_n\ (\mathit{fun}, g) &\iff \text{true} \qquad \text{for } n < 2 \end{aligned}$$

**Related values** $\boxed{d \trianglelefteq^\pm w \qquad \mathit{Pres}(d, d') \qquad \mathit{Reflect}(d, d')}$

$$d \trianglelefteq^\pm w \iff \forall n.\ \pi^\infty_n(d)\ \mathcal{R}^\pm_n\ w \qquad \mathit{Pres}(d,d') \iff d \trianglelefteq^+ d' \qquad \mathit{Reflect}(d,d') \iff d' \trianglelefteq^- d$$

Fig. 7. The relations of the Scott-domain proof of **split closure**, between a source (EnvClos) value $d$ and a target (PairClos) value $d'$. The superscript ± stands for either direction. We read $s\ \mathcal{R}^\pm_n\ w$ as "$s$ and $w$ are related at stage $n$", $\mathit{Pres}(d, d')$ as "$d'$ preserves $d$", and $\mathit{Reflect}(d, d')$ as "$d$ reflects $d'$". Here $g'$ is the strict function that agrees with $g$ on defined values.

*The relations.* Figure 7 defines the relations. At stage $n + 1$, a defined approximated value $s$ is related to $w$ when $w$ is defined too, the two have related tags, and their payloads are related at stage $n$; an undefined approximated value is related to every value. Related values have the same tag, except for closures. Under preservation, a source function is related to a target pair when the function and calls of the pair, on related arguments, give related results; this clause has the shape of the relation of Nielsen [2000] for defunctionalization. Under reflection, the approximated value is the target pair, and its calls show their results two stages down. The clause for a pair at stage $m + 2$ therefore speaks of stage $m$, and it says nothing at stages 0 and 1. A source value $d$ and a target value $d'$ are related for preservation when $\pi^\infty_n(d)\ \mathcal{R}^+_n\ d'$ at every stage $n$, and for reflection when $\pi^\infty_n(d')\ \mathcal{R}^-_n\ d$ at every stage $n$.

*Properties of the stages.* The proof needs four properties of each stage relation, which hold by induction on the stage.

Lemma G.1 (Stages). *For both directions and every $n$:*

(1) *$s\ \mathcal{R}^\pm_n\ w$ is a proposition;*
(2) *it is downward closed in $s$, that is, if $s' \sqsubseteq s$ and $s\ \mathcal{R}^\pm_n\ w$, then $s'\ \mathcal{R}^\pm_n\ w$;*
(3) *it is admissible in $s$, that is, if $s_i\ \mathcal{R}^\pm_n\ w$ for every $i$ of a directed family, then $\bigsqcup_i s_i\ \mathcal{R}^\pm_n\ w$;*
(4) *it survives moving between stages, that is, if $s\ \mathcal{R}^\pm_{n+1}\ w$, then $\pi_n(s)\ \mathcal{R}^\pm_n\ w$, and if $s\ \mathcal{R}^\pm_n\ w$, then $\varepsilon_n(s)\ \mathcal{R}^\pm_{n+1}\ w$.*

*Consequently, if $s\ \mathcal{R}^\pm_n\ w$, then $\varepsilon^\infty_n(s) \trianglelefteq^\pm w$.*

The closure clauses make parts 3 and 4 the hard ones. For part 3 under reflection, $call_m$ is continuous in the pair, so the call of a supremum of pairs is the supremum of their calls. For part 4, $call_m$ commutes with $\pi$ and $\varepsilon$ up to the order of $\mathbb{D}_m$, and part 2 absorbs the difference.

*The fundamental lemma.* Write $\sim$ for either *Pres* or *Reflect*. Both relate $\bot$ to every value in the approximated position, and the operators that EnvClos and PairClos share preserve both.

Lemma G.2 (Operators). *If* $d_i \sim d_i'$ *for each* $i$, *then* $\mathcal{T}(d_1, \ldots, d_n) \sim \mathcal{T}(d_1', \ldots, d_n')$, *and likewise for* $\pi_{tup_i}$, $\mathsf{inl}$, $\mathsf{inr}$, $\mathsf{prim}_p$, *and* $\mathsf{seq}$; *constants and* $\langle\rangle$ *are related to themselves. If the branches of* $\mathsf{cases}$ *map related values to related values, then* $\mathsf{cases}$ *preserves* $\sim$.

The two languages differ only in closures and calls.

Lemma G.3 (Closures and calls). *Let* $t \sim t'$, *and let the bodies* $B$ *and* $B'$ *map related environments and arguments to related values. Then*

$$\mathsf{app}(\Lambda(\boldsymbol{\lambda} X.\, \Lambda(\boldsymbol{\lambda} Y.\, B(X, Y))), t) \sim \mathsf{pair}(\Lambda(\boldsymbol{\lambda} X.\, \Lambda(\boldsymbol{\lambda} Y.\, B'(X, Y))), t').$$

*If* $l \sim l'$ *and* $a \sim a'$, *then* $\mathsf{app}(l, a) \sim \mathsf{app}(\mathsf{app}(\mathsf{car}(l'), \mathsf{cdr}(l')), a')$.

Under preservation, both parts follow from the clause for functions and part 4 of Lemma G.1. Under reflection, the closure case uses the clause for pairs. A call of the approximation at stage $m + 2$ of the target pair is below the approximation at stage $m$ of the call of the pair itself, and part 2 of Lemma G.1 applies. In the call case, the approximation at stage $n$ of the result of a call is the supremum of the calls of the approximations of the curried function, the environment and the argument, so part 3 reduces the call to the calls of approximations, and each of those is below an approximation of a call at a larger stage, of which the clause for pairs speaks.

Lemma G.4 (Fundamental lemma). *For each direction, let* $M$ *be a term of EnvClos and let* $\rho$ *and* $\rho'$ *be environments related at every variable of* $M$. *Then* $[\![M]\!]\rho \sim [\![\boldsymbol{split}(M)]\!]\rho'$.

The proof is by induction on $M$, with the same induction for both directions. Lemma G.2 covers the shared operators, and Lemma G.3 covers the closures and calls.

Lemma G.5 (Correctness of **split closure**, Scott domains). *Let* $M$ *be a closed term of EnvClos. For every constant* $c$, $[\![M]\!]\rho_0 = \lfloor (c, ()) \rfloor$ *if and only if* $[\![\boldsymbol{split}(M)]\!]\rho_0 = \lfloor (c, ()) \rfloor$. *Moreover,* $[\![M]\!]\rho_0$ *is defined if and only if* $[\![\boldsymbol{split}(M)]\!]\rho_0$ *is defined.*

The empty environments are related, so Lemma G.4 gives that $[\![M]\!]\rho_0$ and $[\![\mathbf{split}(M)]\!]\rho_0$ are related by both *Pres* and *Reflect*. In either relation, a defined approximated value has a defined partner, and a defined value related to a constant is that constant. Preservation gives the forward implications and reflection the backward ones.